\documentclass[12pt]{article}

\newcommand{\da}{\dot{\alpha}}
\newcommand{\db}{\dot{\beta}}
\newcommand{\mc}{{\it multi-}\circ}
\newcommand{\om}{\omega}
\newcommand{\di}{{\frac{d-3}{2}}}
\newcommand{\vp}{\varphi}

\newcommand{\we}{\frac{1}{(2\pi\hbar)^d}}
\newcommand{\iww}{{{(2\pi\hbar)}^{-d}}}
\newcommand{\pd}{\pi^{\frac{d}{2}}}
\newcommand\re[1]{{(\ref{#1})}}
\newcommand{\de}{\delta}
\newcommand{\hd}{\frac d2}
\newcommand{\sg}{\sqrt{g}}
\newcommand {\cA}{{\cal A}}

\newcommand {\cC}{{\cal C}}
\newcommand {\cD}{{\cal D}}

\newcommand {\cG}{{\cal G}}

\newcommand {\cM}{{\cal M}}
\newcommand {\cN}{{\cal N}}
\newcommand {\cO}{{\cal O}}
\newcommand {\cP}{{\cal P}}

\newcommand{\p}{\psi}
\def\a{\alpha}
\def\b{\beta}
\def\c{\chi}
\def\d{\delta}
\def\e{\epsilon}
\def\f{\phi}
\def\g{\gamma}
\def\G{\Gamma}

\def\k{\kappa}
\def\l{\lambda}
\def\m{\mu}
\def\n{\nu}
\def\o{\omega}
\def\q{\theta}
\def\r{\rho}
\def\s{\sigma}
\def\t{\tau}

\def\z{\zeta}
\def\D{\Delta}

\def\L{\Lambda}

\def\P{\Pi}

\newcommand{\fr}{\frac{1}{2}}
\newcommand{\be}{\begin{equation}}
\newcommand{\ee}{\end{equation}}
\newcommand{\bea}{\begin{eqnarray}}
\newcommand{\eea}{\end{eqnarray}}

\newcommand{\ba}{\begin{array}{c}}
\newcommand{\ea}{\end{array}}

\newcommand{\ve}{\varepsilon}
\newcommand{\vf}{\varphi}

\newcommand{\ab}{{\a\b}}

\newcommand{\pa}{\partial}
\newcommand{\na}{\nabla}

\newtheorem{teo}{Proposition}
\newtheorem{lem}{Lemma}
\usepackage{latexsym}

\begin{document}
\begin{flushright}
\vspace{1mm} hep-th/0207212                                                                                                                            \\
FIAN/TD/14/02\\
July 2002\\
\end{flushright}
\vspace{10mm}

\begin{center}

{\Large \bf Conformal Higher Spin Theory} \\

\vspace{10mm} Arkady Y. Segal ${}^\dag$\\

\vspace{3mm}{$\dag$ \it I.E.Tamm Department of Theoretical Physics,
Lebedev Physical Institute,\\
Leninsky prospect 53, 119991, Moscow, Russia\\
e-mail: segal@lpi.ru}

\vspace{2mm}

\end{center}
\vspace{5mm}

\begin{abstract}
\noindent We construct gauge theory of interacting symmetric
traceless tensors of all ranks $s=0,1,2,3,...$ which generalizes
Weyl-invariant dilaton gravity to the higher spin case, in any
dimension $d>2$. The action is given by the trace of the projector
to the subspace with positive eigenvalues of an arbitrary
hermitian differential operator $\hat{H}$, and the symmetric
tensors emerge after expansion of the latter in power series in
derivatives. After decomposition in perturbative series around
conformally flat point $\hat{H}=\Box$ with Euclidean metric, the
action functional describes conformal higher spin theory. Namely,
the linear in fluctuation term cancels, while the one quadratic in
fluctuation breaks up as a sum of conformal higher spin theories,
the latter being free gauge theories of symmetric traceless
tensors of rank $s$ with actions of $d-4+2s$ order in derivatives
(in odd dimensions they are boundary terms), for all integer $s$,
introduced in $4d$ case by Fradkin and Tseytlin and studied at the
cubic order level by Fradkin and Linetsky. Higher orders in
interaction are well-defined. The action appears to be the unique
functional invariant w.r.t. general similarity transformations
$\hat{H}'=e^{\hat{\omega}^\dagger} \hat{H} e^{\hat{\omega}}$, the latter
invariance plays the role of gauge symmetry group of the model. In
the framework of the perturbative decomposition, the hermitian
part of $\omega$ gauges away the trace parts of the symmetric
tensors parameterizing the fluctuation, while the anti-hermitian
one provides standard linearized gauge transformations of
conformal higher spin fields. The action can be calculated as a
semiclassical series in $\hbar$ which counts the number of
space-time derivatives and thereby exhibits itself as a parameter
of low-energy expansion, like $\sqrt{\alpha'}$ in string theory,
in so doing the classical term is given by the volume of the
domain $H(x,p)>0$ (where $H(x,p)$ is the Weyl symbol of
$\hat{H}$), it does not contain derivatives and is interpreted as
a cosmological term. At the same time, further terms of the
$\hbar$-expansion are given by integrals of distributions
localized on the constraint surface $H(x,p) =0$, and the conformal
higher spin-$s$ action arises from the
$\hbar^{d-4+2s}$-correction. Next, full gauge invariance of the
model is interpreted as covariance algebra of generalized
Klein-Gordon equation $\hat{H}|\psi>=0$ for complex scalar field
$\psi$, and gives rise to the infinite-dimensional global symmetry
identified with the algebra of observables of the quantized point
particle. Each global symmetry produces a Noether current
constructed out of $\psi$ according to general formula we present
in the paper. In the case $\hat{H}=\Box$ the algebra of
observables is an extension of the conformal algebra decomposed
w.r.t. its adjoint action as a direct sum of finite-dimensional
representations characterized by traceless rectangular two-row
Young tableaux. This infinite-dimensional algebra coincides, in
$d=3,4,6$ with conformal higher spin algebras constructed before
in terms of even spinor oscillators, which origin is due to
twistor reformulation of the dynamics of the massless particle.
The construction of the paper may be a starting point for diverse
conjectures. First, we discuss the extension of the geometry
"quantized point particle + conformal higher spin fields in $d$
dimensions" to the one "tensionless $d-1$ brane + higher spin
massless fields in $d+1$ dimensions", where the phase of $\psi$
appear to describe transverse motion of the brane inside
$d+1$-bulk. This picture arises in the semiclassical approximation
to the quantized particle dynamics, the latter approximation
provides $d$-dimensional generalization of $W$-geometry elaborated
on by Hull in $d=1,2$ case. Next, we propose a candidate on the
role of Higgs-like higher spin compensator able to spontaneously
break higher spin symmetries. At last, we make the conjecture
that, in even dimensions $d$, the action of conformal higher spin
theory equals the logarithmically divergent term of the action of
massless higher spin fields on $AdS_{d+1}$ evaluated on the
solutions of Dirichlet-like problem, where conformal higher spin
fields are boundary values of massless higher spin fields on
$AdS_{d+1}$, the latter conjecture conforms with recent proposal
(for $d=4$) by Tseytlin and provides information on the full
higher spin action in $AdS_{d+1}$.
\end{abstract}

\begin{center}{\it A part of the results of this paper was delivered by the author \\ at
the Third International Sakharov Conference, June 22-29, 2002, Moscow}\end{center}

%
%
%

\tableofcontents

\section{Introduction and Statement of the results}

In this paper, we report on the construction of the
{\it boson conformal higher spin theory}, the gauge theory of infinite
number of symmetric traceless tensor fields of all ranks from $0$ to
$\infty$, in any dimension $d>2$. Each rank-$s$ tensor field enters the
spectrum of the theory one time.

The theory described in the paper appears to be the first example, besides
the string field theory, of a consistent lagrangian model
which includes an infinite number of symmetric tensor fields with arbitrary high rank.
In our case, however, one has the theory of symmetric fields {\it only}.
We construct, proceeding with quite
clear principles, the action of the model and show that, after
expansion into perturbative series near certain vacuum representing
conformally flat space, the quadratic part of the action breaks up as a
sum of {\it conformal higher spin} theories introduced in $4d$ case by Fradkin and
Tseytlin  \cite{FT}. These models appear to be important for diverse reasons.

For the first, conformal higher spin theories in $d$ dimensions appear to present a kind of truncation
(which may arise in a high energy limit)
of higher spin fields theories \cite{fr} in $d$ dimensions,
the latter describe (at the quadratic level) massless particles
of arbitrary spin and could present a new kind of consistent extension of supergravities with
infinite number of gauge symmetries \cite{Fradkin:ah},\cite{Fradkin:bc}.
Therefore, the construction of the consistent lagrangian theory
of conformal higher spin fields (achieved in this paper)
is an important step towards solution of the higher spin interaction problem \cite{ardeser}
(see Sec.\re{sublin1} below for a little
bit more detailed reference to the higher spin problem).

For the second,
in view of the $AdS/CFT$ correspondence \cite{mald},\cite{Liu:1998bu},\cite{sund},\cite{witt} because conformal higher spin
fields in $d$-dimensions arise as boundary
values of higher spin fields on $AdS_{d+1}$ \cite{Metsaev:2002vr},
\cite{Tseytlin:2002gz}, the latter statement generalizes
standard $AdS/CFT$ consideration of e.g. $AdS_{d+1}$ graviton with boundary values described by
$d$-dimensional conformal graviton \cite{mald},\cite{Liu:1998bu}. Therefore, studying
conformal higher spin fields in $d$ dimensions may provide information on
$AdS_{d+1}$ higher spin massless fields.

For the third, conformal higher spin fields appear to be inherently related
to the point particles.
Namely, it will be shown that general coupling of point particle to background fields is
parameterized exactly by conformal higher spin fields and, as a byproduct, point particles play a role of a source
for conformal higher spin fields. Being combined with the just mentioned $AdS/CFT$ interpretation of conformal
higher spin fields this fact
explains the matching "massless higher spin fields in $AdS_{d+1}$ $\leftrightarrow$ all bilinear conserved currents
of a free massless complex scalar field on the boundary of $AdS_{d+1}$" \cite{konsh}
which is in  core of recent conjectures on duality of a large-$N$ limit $d=4, \cN=4$ SYM theory and
tensionless $IIB$ superstring on $AdS_5\times S^5$ with RR flux \cite{sund}, \cite{witt}.
As a byproduct, conformal higher spin fields
are able to mediate interactions of point-like sources, as we show in this paper.

In the rest of this section we briefly describe our approach,
spell main results and describe the structure of the main text, which
appears to be rather extensive. The main reason we do not split the paper into smaller ones
is that all considered topics are closely connected to each other and follow
almost immediately one from another. Here is the very sketchy list of the topics covered:
classical and quantum point particle in general background fields (the latter appear to be
symmetric traceless tensors of all ranks), gauge-invariant action for background fields and
its low-energy expansion, perturbative decomposition of the action around a conformally flat vacuum and
finding that the quadratic action is a sum of conformal higher spin theories with all integer spins,
calculation of the cubic action, global symmetries and infinite-dimensional "conformal higher spin algebras"
in any dimension, all conserved currents of free complex scalar field for an arbitrary wave operator and Noether
interaction, higher spin compensator that can spontaneously break higher spin gauge symmetries, AdS/CFT correspondence
and an extension of geometry "point particle + conformal higher spin fields"  in $d$ dimensions to
"tensionless $d-1$-brane + massless higher spin fields" in $d+1$ dimensions,
interpretation of the conformal higher spin theory action in $d$ (even) dimensions as of logarithmically divergent term
of the action of higher spin massless fields in $AdS_{d+1}$ evaluated on solutions of Dirichlet like problem.

The geometrical setup of the theory springs from the
interpretation of the infinite number of symmetric tensors
\footnote{Although we speak about "tensors", actual behavior of
$H^{m_1...m_s}(x^k)$ w.r.t. $x$-diffeomorphisms presents a
deformation of standard tensor transformations, see Sec.
\re{subs42}.} \be\label{tens1}
\tilde{H}^{m_1...m_s}(x^k)\,\;s=0,1,2,3,...,\;k=0,1...,d-1 \ee as
coefficients of the Hermitian differential
operator\footnote{$\hbar$ is a real constant which serves as an
expansion parameter counting number of $x$-derivatives.}
\be\label{hath} \hat{H}=\sum \limits_{s=0}^{\infty}(-i\hbar )^s
\tilde{H}^{m_1...m_s}(x^k)\;\pa_{m_1}...\pa_{m_s},\;
\hat{H}^{\dag}=\hat{H}, \ee which acts in the linear space of
complex wave functions $\psi(x)$ and governs the generalized
Klein-Gordon equation \be\label{hpsi000} \hat{H}\psi(x)=0 \ee
Actually, the action of the model is defined  as the trace of the
projector to the subspace with positive eigenvalues of $\hat{H}$,
\be\label{theaction000} \cA[\hat{H}]=Tr \pi_+[\hat{H}] \ee or,
what is the same in some cases (e.g. for even $d$ and $\hat{H}$
with positive dilaton), as a time-independent term that appears in
the Schwinger-DeWitt-Seeley like asymptotic decomposition of the
trace of the evolution operator \be\ba\label{weylan}
\cA[\hat{H}]=\a_{0}\\ \\
Tr \exp(-\t\hat{H})=\sum \limits_{n} \t^{n} \a_n, \t \to 0^+.
\ea \ee
The action  \re{theaction000} will be shown to be gauge-invariant w.r.t. transformations
of the form
\be
\hat{H}'=e^{\hat{\o}^{\dag}}\hat{H} e^{\hat{\o}},
\ee
where $\hat{\o}$ is a general operator of the same form \re{hath} as $\hat{H}$,
\be
\hat{\o}=\sum \limits_{s=0}^{\infty}(-i\hbar)^s \o^{m_1...m_s}(x^k)\;\pa_{m_1}...\pa_{m_s}\ee
but without hermiticity restrictions and
with coefficients $\o^{m_1...m_s}(x)$ being smooth functions with a compact support in $x$-space.
The standard gauge symmetries are parameterized by the low-rank $\o$'s. Namely,
the real part of $\o$ is identified with Weyl dilations, while imaginary parts of  $\o,\;\o^m$ parameterize
$U(1)$ transformations and $x$-diffeomorphisms.
When the higher rank
fields $\tilde{H}^{m_1...m_s},\; s>2$ are set zero,
there arises a truncation of the model to a version of {\it Weyl-invariant
dilaton gravity}, in particular, in even dimensions the action is given by
the well-known expression for the time-independent term
of the Schwinger-Dewitt expansion of the trace of the heat kernel \cite{Barvinsky:an}.

When all higher rank tensors are switched on, the action
\re{theaction000} provides a gauge invariant theory of symmetric
tensors \re{tens1}. The functional \re{theaction000} is computed
by using the well known technique of symbols of operators
\cite{berezin}. Given operator $\hat{H}$ \re{hath} one introduces
its Weyl symbol \be H(x,p)=\sum \limits_{s=0}^{\infty}
H^{m_1...m_s}(x^k)\;p_{m_1}...p_{m_s},\; \ee where
$H^{m_1...m_s}(x^k)$ are related to the components of the operator
\re{hath} as
$\tilde{H}^{m_1...m_s}(x^k)=H^{m_1...m_s}(x^k)+o_s(\hbar \pa H,
\hbar^2 \pa^2 H, ...)$, (see App.\re{operweyl} for the exact
formula). Then any expression built from $\hat{H}$ is rewritten in
terms of its Weyl symbol $H(x,p)$. Making use of the Weyl symbol
formalism the action functional \re{theaction000} can be computed
in a form of semiclassical series like \be\label{quasi000} Tr
\pi_+(\hat{H})= (2\pi\hbar)^{-d} \int d^d x \left[\sum
\limits_{k=0}^{\infty} \hbar^{2k} L_{2k}(H, \pa H, \pa^2 H,
...\pa^{2k} H)\right] \ee where $L_{2k}(H, \pa H, \pa^2 H,
...\pa^{2k} H)$ are expressions built from the components of the
tensor fields $H^{m_1...m_s}$ and their $x$-derivatives up to
order $2k$, with total degree in $x$-derivatives $2k$. Therefore,
semiclassical expansion exhibits itself as a low-energy one, with
$\hbar^2$ playing the role similar to that $\a'$ plays in string
theory. The zeroth, "classical", term of the $\hbar$-expansion
does not contain $x$-derivatives and exhibits itself as a
cosmological term. It appears the same term has an interpretation
of the volume of the domain \be H(x,p)\geq 0 \ee in the phase
space of the classical particle in $d$-dimensions, while
$\hbar^{2k}$-terms with $k>0$, have a form of phase space
integrals of densities localized on the "constraint surface" \be
H(x,p)=0 \ee On the other hand, perturbative expansion of the
functional \re{theaction000} around a particular configuration
representing the flat space, \be\label{thevacuum000}
\hat{H}=\hbar^2 \Box \Leftrightarrow H=-p^2, \ee where $\Box$ and
$p^2$ are constructed with flat Euclidean metric, appears to
possess the following properties:

1) term linear in fluctuation cancels for $d>2$, therefore in
$d>2$ \re{thevacuum000} is a solution of the classical equations
of motion of the action \re{theaction000};

2) term quadratic in fluctuation breaks up as a sum of conformal
higher spin theories with conformal spin $s=0,1,2,3,...$, which
were introduced, in $d=4$ case, by Fradkin and Tseytlin \cite{FT}
and studied by Fradkin and Linetsky \cite{linetsky}; with the
actions of the form \be \label{defac456}\cA_s[\vf_s]=\int d^d x
\;\vp^{a(s)}\,\Box^{\hd-2}\, P_{a(s)b(s)} (\pa) \,\vp^{b(s)}, \ee
where $\vp^{a(s)}$ are symmetric traceless tensors made of the
fluctuations of $\hat{H}$ around the vacuum \re{thevacuum000},
while $P_{a(s)b(s)}(\pa)$ is a Poincar\'e-invariant differential
operator of order $2s$ in derivatives, satisfying set of
identities \be \label{P04567}
P_{a(s)b(s)}=P_{b(s)a(s)}\;\;,\;\;{P^c}_{ca(s-2)b(s)}=0\;\;,\;\;
P_{a(s-1)c b(s)}\, \pa^c =0. \ee

3) all higher orders in the fluctuation are well-defined and can
be calculated explicitly. Therefore we interpret the functional
\re{theaction000} as the action of  conformal higher spin theory.
General structure of the vertices is \be \cA \sim \int d^dx\; \sum
\limits_{l=0}^{\infty}  e^l (\hbar\pa)^{d-2l+\sum
\limits_{i=0}^{l} s_i}  \vf_{s_1}...\vf_{s_l}, \ee where partial
derivatives act in a diverse way on the rank-$s_1...s_l$
components of the fluctuation. The summation runs only over that
subset of indices for which the degree of $\hbar\pa$ is
non-negative.

Besides,

4) For any signature of the metric $\eta_{mn}=\mbox{diag}(d-q,q)$,
which enters \re{thevacuum000}, the theory possesses an
infinite-dimensional global symmetry. Namely, the algebra of
observables of the quantized particle being defined as a factor
algebra of the subalgebra of gauge symmetries that preserves the
vacuum \re{thevacuum000}, \be\label{omega000} \hat{\o}^{\dag}
\hat{H} +\hat{H} \o =0 \ee by the ideal of trivial symmetries of
the form \be\label{omega0001} \hat{\o}_{triv} = i\hat{\m}
\hat{H},\;\hat{\m}=\hat{\m}^{\dag}, \ee is, in the case
\re{thevacuum000}, an infinite-dimensional {\it conformal higher
spin algebra $chs(d-q,q)$}, which includes the conformal algebra
$so(d-q+1,q+1)$ as a finite-dimensional subalgebra. For $d=3,4,6$
and Minkowski signature, $chs(d-1,1)$ appear to coincide with
boson higher spin algebras introduced before
\cite{linetsky},\cite{Vasiliev:2001ur}, \cite{Sezgin:2001ij} in
terms of even spinor oscillators, the latter appear to arise in
our construction via twistor reformulation of the massless
particle dynamics in terms of Dirac spinor of the conformal
algebra $so(d,2)$.

The global symmetry $chs(d-q,q)$
acts on fluctuations of $\hat{H}$ linearly,
hence each order of the perturbative
expansion of the action of the theory is invariant w.r.t. infinite-dimensional global symmetry $chs(d-q,q)$.

To each element of the algebra of observables a Noether current is assigned. We find the expression
for all Noether currents of the complex scalar field $\psi(x)$ for an arbitrary background
wave operator $\hat{H}$ provided its algebra of observables is known. The expression reads
\be \ba\label{noet233}
J_\o^m=-\fr \sum \limits_{s=0}^{\infty}\sum \limits_{k=0}^{s}\sum \limits_{l=1}^{s-k} \\ \\
 \left\{\frac{(-i\hbar)^s  (-)^{l-1}}{2^k}C^k_s C^l_{s-k}
 \pa_{m_1}...\pa_{m_{l-1}}\left[\psi^* H^{m(k)m(s-k)}_{,m(k)}(x)  (\hat{\o}\psi)_{,m(s-k-l)}\right]\right\}+c.c.,
\ea\ee
where $C^k_s$ are binomial coefficients and $H^{m(s)}(x)$ are coefficients of the decomposition of the Weyl
symbol of the operator $\hat{H}$ in power series in momenta. In the case \re{thevacuum000} our
set of Noether currents coincides, up to a choice of a basis, with the one obtained by the authors
of \cite{konsh} by direct calculation.

The exposure of the main results of the paper has appeared possible due to the extensive use
of the notion of a {\it geometry with a source} and of its {\it covariance algebra}.
Let $O$ be a localized object in $d$-dimensions, whose dynamics is parameterized by a set of variables $o$, and
let $S_H[o]$ be an action that governs the dynamics of $o$, while $H$ in the subscript stands for the set of all
possible functionals of $o$ that determine general form of the action. By definition,
the algebra $\cG$ of covariance transformations consists of all infinitesimal transformations $g$
that act simultaneously on $o$ and on the set of $H$ in such a way that the
action is invariant modulo terms which do not affect classical equations of motion,
\be
S_{H+\d_g H}[o+\d_g o]=S_H[o] +\{...\},
\ee
where $\{...\}$ stands for terms which do not affect classical equations of motion.
Equivalently, the variation of the action due to change of the functionals $H[o]$ is {\it compensated} by
transferring to the new variables $o'(o)$.
We say the last formula determines  {\it the geometry with the source (or ``test object'') $O$,
geometric structure (or ``background field'') $H[o]$, coupling $S_H[o]$, and the covariance algebra $\cG$.}

Given a geometry with a source, one interprets covariance transformations
for $H$ as gauge transformations and looks for a gauge-invariant functional of $H$
\be\label{inducedac}
\cA[H+\d_g H]=\cA[H]+\{...\},
\ee
where $\{...\}$ stands for terms which do not affect classical equations of motion.
Let us call any solution of the last equation as {\it invariant induced action}
\footnote{The relation of invariant induced action
to the ordinary induced action which looks like $-\fr Tr Ln \hat{H}$ will become clear from
the below treatment.}.

In the case of conformal higher spin theory,
$\cO$ will be the {\it quantized point particle},
$H=\hat{H}$ be an operator \re{hath} governing
evolution of the wave function of the particle by means of the wave equation \re{hpsi000}.
The coupling is provided by the quadratic action
\be\label{coupling09}
S_H[\psi]=  \int d^d x \;\psi^*(x)\hat{H} \psi(x),
\ee
while the covariance algebra consist of all transformations of the form
\be\label{omegah000}
\d \hat{H}=\hat{\o}^{\dag} \hat{H} +\hat{H} \hat{\o},
\ee
which act on the wave function by the rule
\be
\d \psi(x)=-\hat{\o}\psi(x)
\ee
The main result of this paper, the invariant induced action $\cA[H]$ has a form \re{theaction000}, and
admits the formal path-integral representation
\be
\cA[H]=Tr \theta(H*)= \int \cD x \cD p  \frac{\cD \l}{\l} \d(\dot{\l})
\exp \{\frac{i}{\hbar} S_H[x,p,\l]\},
\ee
where $S_H[x,p,\l]$ is Hamiltonian action of the classical particle,
\be
S_H[x,p,\l]=\int d\t \left(p_m \dot{x}^m - \l  H(x,p)\right),
\ee
where $(x^m(\t), p_m(\t), \l(\t))$ is the set of particle's phase-space variables
which include the Lagrange multiplier $\l$ to the first-class constraint $H(x,p)$ being a Weyl symbol of
the operator $\hat{H}$.
The integration contour for $\l$ is not that providing the one-loop effective action $-\fr\,Tr Ln \hat{H}$, $\l \in [\m, \infty)$
but another one,
\begin{equation}  \label{optionC}
\lambda(\tau) =\tau -i\varepsilon ;\;\tau \in {\bf R},\;
\varepsilon \rightarrow 0^+.
\end{equation}

Thus the conformal higher spin fields exhibit themselves as background fields of the
point particle. The united action "invariant induced action+coupling",
\be
\cA[\hat{H},\psi]=Tr \;\pi_+[\hat{H}]+
e\int d^d x \;\psi^*(x)\hat{H} \psi(x),
\ee
where $e$ is  a coupling constant describes a joint dynamics of
conformal higher spin fields and quantized point particles. By construction
$\cA[\hat{H},\psi]$ is gauge-invariant w.r.t. covariance transformations \re{omegah000}.
A vacuum is provided by the configuration
\be
\hat{H}=\hbar^2\Box,\;\psi(x)=0,
\ee
which is preserved by the infinite-dimensional global symmetry $chs(d,0)$ (\ref{omega000},
\ref{omega0001},\ref{thevacuum000}). After analytic continuation
to Minkowski space, fluctuations around this vacuum describe interaction of sources built from complex massless field (quantized particle)
by means of exchange by conformal higher spin fields, interactions due to exchange by $s=1$ quanta in $d=4$ are
familiar Maxwell interactions of charged matter.

This $d$-dimensional system possesses classical conformal invariance and may get $AdS_{d+1}$ dual interpretation.
Without doubt, conformal higher spin fields are interpreted as boundary values of massless
higher spin fields in $AdS_{d+1}$ (see Sec. \re{adscft}). On the other hand, we
show that in a particular low-energy limit the theory of the complex scalar field $\psi(x)$
exhibits itself as a theory of a {\it tensionless $d-1$-brane} in $AdS_{d+1}$, where
the phase of the wave function is a field that describes position of the $d$-dimensional world volume inside
a $d+1$-dimensional bulk. Thus, we conjecture
there exists an extension of the geometry "conformal higher spin fields+point particle in $d$ dimensions"
to the geometry "massless higher spin fields + tensionless $d-1$-brane in $d+1$ dimensions".

We also show how one could spontaneously break higher spin gauge symmetries of conformal higher spin models
by finding a natural candidate on the role of compensator of gauge transformations \re{omegah000}.

Let us describe the structure of the paper along with providing more comments.

Lengthy sections \ref{slin} and \ref{stech} are preliminary ones.

In Sec. \ref{slin}, we recall the "initial data" the full model is able to reproduce, namely, the
Weyl-invariant dilaton gravity and linearized conformal higher spin theories. Also we describe free
Fronsdal theories of higher spin massless fields and show how conformal higher spin theories may arise
either as a high-energy truncation of Fronsdal models or as holographic images of the latter via $AdS/CFT$.

In Sec. \ref{stech}, we provide another preliminary data which we call
"deformed conformal higher spin theories", the latter being  free Poincar\'e-invariant models constructed out of infinite
set of symmetric tensors, they appear to be related to free conformal higher spin theories
by a field redefinition. In our treatment, the latter models
will always arise as deformed conformal higher spin theories.

In Sec. \ref{sgeom}, we specify geometry that will govern our consideration in the paper and describe in detail
the geometry of classical point particle in general background fields, its
quantum deformation and semiclassical picture of the quantum geometry, with the coupling
provided by a "Hamilton-Jacobi" sigma-model
which appear to be a
$d$-dimensional generalization of $W$-geometry sigma-model introduced by Hull \cite{Hull:1992vj}. In the end of the section,
we prove an important theorem that
states that once one has constructed an invariant induced action with the vacuum of the form
\re{thevacuum000} the quadratic expansion
of the invariant induced action around this vacuum should be a direct sum of conformal higher spin theories with a special value of
the deformation parameter.

In Sec. \ref{secaction}, we find a functional of background fields which is invariant w.r.t. covariance transformations, i.e.
an invariant induced action. We show it can be expanded in a semiclassical ("low-energy") series
\re{quasi000} and
ground its interpretation of the "quantized volume" of a domain in the phase space being
bounded by the constraint surface.

In Sec. \ref{srest}, we show that in the case higher rank fields (with rank $s>2$) are zero,
the "quantized volume" action reproduces the Weyl-invariant dilaton  gravity. We also
compare our calculations with the ones that follow from Schwinger-Dewitt method of evaluating traces of
integral kernels of evolution operators, find the precise agreement, and formulate the
conjecture that, in even dimensions, our action is nothing but the time-independent term that appear in the
asymptotic decomposition of the trace of the heat kernel of evolution operator $\exp (i\t \hat{H})$.
This conjecture is in accord with recent reasoning by Tseytlin \cite{Tseytlin:2002gz} which is
$d=4$, quadratic in fluctuation of $\hat{H}$ version of our one.

In Sec. \ref{scfv}, we show the configuration \re{thevacuum000} is a vacuum solution of the equations of motions for the "quantized volume"
action. Therefore, by the theorem of Sec. \re{sgeom}, the quadratic term of the expansion of the action around
this vacuum should be a sum of conformal higher spin theories.

In Sec. \ref{corollary}, we show by direct calculation that the quadratic term of the perturbative expansion of the "quantized volume" action
around the vacuum \re{thevacuum000} is nothing but the sum of conformal higher spin theories with conformal spins $s=0,1,2,3,...$ arising
in the form of "deformed conformal higher spin models" with a special deformation parameter. The result of this
calculation is the main technical result of the paper. Furthermore, we calculate the cubic term of the perturbative expansion
of the invariant induced action.

In Sec. \ref{sequat}, we point out once again the conformity of point particles and conformal higher spin fields, and
outline the procedure of finding conformal higher spin interactions of point particles.

In Sec. \ref{sgl}, we study algebra of global symmetries of the model. We prove an important lemma
that the representation of the conformal algebra on
higher spin fields is diagonalized w.r.t. conformal spin and therefore each "conformal higher spin-$s$"
model is conformally invariant indeed. Then we identify algebra of global symmetries of the model
with the algebra of observables of
a quantized particle. With each global symmetry a Noether current is associated, being a bilinear in the particle's
wave function. We find general expression
for all Noether currents of a complex scalar field for any wave operator $\hat{H}$.
In the case $\hat{H}=-\hbar^2\Box$ the algebra of observables $chs(d-q,q)$ contains conformal algebra
$so(d-q+1,q+1)$ while the full algebra of observables is decomposed w.r.t. conformal algebra as a sum of
finite-dimensional representations described by traceless two-row rectangular Young tables, which is exactly
the structure anticipated from the comparison with description of free massless fields on $d+1$-dimensional
anti-de Sitter space.
We conclude we have constructed boson conformal higher spin algebras in an arbitrary dimension $d$, or,
what is the same, the higher spin algebras in dimension $d+1$ (previously,
they were known for Minkowski signature and dimensions (if $d>2$) $d=3,4,6$ only).
We also exhibit the classical limit of the algebra of observables and of the Noether
currents.

In Sec. \ref{sNint}, we show that Noether  interaction "Noether current-gauge field" of the complex scalar field to
one-forms which take values in the algebra of its global symmetries, is a superfluous formulation of the interaction
"$\int d^d x\, \psi^* \hat{h} \psi$," where $\hat{h}$ is a general fluctuation of the wave operator,
expressed via set of one-forms  by certain projection, which in fact
expresses conformal higher spin fields via one-forms of
"gauge description" of conformal higher spin fields introduced by Fradkin and Linetsky in the Vasiliev's-like approach \cite{linetsky}.

In Sec. \ref{stwist}, we show that the previously known
conformal higher spin algebras in $d=3,4,6$ (=higher spin algebras
in $d+1=4,5,7$) constructed in terms of power series of even spinor oscillators, do coincide with $chs(d-1,1)$,
while the origin of even spinor oscillators is precisely due to the twistor reformulation (available in $d=3,4,6,\, q=1$)
of the $d$-dimensional massless particle dynamics in terms of
Dirac spinor of the conformal algebra $so(d,2)$.

In Sec. \ref{spec}, we speculate on two diverse subjects, each being related to an extension of the conformal higher
spin theory.

In subsec. \ref{spec1}, we propose a candidate on the role of higher spin compensator which
is a Higgs-like object able to spontaneously break higher spin gauge invariance down to low spin
algebra consisting of $x$-diffeomorphisms only, thus we
present a first step to the program of breaking higher
spin symmetries discussed in \cite{witt} .

In subsect. \ref{spec2}, we
show that the semiclassical approximation to the action of the complex scalar field, having a form of
"Hamilton-Jacobi" sigma model of Sec. \ref{sgeom}, is interpreted as the action of tensionless
$d-1$-brane in $d+1$ dimensions, with the phase of the complex scalar field playing the role of the
transverse position of the $d-1$-brane world volume inside a $d+1$-bulk. Then we speculate on the
existence of $d+1$-dimensional extension of the picture
"conformal higher spin fields+quantized point particle in $d$ dimensions" to a system "massless higher spin fields
 + tensionless $d-1$-brane in $d+1$ dimensions.

In Conclusion, we discuss some open problems and perspectives.

In Appendix A we collect general facts on perturbative expansion in gauge theories.
In App. B, the equations of motions are analyzed in the formal operator approach.
App. C contains machinery  for handling with infinite collections of symmetric tensors and
extracting their traceless parts. App. D provides generating functions for the integrals
either over a ball or over $d-1$-spheres in $d$-dimensions. App. E contains necessary formulae
on Bessel's functions. App. F deals with fields redefinitions that allow to pass from
coefficients of the Weyl symbol of the low-spin wave operator to familiar
tensor fields of Weyl-invariant dilaton gravity. App. G contains "two-time" description of
massless scalar in $d$ dimensions in terms of a field on the light-cone in $d+2$ dimensions, and
calculation of the special operator identity valid for generators of conformal group in this representation.
App. H provides the expression for the differential operator with a given Weyl symbol being
a power series in momenta. App. I contains Weyl spinor formalism in six dimensions.

On our notation: whenever the indices denoted by the same letter appear their full symmetrization is implied (the full
symmetrization is a projector), the symbols like $m(s)$ stand for $m_1...m_s$. We also use
sometimes the notation $AB$ or $(AB)$
for the contraction $A^k B_k$ of  vector $A^k$ and
covector $B_k$.

Let us make a remark on bibliography. The literature on higher spin problem is very extensive, therefore we did not
try to present an exhaustive citation of any kind, but rather have drawn  most immediate links. The author
apologizes for incomplete or missed citations, and would appreciate criticism on this point.

\section{Low spin data. Linearized data}\label{slin}
The theory we construct in the paper possesses ``initial data'' of two kinds.

One set of initial data is provided by the
low-spin truncation, when one sets zero all higher rank tensors (with rank $s>2$).
In this case the theory reduces to the Weyl-invariant dilaton-Maxwell gravity with cosmological term
given by the dilaton's potential.

Another set of initial data is provided by free theories of ``conformal higher spin fields'' on
Minkowski space (or its conformal completion), they are gauge theories
of symmetric {\it traceless} tensors, where for each given rank $s=1,2,3...$ the theory possesses gauge invariance
w.r.t. transformations with a parameter being  a symmetric {\it traceless} tensor of rank $s-1$. In $d=4$,
these free models were described by Fradkin and Tseytlin \cite{FT} ("pure
spin" models) in the Lorentzian signature case, and were conjectured to be
invariant w.r.t.  full conformal algebra $so(4,2)$. They possess supersymmetric
extensions, studied by Fradkin and Linetsky \cite{linetsky}.
The ``pure spin''-$s$ theory is described by an action with $2s$ derivatives, up to arbitrariness in multiplying the
wave operator by an arbitrary function of $\Box$, $f(\Box)$.
It appears $4d$ models may be easily generalized to arbitrary dimension and, as we show in the paper,
for any $d$ there exists the choice of $f(\Box)$ when the ``conformal higher spin theories'' are conformally
invariant indeed.

One of the main messages of the paper is that these
two types of data are nothing but different limits of one theory. After constructing the full theory in Sec.
\re{action111} we will recover both sets of data by direct calculations.
Below we describe them in more detail.

\subsection{Low spin data. Weyl invariant dilaton  gravity}\label{weylg}
The theory we are going to construct in the paper may be viewed as a higher spin extension of the Weyl-invariant
dilaton gravity, the latter is formulated in terms of low spin symmetric tensors
$D(x), A_m(x), g_{mn}(x)$. The action is required to be invariant w.r.t.
gauge transformations
\be\ba  \label{qtrans25}
\delta _\varepsilon D =-\xi ^m \partial _m \vf
+2 \a D \nonumber
\\ \\
\delta _\varepsilon A_m =-\xi ^k \partial
_k A_m -A_k \partial _m \xi ^k +\partial _m \ve, \\ \\
\delta _\varepsilon g^{m n }=-\xi ^k \partial _k g^{m n
}+g^{m k }\partial _k \xi ^n +g^{n k }\partial _k \xi ^m +2\a g^{mn}
\nonumber \\
\ea\ee
where $\xi^m,\a,\ve$ parameterize diffeomorphisms, Weyl dilations and $U(1)$ ``gradient'' transformations,
respectively. The inverse metric $g^{mn}$ and the dilaton $D$ are subject to Weyl dilations with Weyl weights equal to $2$,
while $A_m$ is Weyl-invariant.

Below we analyze
general action invariant w.r.t.
these gauge transformations, as in linearized approximation
around a flat background as at the full nonlinear level, to exhibit structures that will arise from
the action of the conformal higher spin theory.

Let us describe general structure of invariant action.
It has a form like
\footnote{We do not pursue complete classification of all possible invariant actions but rather
draw general picture of possible structures. This is sufficient for our purposes in the paper.}
\be\ba\label{weylgrav}
\cA[D,A,g]=w_d
\int d^d x \sg \left\{\g_{1,0}\; D^{\hd}+\right.\\ \\
+\g_{1,1} \;\left(D^{\hd-3} g^{mn} D_m D_n +\frac{4}{(d-1)(d-2)}D^{\hd-1} R \right)+
\ea \ee
$$\ba +
\sum \limits_{k=2}^{\infty}[\g_{1,k}\; D^{\hd-1-k} (\Box^{k}+ o_{1,k}(R))] D +
\\ \\+ \left. g^{mk} g^{nl} F_{mn}
\sum \limits_{k=0}^{\infty}[\g_{2k} \;D^{\hd-2-k} (\Box^{k}+ o_{2,k}(R))] F_{kl} +  \right. \\ \\ \left.
+ C_{mnkl}\sum \limits_{k=0}^{\infty}[\g_{3,k}\; D^{\hd-2-k} (\Box^{k}+ o_{3,k}(R))] C^{mnkl}  +...\right\},
\ea
$$
where $w_d, \g_{1,k}, \g_{2,k}, \g_{3,k}, k=0,1,2,3,...$ are constants, $F_{mn}=\pa_m A_n -\pa_n A_m$ is the Maxwell field strength, $R$ is scalar curvature,
$C_{mnkl}$ is the Weyl tensor, and $"+..."$ stands for all other Weyl-invariant scalars built from $D, A_m, g_{mn}$, in particular,
terms of higher degree in curvatures $F_{mn},C_{mnkl}$.
$\Box$ is the covariant d'Alambert operator which scales as $\d \Box =2\a \Box$ w.r.t. Weyl dilations with constant $\a$.
Whenever they exist, $(\Box^k+ o_{1,k;2,k;3,k}(R))$
denote Weyl-covariant operators of Weyl weight $2k$,
where $o(R)$'s stand for appropriate curvature corrections which scale by the same law as $\Box^k$ w.r.t. Weyl dilations
with a constant parameter, like
\be \label{ottt}
o_k \sim  \sum \limits_{l=0}^{k-1} (g^{mn})^{k+l}(\na_r)^{2k-2l} (R_{efgh})^l , \ee
where upper and lower indices are to be contracted in diverse way.
Note that as the scalar curvature $R$ scales by the same law as
$\Box$ w.r.t. Weyl dilations with constant parameter, $\d R =2\a r, \a=const$,
$o_k(R)$ is allowed to contain degrees of $R$ of only $k$-th order. This remark plays an important role below.

The degrees of $D$ and $\Box$ in \re{weylgrav} are controlled by the requirement that Lagrangian should have zero Weyl weight,
therefore the total degree of $D$ and $\Box$ in square brackets in the third row of \re{weylgrav} is equal to $\hd-1$ and in the
fourth and fifth row is $\hd -2$.
Therefore, in even dimensions, the action \re{weylgrav} contains only integer powers of $D$, while for odd dimensions the action \re{weylgrav}
contains only half-integer powers of $D$. It should be noted that
the model that arises as a low spin truncation of the full theory of the
paper possesses, in even dimensions, only non-negative powers of $D$.

The first term $\g_{1,0}(...)$ plays the role of cosmological term.
The term $\g_{1,1}(...)$ is the standard Weyl-covariant coupling of dilaton to gravity.
Indeed, this may be checked either by direct calculation or by making change of variables
\be D=\vf^{\frac{4}{d-2}},\ee that leads
to
\be  \ba
\int d^d x \sg \left\{\g_{1,0} D^{\hd}+\g_{1,1} \left(D^{\hd-3} g^{mn} D_m D_n +\frac{4}{(d-1)(d-2)}D^{\hd-1} R \right)\right\}=\\ \\
=\int d^d x \sg \left\{\g_{1,0}\vf^{\frac{2d}{d-2}}
+\g_{1,1}  (\frac{4}{d-2})^2\left(g^{mn} \vf_m \vf_n +\frac{d-2}{4(d-1)}
R \vf^2\right)\right\}
\ea \ee
which is the more conventional form of coupling of a conformal scalar to gravity \cite{bireldav}.
Note also that the well-known conformally -invariant potential for the dilaton $\vf$ plays the role of
cosmological term.

Let us
study expansion of the general action  \re{weylgrav} around the following configuration of dynamical variables:
\be\label{b11111}
D=\bar{D}=const,\;A_m=0,\;g_{mn}=\eta_{mn},
\ee
where $\eta_{mn}$ is the Minkowski metric and show that the case $\bar{D}=0$
corresponds, in $d>2$, to a solution of equations of motion of general action \re{weylgrav}.
Introduce fluctuations $h,h^m,h^{mn}$ of $D, A^m, g^{mn}$ around the background
\re{b11111} and study the first, linear in fluctuation, term in the expansion of the action \re{weylgrav}.
Let us restrict to the case when only non-negative powers of $\Box$ enter the action
(that will always be the case in our treatment) . Then,
as the metric is flat, all background curvatures equal zero. As dilaton is constant, $\Box^k {\bar D}=0$, the same is true for
all terms of the sort $\pa_m...\pa_m {\bar D}$.
Thus, the second and the third rows of \re{weylgrav} give zero contributions to the linear action.
The only combinations surviving at the linear level
come from the first row of \re{weylgrav}: from $k=0$ term  (the cosmological term)
\be
w_d\int d^d x \; (\sg \bar{D}^{\hd}) =
w_d\int d^d x \;  \left(\fr {h^m}_m \bar{D}^{\hd} + \hd \bar{D}^{\hd-1} h \right) +
O(h,h^m, h^{mn})
\ee
and from $k\neq 0$ terms of the form
\be
w_d\int d^d x \;\bar D^{\hd-k} \b_k r(h_{mn})+O(h,h^m, h^{mn})
 \ee
where $\fr {h^m}_m$ is the fluctuation of $\sg$,
$\b_k$ is the multiple of $R$-linear term in $o_{1k}$,
and $r(h_{mn})$ is the fluctuation of the scalar curvature.
Recalling the remark after the Eq. \re{ottt} one proves that
\be
\b_k=0, k>1.
\ee
Therefore,
the linear part of the action vanishes for $d>2$ provided  \be {\bar D}=0.\ee
This means that in $d>2$, the background configuration
\be\label{b22222}
D=0,\;A_m=0,\;g_{mn}=\eta_{mn},
\ee
is a vacuum, i.e. a solution of the classical equations of motion of the action \re{weylgrav}.

Useful information may be obtained by studying
the linearized gauge transformations. To this end one has to rewrite the gauge transformations
\re{qtrans25}
in terms of fluctuations and keep only terms of zero order in fluctuations.
One gets
\be\ba
\d h^{mn} = -\pa^m \xi^n -\pa^n \xi^m -2\a {\bar g}^{mn} \\ \\
\d h^m = \pa^m \ve\\ \\
\d h = 0
\ea\ee
As it is  recalled in App. \re{ApA}, after expansion around the vacuum \re{b22222},
the quadratic part of the action is
invariant w.r.t. linearized gauge transformations.

The scalar fluctuation $h$ is Weyl-inert, and Weyl transformations reside in the
rank-$2$ sector, where they can be used to set $h^{mn}$ traceless,
\be
{h^m}_m=0.
\ee
The remaining gauge transformations are
\be\ba
\d h^{mn} = -\mbox{Traceless part of} (\pa^m \xi^n +\pa^n \xi^m )\\ \\
\d h^m = \pa^m \ve\\ \\
\d h = 0.
\ea\ee
This means the quadratic action should be a sum of linearized {\it conformal gravity} + Maxwell theory+
scalar theory, where the action of conformal gravity is built from the linearized {\it Weyl tensor} and its derivatives.
Thus the general structure of the quadratic action in this case should be
\be
\cA_2[h,h^m,h^{mn}]=\int d^d x \left\{h f_1(\Box) h + F_{mn} f_2 (\Box) F_{mn} + c_{mnkl} f_3 (\Box) c^{mnkl} +...\right\},
\ee
where $f_{1,2,3}$ are some functions of $\Box$, $F_{mn}=\pa_m h_n -\pa_n h_m$,
$c_{mnkl}$ is the linearized Weyl tensor. This is confirmed by studying
 the quadratic terms in the perturbative expansion of the action \re{weylgrav} around the vacuum \re{b22222}.
The only nonvanishing terms come from $k=\hd -2$ in the first, second and the third rows of \re{weylgrav},
Therefore, around the vacuum \re{b22222}, the quadratic part of the action \re{weylgrav} reads
\be\label{cs012}
\cA_2[h,h^m,h^{mn}]=\int d^d x \left\{\g_{1,\hd-2} h \Box^{\hd-2} h + \g_{2,\hd-2} F_{mn}  \Box^{\hd-2} F_{mn} +
\g_{3,\hd-2} c_{mnkl} \Box^{\hd-2} c^{mnkl}\right\},
\ee
provided the dimension $d$ is even, while in odd dimensions $\cA_2$
is zero. Nevertheless, the formula \re{cs012} may be considered as a definition
which is valid in odd dimensions either, if one defines the nonlocal operators
$\Box^{\frac{k}{2}},\, k=1,3,5,...$ to be antisymmetric in the sense
  \be\label{antod}
\int d^dx\; A \Box^{\frac{k}{2}} B =-\int d^dx\; B \Box^{\frac{k}{2}} A
+\mbox{boundary terms},
\ee
then the formula \re{cs012} provides a boundary term.

The $h,h^m,h^{mn}$-terms are what we call {\it conformal spin-$0,1,2$ theories in $d$ dimensions}, respectively.
In $d=4$, spin-$1$ and spin-$2$ theories are the linearized Maxwell and Weyl gravity actions.
Note also that, in $d=4$, the scalar $h$ is auxiliary. At the quadratic level $h=0$ on-shell, while at
nonlinear level $h$ is expressed via the higher rank fields and their derivatives.

Let us make a remark. The general action \re{weylgrav} is
constructed in such a way that the Weyl weight of the monomials in
curvatures $F_{mn},C_{mnkl}$ is compensated either by the Weyl
weight of the powers of the dilaton $D$ or by the Weyl weight of
the operators which contain covariant derivatives, like
$\Box^k+o_k(R)$. These are the latter "derivative" terms that
survive in the limit $D\to 0$ and which expansion gives rise to
the quadratic conformal theories \re{cs012}. In particular, the
$s=0$ quadratic action comes not from the "conventional dilaton's
kinetic term" $k=1$ in the first row of \re{weylgrav} but rather
from $k=\hd-2$ one (the only case when these terms coincide is
$d=6$). On the other hand, let us redefine the dilaton as
\be\label{red654} D(x)=\f^l(x). \ee This change of dynamical
variables is degenerate at the vacuum point $D=0$, therefore, the
quadratic theories of the fluctuations of $D$ and $\f$ are in
general non-equivalent. The mechanism that ensures the lack of
equivalence at the quadratic level is that, for different $l$, the
quadratic actions for the fluctuations of $\f$ originate from
different terms in \re{weylgrav}. In fact, the terms which give
the non-vanishing contribution to the quadratic action of the
fluctuation of $\f$ are those with $k=\hd-\frac{2}{l}$, then the
quadratic action is proportional to $\int d^d x \f \Box^{k}\f$. In
particular, for the redefinition \re{red654} with
$l=\frac{4}{d-2}$ one has $k=1$ and the "conventional" action of
the massless scalar.

The last issue we would like to discuss in this section is the global invariance.
Usually the global symmetries arise as a subgroup of gauge transformations preserving the vacuum.
In the case of gauge transformations \re{qtrans25} and the vacuum \re{b22222}, the global transformations are
{\it conformal transformations},
The third row is the definition of the {\it conformal Killing vector} $\xi^m$, with general solution of the form
\be\ba\label{glob0135}
\xi^a(x)= b^a + 2 b^{ab} x_b + c x^a + x^2 c^a-2 x^a  x^b c_b,\;
\a= -c+2c^a x_a,
\ea\ee
where $b^a, b^{ab}=-b^{ba}, c, c ^a$ parameterize Poincar\'e translations, Lorentz rotations, dilations and
special conformal transformations, which altogether generate the Lie algebra
$so(d-q
+1,q+1)$. These transformations
do not change the vacuum \re{b22222} because  dilation of the flat metric due to diffeomorphisms is compensated by
a Weyl dilation.

\subsection{Linearized data.}\label{sublin1}

\subsubsection{Free conformal higher spin theories}\label{fchst}
After the perturbative expansion over a conformally flat vacuum, the quadratic part of the action of
our model breaks down as theory of free scalar plus a sum of {\it conformal higher spin} actions for conformal spin
$s=1,2,3,...$

Let us describe these actions in more detail.

Given $s=1,2,3,...$ consider symmetric {\it traceless} tensor field
\be \vf^{a(s)}(x) \;\;,\;\; \vf^{a(s-2)b}{}_b=0 \ee on a $d$-dimensional
flat space ${\bf R}^{d-q,q}$, equipped with a flat metric $\eta_{ab}$.
Consider gauge transformations \be \label{gtr1128}\d \vf^{a(s)}=
\mbox{Traceless part of} \;\partial^a \ve^{a(s-1)}(x), \ee where
$\ve^{a(s-1)}(x)$ is an arbitrary smooth {\it traceless} tensor field with
a compact support, and wonder if there exists a gauge-invariant and
Poincar\'e-invariant quadratic action $\cA_s[\vf_s]$.  The answer is yes,
and the action has the form \be \label{defac}\cA_s[\vf_s]=\int d^d x \vf^{a(s)}
P_{a(s)b(s)} (\pa) \vf^{b(s)}, \ee where $P_{a(s)b(s)}(\pa)$ is a
Poincar\'e-invariant differential operator, satisfying set of identities
\be \label{P01}
P_{a(s)b(s)}=P_{b(s)a(s)}\;\;,\;\;{P^c}_{ca(s-2)b(s)}=0\;\;,\;\;
P_{a(s-1)cb(s)} \pa^c =0.
\ee
The Poincar\'e-invariant solution to the equations \re{P01} appears to
have the form
\be\label{f0}
P_{a(s)b(s)}(\pa)=\cP_{a(s)b(s)}(\pa) f(\Box)
\ee
where operator $\cP_{a(s)b(s)}(\pa)$ is of
$2s$-order in $x$-derivatives and $f$ is an arbitrary function of
$\Box=\pa_a \pa^a$. The generating formula \re{baseq102} for all $s$ is exhibited in
Sec.\re{dthst}.

As far as one deals with boson actions the choice of the metric signature is
inessential, and the actions look identical for any signature, being hidden in
the definition of the metric.

Needless to say, the gauge invariance
of the actions (\ref{defac},\ref{P01}) holds in arbitrary
dimension, as a consequence of identities \re{P01}. In this paper we demonstrate that, in even
dimensions $d>2$, the actions (\ref{defac},\ref{P01},\ref{f0}) are invariant  w.r.t. conformal algebra in
$so(d-q+1,q+1)$, under the choice
\be
\label{f1} f(\Box)=\Box^{\frac{d}{2}-2}  \ee
Therefore, for any even $d$, it is reasonable to
name the theories (\ref{defac},\ref{P01},\ref{f0},\ref{f1}) spin-$s$
{\it conformal higher spin theories}. If the factor $f(\Box)$ is arbitrary we refer corresponding model as
{\it traceless higher spin theory}.
 Note that in $d=4$ the conformal spin-$0$ field is an
auxiliary field, the fact that will play an important role below.
For odd $d$, it is worth defining operators $\Box^{\hd}$ to be antisymmetric in the sense
of Eq. \re{antod} then the quadratic action \re{defac} is a boundary term.

In $d=4$, these models were introduced by Fradkin and Tseytlin \cite{FT}. Their supersymmetric extensions
and cubic order interactions were studied by Fradkin and Linetsky \cite{linetsky}.

\subsubsection{Free higher spin theories and the higher spin problem}\label{hsfields}
The conformal higher spin theories may be considered as a truncation of
{\it higher spin theories}. The latter are formulated, at the linearized level,
in terms of two symmetric traceless tensor fields $\vf^{a(s)}(x),\chi^{a(s-2)}$
subject to the following gauge transformations
\be
\label{gi0}
\d \vf^{a(s)}=
\mbox{Traceless part of} \;\partial^a \ve^{a(s-1)}(x)
\ee
\be\label{gi1}
\d \chi^{a(s-2)}=\frac{s-1}{2s-4+d}\,
\partial_b \;\ve^{ba(s-2)}(x),
\ee
where the gauge parameter $\ve^{a(s-1)}(x)$ is an arbitrary {\it traceless}
tensor field, just like in the conformal higher spin theory case.
The Poincar\'e-invariant and gauge-invariant action
$\cA_s[\vf_s, \chi_{s-2}]$ is fixed unambiguously
by the requirement of absence of higher (more than $2$) derivatives in the Lagrangian.
\be  \label{fraction}
\begin{array}{cllc}
{\cal A}_s[\vf_s,\c_{s-2}]= \frac{(-)^s}{2} \int d^d x \bigl\{
\pa_{n} \f_{m_1...m_s} \;\pa^{n}  \f^{m_1...m_s}& &\\ & &\\
-\fr s(s-1) \pa_{n} {\f^{k}}_{km_1...m_{s-2}}\; \pa^{n}{{\f^{k}}_k}^
{m_1...m_{s-2}} & &\\ & &\\
+ s(s-1) \pa_{n} {\f^{k}}_{km_1...m_{s-2}}\; \pa_{l} \f^{nlm_1...m_{s-2}}
-s \,\pa_{n} {\f^{n}}_{m_1...m_{s-1}}\; \pa_{k} \f^{km_1...m_{s-1}}& & \\ & &\\
-\frac{1}{4} s(s-1)(s-2)
\pa_{n} {{\f^{k}}_k}_{nm_1...m_{s-3}}\; \pa_{r}
{{\f^{l}}_l}^{rm_1...m_{s-3}}\bigr\},& &
\end{array}
\ee
where double-traceless field $\f^{m(s)}$ is
\be
\f^{m(s)}=\vf^{m(s)}+\eta^{m(2)}\c^{m(s-2)}
\ee
These models were introduced by Fronsdal in $4d$ \cite{fr}, on the other hand,
the form of the actions appears to be independent of the dimension \cite{Aragone:yx}.
They admit generalization to $d$-dimensional anti de Sitter space \cite{Lopatin:hz},
\cite{Metsaev:1997nj},
\cite{Buchbinder:2002ry},
\cite{Biswas:2002nk},
\cite{Segal:2001qq}.
We will refer these models as {\it Fronsdal theories of spin $s$}.
For Minkowski signature, for any $s$
Fronsdal theories describe on-shell the unitary irreducible representation
of the Poincar\'e (anti- de Sitter) group and therefore some free particle, and
for $s=0,1,2,$ they are just free scalar,
photon and graviton, respectively. A natural question arises is there a
consistent nonlinear theory, describing interaction of higher spin
particles $s \geq 2$ to low spin particles ($s\leq 2$: scalars, photons, gravitons)
and among themselves. This task is highly nontrivial,
as almost any naive attempt of introducing interaction
in Fronsdal theories breaks down their extended gauge invariance \re{gi0},\re{gi1}
thus making the nonlinear models inconsistent.
Despite the  progress in the last two decades  \cite{Vasiliev:en}-\cite{Vasiliev:2001wa}, \cite{Metsaev:1996pd},
\cite{Engquist:2002vr} this problem has not got the convincing solution.

We see that, at the linearized level, gauge parameters
in the spin-$s$  conformal higher spin theories
and spin-$s$  Fronsdal theories are the same, moreover, the gauge transformations are the same for $\vf_s$
field. The difference
is that Fronsdal models possess
an additional symmetric traceless tensor field $\chi^{a(s-2)}$ which we call
{\it compensator}. In this sense,
{\it the conformal higher spin theory arises
as a limit of higher spin theory when the compensator decouples.}

More precisely,
it is natural to anticipate existence of a united nonlinearized model which, being expanded around certain vacuum,
reproduces in quadratic approximation the sum of actions like
\be
\cA=\sum \limits_{s=0}^{\infty}\left( l^2 \cA_s[\vf_s,\c_{s-2}]+ \{...\} +l^{d-4+2s} \cA_s[\vf_s]\right)
\ee
where $l$ is a parameter counting number of derivatives and $\{...\}$ stands for
another higher derivative gauge-invariant terms depending on $\vf_s,\c_{s-2}$,
with number of derivatives being less than $d-4+2s$.
Provided $s$ is fixed, in the limit $l\to\infty$ the $l^{4-d+2s}$ terms which depend solely on $\vf_s$ dominate,
and the theory describes conformal higher spin fields. What we have constructed in this paper is
the nonlinear generalization of the sum of these terms. We suppose analogous decoupling
phenomena arise at the nonlinear level.
In this sense, the construction of the full conformal higher spin theory, achieved in this paper,
is a step towards the solution of the higher spin problem.

Note that higher spin theory is believed not to admit flat vacua, and higher spin interactions
are being formulated in anti-de Sitter space \cite{Fradkin:ah}.
As we will show in the paper the conformal higher spin theory possesses conformally flat
vacua with an arbitrary value of the conformal factor of the metric. As anti-de Sitter space is
conformally flat, our results do not come to contradiction with existing results
on higher spin interactions.

\subsubsection{On AdS/CFT "higher spin massless fields in $AdS_{d+1}$$/$conformal higher spin fields
in $d$-dimensional Minkowski space"}\label{adscft}
Conformal higher spin fields and Fronsdal fields have one more important interrelation. Namely,
conformal higher spin fields in $d$ dimensions arise as boundary values of $AdS_{d+1}$ Fronsdal fields.
Let us comment on this point, following Refs \cite{Metsaev:2002vr}, \cite{Tseytlin:2002gz}.
Let $d$ be even. Consider free massless spin-$s$ theory in  $AdS_{d+1}$ and
evaluate the action of this model in terms of solutions of Dirichlet like problem in terms of
the "boundary values" of $\Phi_s$. According to general $AdS/CFT$ principles \cite{mald},
one arrives at a quadratic nonlocal
functional of "boundary values" $\vf_s(x)$ of $\Phi_s (x,z)$, which is a generating functional of the two-point function
of primary operators $\cO_s$ in a $d$-dimensional conformal field theory, with mass dimension $\Delta=E$, where $E$ is a lowest
eigenvalue of the $AdS$ energy operator in the bulk theory. For the bulk theory in question one has \cite{Metsaev:1997nj}
\be
E=s+d-2
\ee
and therefore the coupling
\be
S=\int d^d x \,\vf_s(x)\, \cO_s(x)
\ee
is conformally invariant provided conformal dimension of $\vf_s$ is, in mass units,
\be\label{massun}
[\vf_s]=2-s,
\ee
wherefrom one expects the result of calculation of $AdS$ action to have a form
\be
\tilde{\cA}_s[\vf_s]= \int d^d x \,d^d x' \,\frac {\vf_s (x) \,P(x-x') \,\vf_s(x')}{((x-x')^2+\e^2)^{s+d-2}},
\ee
where $\e$ is a regulator which restricts the limits of integration in $AdS_{d+1}$ and $P(x-x')$ is
a kernel of an operator of zero mass dimension.

In the limit $\e \to 0$ one extracts the logarithmically divergent part of this expression to be, schematically,
\be
\sim\ln\e^2\, \int d^d x \,\vf_s (x) \,\Box^{\hd-2+s}\, P(\pa)  \,\vf_s(x),
\ee
that is in accord with the structure of conformal higher spin theory action (\ref{defac}-\ref{f0}),\re{f1}.
Actually, the last expression is gauge invariant w.r.t. transformations
\re{gtr1128} (the latter arise as a consequence of gauge transformations for $AdS_{d+1}$ massless fields) and
coincides with the action of free spin-$s$ conformal higher spin theory.
Thus, in even dimensions $d$,
free conformal higher spin theories action present the logarithmically divergent terms in the
free actions of $AdS_{d+1}$ massless fields expressed in terms of the solutions of Dirichlet like problem.

\section{Technical digression. Deformed traceless higher spin theories} \label{s2}\label{stech}

As we stated above, our theory reproduces the sum of conformal higher spin models in the framework of a
perturbative expansion around a conformally flat vacuum.
In that framework, conformal higher spin  models reveal themselves
not in their original form described in the previous subsection
but in the form of {\it deformed traceless higher spin theories},
the latter being equivalent to the former.

To show this equivalence and to exhibit the map ``deformed traceless higher spin theories $\leftrightarrow$
conformal higher spin theories'' we have to provide a rather extensive technical digression we devote this section to.
We describe deformed traceless higher spin theories and show that,
in an appropriate basis of fields, the deformed traceless higher spin theories are just the
conformal higher spin theories of the previous subsection.

The results of this section will be very useful in analyzing perturbative expansion of our theory
around a conformally flat background.

\subsection{Deformed traceless higher spin theories.}\label{dthst}

In fact, here we recall and update the results of our paper \cite{Segal:2001di}.
By definition, a deformed traceless higher spin theory is
characterized by a Poincar\'e-invariant quadratic action
built from the infinite number of symmetric tensor fields \be \label{field6}h^{m_1...m_k}(x), k=0,1,2,3,...,\ee
subject to gauge transformations of the form ($\eta^{mn}$ is Minkowski metric)
\be \label{ga456}
\d h^{m(k)}(x) =-2 \m^2 a^{m(k)} (x) - 2\eta^{m(2)} a^{m(k-2)}(x) -2\pa^m
\e^{m(k-1)}(x),
\ee
with the infinite set of gauge parameters, symmetric tensors
\be\label{ga4567}
\e^{m_1...m_k},\;a^{m_1...m_k}, k=0,1,2,3,...,
\ee
which are required to have a compact support in $x$-space.
$\m^2$ is a general Poincar\'e-invariant scalar operator, being a general function of $\Box$:
\be
\m^2 =\m^2(\Box)
\ee
Deformed higher spin theories are parameterized by integer $s=0,1,2,...$ up to
the arbitrariness in multiplying the wave operator by an arbitrary function of
$\Box$.

The gauge transformations may be written in generating form if one introduces power series,
depending on a covector variable $p_m$,
\be\ba
h(x,p)=\sum \limits_{k=0}^{\infty} h^{m_1...m_k}(x) p_{m_1}...p_{m_k},\\ \\
\e(x,p)=\sum \limits_{k=0}^{\infty} \e^{m_1...m_k}(x) p_{m_1}...p_{m_k},\\ \\
a(x,p)=\sum \limits_{k=0}^{\infty} a^{m_1...m_k}(x) p_{m_1}...p_{m_k}.\\ \\
\ea\ee
The gauge transformations \re{ga456} read
\be \label{gtr111}
\d h(x,p)=-2(p^2+\m^2)a(x,p) -2 p\pa_x \e(x,p)
\ee
Note that provided $\m^2=0$ the $a$-gauge transformations in \re{ga456} allow one to gauge away all the traces of tensors
$h^{m_1...m_k}$ and to set $h^{m_1...m_k}$ {\it traceless},
\be
h^{m_1...m_k}=\vf^{m_1...m_k},\;{\vf^{m_1...m_{k-2}n}}_n=0
\ee
Then the invariant action should be expressed in terms of $\vf$'s. In terms of $\vf$'s, the gauge transformations \re{ga456}
read
\be\ba \label{ga45611}
\d \vf^{m(k)}(x) = \mbox{Traceless part of}\;  \pa^m \ve^{m(k-1)}(x),\\ \\
\ve^{m(k-1)}(x)=-\fr \;\mbox{Traceless part of}\;\e^{m(k-1)}(x),
\ea\ee
that coincides with gauge transformations for free conformal higher spin theories of the Sec. \re{fchst}.
Therefore, {\it the quadratic gauge-invariant Poincar\'e-invariant
action which depends on fields (\ref{field6}) subject to the gauge
transformations (\ref{ga456}, \ref{ga4567}) is, in the case $\m^2=0$, the sum of free conformal higher spin
theories \re{defac}}.

Let us see what happens when $\m^2\neq 0$.
Consider a general Poincar\'e-invariant quadratic action for the fields \re{field6}.
The general form of the action is
\begin{equation}\label{Saction}
  A_P[h]=\sum \limits_{k=0,k'=0}^{\infty} \int d^d x
  h^{m_1...m_k}(x) P_{\{m_1...m_k | n_1...n_{k'}\}} (\partial_l)h^{n_1...n_{k'}} (x),
\end{equation}
where $P_{ \{ m_1...m_k | n_1...n_k \} } (\partial_l)$ are some
(pseudo)differential operators constructed from the partial
derivative $\partial_m$ and the Minkowski metric, they are also
allowed to contain any function of $\Box$. The operator $P_{ \{
m(k)| n(k) \} } (\partial_l)$ is to be symmetric,
therefore
\begin{equation}\label{Psym}
P_{ \{ m(k)| n(k') \} } (-\partial_l)=P_{ \{ n(k')| m(k) \} }
(\partial_l).\end{equation}
Let us require the action \re{Saction} to be invariant w.r.t.
gauge transformations (\ref{ga456},\ref{gtr111}).
The gauge invariance of the action \re{Saction} is
equivalent to the following set of identities on the components of the wave
operator:  \begin{equation}\label{constr} \begin{array}{c} P_{\{l_1,l_2...l_k |
   n_1...n_{k'-1}r\}}\,\pa^r =0; \\ \\ \eta^{ab} P_{\{l_1...l_k |
 n_1...n_{k'-2}a\,b\}} +\m^2 P_{\{l_1...l_k | n_1...n_{k'-2}\}}=0;\,\forall
k,k', \end{array} \end{equation} and analogously for $n\leftrightarrow l$.

This infinite system of identities may be easily solved in
a generating framework. Introduce the power series of two
variables $q^m, q'^m$
\begin{equation}\label{generP}
P(q,q',\pa)=\sum \limits_{k=0,k'=0}^{\infty} \frac{1}{k!k'!}
q^{n_1}...q^{n_{k'}} q'{}^{m_1}...q'{}^{m_{k}} P_{\{m_1...m_k |
n_1...n_{k'}\}} (\partial_l).
\end{equation}
This function of three vector-like variables
encodes the full information about
the quadratic action \re{Saction} and moreover, given $P(q,q',\pa)$ the action
is easily recovered by means of the formula
\be
A_P[h]=\int d^dx\; \left\{ h(x,p)P(
\stackrel{\leftarrow}{\frac{\pa}{\pa p}},
\stackrel{\rightarrow}{\frac{\pa}{\pa p'}}, \pa) h(x,p')
\right\}_{p=p'=0} \ee
The infinite system of identities \re{constr} is equivalent
to two equations on $P(q,q')$:
\begin{equation}\label{baseq1}
\eta^{mn}
\frac{\partial}{\partial q^m} \frac{\partial}{\partial x^n} P(q,q') =0,\;
( \eta^{mn}
\frac{\partial}{\partial q^m} \frac{\partial}{\partial q^n} +\m^2)\,P(q,q' ) =0,
\end{equation}
The equation \re{Psym} turns into
\begin{equation}\label{Psym1}
P(q,q',\partial)=P(q',q,-\partial),
\end{equation}
and therefore one gets also
\begin{equation}\label{baseq3}
\eta^{mn}
\frac{\partial}{\partial q'^m} \frac{\partial}{\partial x^n}P(q,q') =0,\;
(  \eta^{mn} \frac{\partial}{\partial q'^m} \frac{\partial}{\partial q'^n} +\m^2 )\,P(q,q' ) =0.
\end{equation}
The equations \re{baseq1}-\re{baseq3} constitute the full set of
conditions for gauge invariance of the action \re{Saction}.

The general solution of the gauge invariance constraints (\ref{baseq1}-\ref{baseq3})
has been obtained in \cite{Segal:2001di}. It has the form
\begin{equation}\ba\label{baseq101}
P(q,q',\pa)=\sum \limits_{s=0}^{\infty} {\tilde\a}_s(\Box) P^{(s)}_{\m}(q,q',\pa)\\ \\
P^{(s)}_{\m}(q,q',\pa) = (\rho\rho')^{-\di}
J_{s+\di} (\frac{\m}{\sqrt{\Box}} \rho) J_{s+\di}
(\frac{\m}{\sqrt{\Box}}\rho')
\sum \limits_{k=0}^{[\frac{s}{2}]}(\frac{\tau}{\rho\rho'})^{2k+\varsigma}
r_{2k+\varsigma,s+\di}\\ \\
\r^2=\Box\,q^2 -(q\pa)^2\;;\; \r'^2=\Box \,q'^2-(q'\pa)^2\;;\;
\tau  = \Box \,(qq') -(q\pa)(q'\pa),
\ea \end{equation}
where ${\tilde\a}_s$ are arbitrary functions of $\Box$ and $J_\n(z
)$ are Bessel's functions of the first kind (see App. \re{ApE}).
Here $\varsigma=0,1$ if $s$ is even or odd, correspondingly,  while
\be\ba\label{sol1}
    r_{2k+\varsigma,s+\di}=\frac{(-1)^k}{(2k+\varsigma)!}
    \frac{(s-\varsigma)!!
    (s+\varsigma +d+2k-5)!!}{(s-\varsigma-2k)!!(s+\varsigma +d-5)!!}\;
    r_{\varsigma,s+\di}\equiv\\ \\\equiv
    c_{2k+\varsigma,s+\di}\;r_{\varsigma,s+\di},\\
    \ea
\end{equation}
$r_{\varsigma,s+\di}$ is an arbitrary "constant",
i.e. arbitrary function of $\Box$, which is to be absorbed by
redefinition of ${\tilde\a}_s$, so without loss of generality
we set it equal to $1$.

Note that the solution contains only even powers of $\rho,\rho'$
as it should.
It appears the sums by $k$ in \re{baseq101} are expressed in terms of Gegenbauer polynomials
\cite{vilenkin} as follows
\be\ba
\sum \limits_{k=0}^{[\frac{s}{2}]}(\frac{\tau}{\rho\rho'})^{2k+\varsigma}
r_{2k+\varsigma,s+\di}=
\frac{(-)^{[\frac{s}{2}]}2^{-s}(s-\varsigma)!!(2s+d-5)!!\Gamma(\di)}
{(s+\varsigma+d-5)!!\Gamma(s+\di)} \cC^{\di}_{s} (\frac{\tau}{\rho\rho'})
\ea\ee
where Gegenbauer polynomials $\cC^{\di}_{s}$ are described in Appendix \re{ApE}.
The overall factor is to be absorbed by ${\tilde\a}_s$ so the general solution reads
\begin{equation}\ba\label{baseq102}
P(q,q',\pa)=\sum \limits_{s=0}^{\infty} \a_s(\Box) P^{(s)}_{\m}(q,q',\pa)\\ \\
P^{(s)}_{\m}(q,q',\pa) = \frac{\G(\di)}{\G(s+\di)}\;(\rho\rho')^{-\di}
J_{s+\di} (\frac{\m}{\sqrt{\Box}} \rho) J_{s+\di}
(\frac{\m}{\sqrt{\Box}}\rho') \cC^{\di}_{s} (\frac{\tau}{\rho\rho'}),
\ea \ee
that presents a refinement of the results of our paper \cite{Segal:2001di}.

To clarify the
meaning of the parameter $s$ it
is worth taking the limit $\m^2 \rightarrow 0$.
As it is argued above, in this limit the action should describe conformal higher spin theory in the
original form (\ref{defac}-\ref{f0}). This is really the case. Indeed, as $\m^2=0$,
only first term of the Bessel's series in \re{baseq102} survives, so one gets,
\begin{equation}\label{baseq11}
 P^{(s)}_{0}(q,q,\pa') \sim
\sum \limits_{k=0}^{[\frac{s}{2}]} {\tau}^{2k+\varsigma}\;
(\rho\rho')^{s-\varsigma-2k} c_{2k+\varsigma,s+\di}.
\end{equation}
As it is clear from the very definition \re{generP}, a term
$I_{a,b,c}\sim \tau^a \rho^{2b} \rho'^{2c}$ in the solution leads
to the corresponding operator in the action \re{Saction}, which
contains $2a+2b+2c$ $x$-derivatives, and involve tensor fields of
ranks $a+2b$ and $a+2c$. Applying this account to the last
equation we see that the corresponding theory contains terms
$I_{2k+\varsigma,[\frac{s}{2}]-k,[\frac{s}{2}]-k}$ for $k=0,...,[\frac{s}{2}]$
and thereby describes
theory of symmetric tensor fields of rank-$s$
\textit{only}, with operators of \textit{only}
$2s$-th order in $x$-derivatives (for $n=\varsigma=0$,
the solution is just a constant).

Let us make an important remark. Due to invariance w.r.t.
$a$-transformations, the actions \re{Saction},\re{baseq10} depend
only upon the special {\it traceless} combinations of $h^{m(k)}$,
described in the Appendix \re{ApC}(one traceless tensor for each rank
$s$). Specifically, $h(x,p)$ may be represented as
\begin{equation}\label{hexp}
h(x,p)=\varphi(x,p)+(p^2+\m^2)\chi(x,p),
\end{equation}
where $\chi(x,p)$ is arbitrary power series in $p_m$ while
$\varphi$ is a traceless power series:
\begin{equation}\label{hexp2}
\varphi(x,p)=\sum \limits_{s=0}^{\infty} \varphi^{m(s)}(x)
p_{m_1}...p_{m_s} \;;\;{\varphi_n}^{nm(s-2)}=0.
\end{equation}
Then it is clear the action may be written in terms of $\varphi$
by making substitution $h(x,p)\mapsto \varphi(x,p)$ in the action
\re{Saction}, after the substitution, the terms $q^2,q'^2$ in the
generating function $P(q,q',\pa)$ \re{baseq10} may be dropped.
This simplify the form of
$\r\mapsto \pm i(q\pa), \r' \mapsto \pm i(q'\pa)$. In this basis,
gauge transformations of $\varphi^{m(s)}(x)$ appear to depend
only upon the special traceless parts of $\epsilon$
(\ref{etr1},\ref{gtra}).

\subsection{Dressing and Undressing maps}\label{dresundres}

The general solution of gauge invariance constraints \re{baseq102} exhibits what we
call dressing. Given  a function of two vector-like variables
$a^m,b_m$ $U(a,b)$,
if one "dresses" $P(q,q',\pa)$ like
\begin{equation}\label{dress}
  P_U(q,q',\pa)=U(q,\pa)\;U(q',\pa)\;P(q,q',\pa) \,\Leftrightarrow\,
  \end{equation}
then new operator $P_U(q,q',\pa)$ already
does not satisfy the same equations as for $P(q,q',\pa)$. However, if
this change is accompanied by the following "undressing" of gauge
fields
\begin{equation}\label{dress1}
h(x,p)=U(\frac{\pa}{\pa p},\pa) h_U(x,p)
\end{equation}
then it results in the same theory, i.e.
\begin{equation}\label{dress2}
 A_{P_U} [ h_U ] = A_P[h].
\end{equation}
This property of "dressing" plays an important role in our considerations.

Consider spin-$s$ deformed model with generating function $P^{(s)}_{\m}$  \re{baseq102}.
It is seen that
\be\ba\label{dthst3}
P^{(s)}(q,q',\pa)=U_{s+\di}(\frac{\m}{\sqrt{\Box}} \rho)\; U_{s+\di}(\frac{\m}{\sqrt{\Box}}
\rho') \;\cP^{(s)}\\ \\ \cP^{(s)}=\frac{\G(\di)}{\G(s+\di)}\Box^{\frac{d}{2}-2} (\r\r')^s
\cC^{\di}_{s}(\frac{\tau}{\rho\rho'}), \ea \ee where \be U_{\n} (z) \equiv
z^{-\n} {J_{\n}(z)},\;\; U_\n(0)=(\frac{1}{2})^{\nu}  \frac{1}{\Gamma(\nu+1)},
\ee
and therefore this model is related to ordinary conformal higher spin
model by the dressing map with the dressing operator
\be
U=U_{s+\di}(\frac{\m}{\sqrt{\Box}} \r)
\ee
Note that this operator contains only even powers of $\r$ so the corresponding
dressing map is well-defined in the space of power series in momenta.

It is seen that up to a constant multiplier
$\cP^{(s)}$ is the $\m \to 0$ limit of
$P^{(s)}$, therefore, it satisfies the set of constraints obtained as
$\m \to 0$ limit of the constraints (\ref{baseq1}-\ref{baseq3}):
\begin{equation}\label{baseq25}
\pa_q^2\,\cP^{(s)}(q,q',\pa ) =0 \;;\; \pa_{q'}^2\,\cP^{(s)}(q,q',\pa ) =0
\end{equation}
\begin{equation}\label{baseq35}
\pa_q\pa \;\cP^{(s)}(q,q',\pa) =0\;;\;\pa_{q'}\pa \;\cP^{(s)}(q,q',\pa) =0
\end{equation}
Besides this, $\cP^{(s)}$ is a homogeneous polynomial of order $s$ in $q$
and, separately, in $q'$,
\be
q\pa_q\; \cP^{(s)}(q,q',\pa)=s \cP^{(s)}(q,q',\pa)\;;\;q'\pa_{q'}\;
\cP^{(s)}(q,q',\pa)=s \cP^{(s)}(q,q',\pa). \ee
Summing up, the sum of deformed spin-$s$ theories \re{baseq102} may be rewritten
as follows
\be
\ba\label{baseq1023}
A_P[h]=
\sum \limits_{s=0}^{\infty} \int d^dx\;
\left\{ \c^{(s)}(x,p) \a_s(\Box) \cP^{(s)}(\stackrel{\leftarrow}{\frac{\pa}{\pa p}},
\stackrel{\rightarrow}{\frac{\pa}{\pa p'}}, \pa) \c^{(s)}(x,p')
\right\}_{p=p'=0} \\ \\ \c^{(s)} =
U_{s+\di}(\frac{\m}{\sqrt{\Box}} \stackrel{\rightarrow}{\r})h(x,p), \ea\ee where \be
\stackrel{\rightarrow}{\r}=\r\;(q \rightarrow
\stackrel{\rightarrow}{\frac{\pa}{\pa p}}).  \ee
Furthermore one notes that each spin-$s$ contribution
$\int...\c^{(s)}...\c^{(s)}$ in \re{baseq1023} depends on only rank-$s$ component of
$\c^{(s)}$, because $\cP^{(s)}$ is a homogeneous polynomial of order $s$ in $q$
and, separately, in $q'$.
Therefore, the sum of spin-$s$ actions is rewritten
in the manner \be \ba\label{baseq1024} A_P[h]= \sum \limits_{s=0}^{\infty}
\int d^dx\; \left\{ \f_s(x,p) \a_s(\Box) \cP^{(s)}(
\stackrel{\leftarrow}{\frac{\pa}{\pa p}},
\stackrel{\rightarrow}{\frac{\pa}{\pa p'}}, \pa) \f_s(x,p') \right\}_{p=p'=0}=
\sum \limits_{s=0}^{\infty} \cA_s[\phi_s],
\\ \\
\f_s\equiv \c^{(s)}_s =\mbox{projection to the rank-$s$
subspace of the power series} \;\;\c^{(s)}(x,p)\\ \\
p\frac{\pa}{\pa p}\; \f_s=s\f_s.
\ea\ee
One observes that the sum of the deformed conformal higher spin models is
always expressed via the set of  symmetric tensor fields
$\f^{m_1...m_k} (x)$ which enter the action via
\be
\f_s(x,p) =\f^{m_1...m_k} (x) p_{m_1}...p_{m_s}
\ee
This set of fields is the substitute of the initial one $h^{m_1...m_k} (x)$
which makes up the power series $h(x,p)$,
in the sense that the quadratic action
is expressed entirely in terms of the "undressed" variables
$\f^{m_1...m_k} (x)$ expressed via original fields $h^{m_1...m_k} (x)$
and their $x$-derivatives up to an infinite order by means of the formulae
(\ref{baseq1023}, \ref{baseq1024}). In terms of the undressed fields the
actions \re{baseq1024} have the simple form of conformal  higher spin theories. Indeed,
consider the action $\cA_s[\f_s]$ separately. The constraints
(\ref{baseq25}-\ref{baseq35}) ensure gauge invariance of the action w.r.t.
gauge transformations
\be
\d \f_s(x,p)=-2p^2 \;a_{s-2}(x,p) -2 p\pa_x \e_{s-1} (x,p),
\ee
where $a_{s-2}, \e_{s-1}$ are homogeneous polynomials in momenta of degrees
$s-2, s-1$, respectively, or , in terms of components,
\be
\d \f^{m(s)} (x)= -2 \eta^{m(2)} a^{m(s-2)}(x) - 2 \pa^m \e^{m(s-1)} (x).
\ee
This is nothing but the gauge laws of $\m=0$ theories. As we noted above,
the algebraic $a_{s-2}$-invariance manifests the decoupling of the trace part
of $\f^{m(s)}$, in other words, the action $\cA_s[\f_s]$ depends on the
traceless part of $\f^{m(s)}$ only,
\be
\cA_s[\f_s]=\cA_s[\vp_s], \;\; \vp^{m(s)}=\mbox{Traceless part of}\;
\f^{m(s)}(x).  \ee
In terms of $\vp_s$, the remaining $\e_{s-1}$-gauge invariance reads
\be\ba
\d \vp^{m(s)}=\mbox{Traceless part of}\; \pa^m \ve^{m(s-1)}(x)\\ \\
\ve^{m(s-1)}=-\fr\mbox{Traceless part of}\; \e^{m(s-1)}.
\ea\ee
These gauge transformations coincide with \re{gtr1128}, which is the starting
point of the undeformed conformal higher spin theories we have started with.
Needless to say, the action $\cA_s[\vp_s]$ is
nothing but the action of $\m^2=0$ conformal higher spin theories
(\ref{defac}-\ref{f0}).
Therefore, {\it the deformed traceless higher spin
theory, expressed in terms of  the ``undressed'' variables $\vp_s$ is just
free conformal higher spin theory.}

\subsection{Reconstruction map}\label{reconstruct}

The quadratic approximation of the conformal higher spin theories looks
most simply in terms of the undressed variables $\f_s$. On the other hand,
for our purposes it is vital
to find the inverse map from the data provided by the set of undressed ``pure
spin'' variables $\f_s$ to the original ones contained in the
fluctuation $h(x,p)$.
The matter
is that the full action as well as closed nonlinearized gauge transformations
will be formulated below in terms of the dressed fields $h$'s, on the other hand, the original
"conformal higher spin fields" are the undressed ones $\f$'s. In order to
get interaction vertices in terms of the undressed fields one has to find the vertices
in terms of $h$ and then to represent $h$ via $\f$.
\begin{teo}
Given the ``undressing'' map
\be\ba \label{undr44}
\f_s\equiv \c^{(s)}_s =
\mbox{projection to the subspace of $s$-th degree in $p_m$ of}
\;\; U_{s+\di}(\frac{\m}{\sqrt{\Box}} \stackrel{\rightarrow}{\r})
h(x,p),
\ea \ee
the equation holds
\be\label{812}
\sum  \limits_{s=0}^{\infty}\frac{i^s}{s!}\;h_s=
2^{\di}\G(\di)\sum \limits_{s=0}^{\infty} i^s
(s+\di)(\frac{\m}{\sqrt{\Box}}\stackrel{\rightarrow}{\r})^s\;
\cC_s^{\di}(\frac{\sqrt{\Box}}{\m\stackrel{\rightarrow}{\r}}) \,\f_s,\ee
where $h_s$ is rank-$s$ component of $h(x,p)$.\end{teo}{\it Proof.}
Represent $\f_s$ as
\be
\f_s=\frac{1}{2\pi i} \oint \frac{d\l}{\l^{s+1}}\; \l^{p\pa_p}\;
U_{s+\di} (\frac{\m}{\sqrt{\Box}} \stackrel{\rightarrow}{\r})h(x,p),
\ee
substitute this to \re{812}, transfer $\l^{p\pa_p}$ to the right by using
\be
\l^{p\pa_p}\;U_{s+\di} (\frac{\m}{\sqrt{\Box}} \stackrel{\rightarrow}{\r})=
U_{s+\di} (\frac{\m}{\sqrt{\Box}}\l^{-1}
\stackrel{\rightarrow}{\r})\;\l^{p\pa_p}
\ee
and use the identity \re{form25} for Bessel's functions in the case $n=d-1$,
to
get the r.h.s equal to \be \frac{1}{2\pi i} \oint \frac{d\l}{\l} e^{i\l^{-1}}
\l^{p\pa_p} h = \sum \limits_{s=0}^{\infty}\frac{i^s}{s!}h_s \ee
The formula \re{812} provides the inverse map to the ``dressing''.
It unambiguously reconstructs the original fluctuation $h(x,p)$ in terms of
``pure spin'' fields $\f_s$ and its $x$-derivatives.
As the function $z^s\cC_s^{\di}(z^{-1})$ is a polynomial
in $z^2$ of maximal degree $2[\frac{s}{2}]$, the map \re{812}
is well-defined in the
space of power series in momenta.  On the other hand, these series in
derivatives terminate and every $\f_s$ enters $h(x,p)$ with its $x$-derivatives
only up to $s$-th order.

Furthermore, as only traceless parts of $\f_s$ contributes the actions \re{baseq1024}
at the linearized
level, one can restrict $\f_s$ to be traceless. Then $h(x,p)$ is traceless
either, because the operator $\pa_p^2$ commutes with
$\stackrel{\rightarrow}{\r}$. Then the ``undressing'' and ``reconstruction''
maps \re{undr44} and \re{812} simplify because $\stackrel{\rightarrow}{\r}$ takes the
form
\be
\stackrel{\rightarrow}{\r}=\pm i\pa_x\pa_p.
\ee
The $\pm$ ambiguity is inessential as all expressions contain only even powers of
$\rho$.
In summary, for traceless "dressed" fields $h^{m(k)}, k=0,1,2,...$ and
"undressed" ones $\f^{m(k)}, k=0,1,2,...$ the undressing \re{undr44} and reconstruction {812}
maps are given by similar expressions like
\be\ba
h^{m(k)}=\f^{m(k)}+\sum \limits_{r=1}^{\infty} c_r \pa_{m_1}...\pa_{m_{2r}} \f^{m(k+2r)}\\ \\
\f^{m(k)}=h^{m(k)}+\sum \limits_{r=1}^{\infty} \tilde{c}_r \pa_{m_1}...\pa_{m_{2r}} h^{m(k+2r)}\\ \\
\ea\ee
where $c_r, \tilde{c}_r$ are some real analytic functions of $\frac{\mu}{\sqrt{\Box}}$.

\section{Geometry}\label{sgeom}

We begin deriving the theory which will reproduce the initial data described in the previous sections.
Our guide is the notion of {\it geometry with  a source} given in Introduction.
This notion is based on a rather general reasoning, which presents nothing
essentially new as compared to the bunch of habitual physical conceptions
of geometry but rather provides some useful systematics which we will be
using extensively. Perhaps, a slightly novel feature of our approach is
the persistence in the joint consideration of what is usually called
geometric structures (like tensor fields, connections etc.) on a manifold
along with objects (in particular, the localized objects) that can
fill an influence of those geometric structures because they determine the
object's dynamics. From physical point of view such a joint analysis is
natural as there
is no "geo-metry" without a "meter" device.

In terms of the definitions of Introduction, the main results of the paper may be spelled as follows.
\be\ba
\begin{array}{rl}\mbox{test object}\; O =&\mbox{quantized point particle}\\
&\mbox{(with the classical action without higher time derivatives),}\\  \\
o=&\mbox{complex wave function of the particle} \;\psi(x)\\ \\
\mbox{background field}\; H=& \mbox{general differential operator}\; \hat{H}\\ \\
\mbox{coupling} \;S_H[o]=& \mbox{action} \;\int d^d x \;\psi^* \hat{H} \psi \;\;\mbox{of the quantized particle}\\ \\
\cG=&\mbox{complexification of the algebra of} \\ &\mbox{quantized canonical transformations}\;\Rightarrow
\\ \\ \Rightarrow H=&\mbox{full set of conformal higher spin fields with}\; s=0,1,2,3,...  \\ \\
\mbox{invariant induced action}\; \cA[H]=&Tr \;\pi_+(\hat{H})=\mbox{action of Conformal Higher Spin Theory},
\end{array}\ea\ee
where $\pi_+(\hat{H})$ is the projector to the subspace with positive eigenvalues of $\hat{H}$.

Below we derive the classical and the quantum
geometry corresponding to the ordinary point particle
in general background fields, with the classical
action without higher time derivatives, and show the emergence of conformal
higher spin fields as background fields of the point particle.
The classical treatment
is a preliminary one, while the true geometry is being uncovered in
considering quantum case.

\subsection{Classical geometry}\label{classgeo}
Consider  classical $d$-dimensional
point particle in general background fields.
Its dynamics is set by writing down the hamiltonian action
\be \label {ham00} S_H [x(\tau), p(\tau),
\lambda(\tau)]=\int\limits_{\t_1}^{\t_2}
 d\tau  \{p_m \dot{x}^m -\lambda H (p, q)\},
\ee where $x^m(\tau),\, m=0,...,d-1$ are the coordinates of the
particle's world line, $p_m(\tau)$ are the momenta and $\lambda$
is a Lagrange multiplier to the unique first class constraint $H
(x^m, p_m) \approx 0 $ which we shall call Hamiltonian. The
Hamiltonian is supposed to be a power series in momenta, \be
\label{hclass00} H =
\sum\limits_{k=0}^\infty H^{m_1...m_k} (x)\,p_{m_1}...p_{m_k} = \sum\limits_{k=0}^\infty H_k \ee where $H_k$
denotes the homogeneous polynomial of $k$-th degree in momenta.

Classical trajectories of the particle are bind to the constraint surface,
\be   \label{mshell}
H(x(\t),p(\t))=0
\ee
The action has the local symmetry with the parameter $\n(\t)$, being the Hamiltonian counterpart of
world line reparametrizations $x'(\t'(\t))=x(\t),$
\be  \label{invar} \delta{x^m(\tau)}=\n(\tau) \{x^m,
H\} , \qquad \delta{p_m(\tau)}=\n(\tau) \{p_m, H\}
,\qquad \delta \lambda=-\dot{\n}. \ee
This local symmetry may be fixed by implying the gauge condition
\be \label{gaugecond}
\dot{\l}=0 \;\;\Rightarrow \ddot{\n}=0
\ee
after that only global modes of gauge transformations survive
\be\label{globmodes}
\n(\tau) = \nu_1 \tau + \nu_2. \ee These global modes may be used to
set the limits of $\t$-integration to $\t_1=0,\;\t_2=1$.

When $H_k=0$ for $k>2$ (i.e. when $H$ is a second order function),
the model describes a particle in general
gravitational + Maxwell background.
It may be seen by writing down the Hamiltonian in the form
\be\label{lspinham}
\tilde{H}(x,p)=
\sum\limits_{k=0}^2 H^{m_1...m_k} (x)\,p_{m_1}...p_{m_k} =
D(x)-g^{mn}(x) (p_m-eA_m(x))(p_n-eA_m(x)),
\ee
and excluding the momenta $p_m$
by means of their equations of motions
$\frac{\d S}{\d p_m}=0$ from the action \re{ham00}, to get
\be\label{masslp}
S=\int d\t (-\frac{\dot{x}^2}{4\l}-\l D +eA_m \dot{x}^m),
\ee
and then, in the case $D\neq 0$, to exclude $\l$ by means of its equations
of motion, to get the conventional Lagrangian action of the massive particle
\be
S=\int d\t \left(-\sqrt{D\dot{x}^2} + eA_m \dot{x}^m\right),
\ee
while in the case $D=0$ the action \re{masslp} describes massless particle in
general gravitational +Maxwell background.

In the case Hamiltonian possesses nonvanishing components $H_k,\, k>2,$
the particle experiences the influence of  higher rank symmetric tensors
$H^{m_1...m_k}(x)$. To get an idea how this interaction looks like in the Lagrangian
setup it
is worth expanding general Hamiltonian around the low-spin one \re{lspinham}  with
$D=-m^2,\;A_m=0$
like
\be
H(x,p)=-g^{mn}(x) p_m p_n -m^2 -2{\bar e} \sum \limits_{s=0}^{\infty}
 h^{m_1...m_k}(x) p_{m_1}...p_{m_k},
\ee
where $h^{m_1...m_k}(x)$ is a fluctuation and $e$ is the expansion parameter,
and exclude the momenta $p_m$ and the Lagrange multiplier $\l$
by means of their equations of motion in the framework of perturbations in $e$.
This is done in \cite{Segal:2000ke}, and the result is
\be \label {hamBdW1}
S_H[x(\tau)] =- \int d\tau \{ \sqrt{-m^2\dot{x}^2}
\left(1+ \sum \limits_{s=0}^{\infty}\frac{\bar e}{m^2} h_{m_1...m_s}(x)\,
\dot{x}^{m_1}...\dot{x}^{m_s}
(-\frac{m^2}{\dot{x}^2})^{\frac{s}{2}} \right)+o(e^2)\}. \ee
Here
each spin-$s$ contribution presents the action being equivalent to that
introduced long ago in \cite{deWit:pe}.
\footnote{On our notation: we use signature
$(-++...+)$, alternative as compared to De Wit and Freedman's paper
and re-introduce the mass parameter $m$ explicitly. The action
\re{hamBdW1} coincides with the De Wit-Freedman ($DW-F$) one
\cite{deWit:pe} after the identification
$\varphi^{DW-F}_{m_1...m_k} = - h_{m_1...m_k}$ and setting
$m^2=-1$ (the negative sign of $m^2$ just accounts the difference
in metric's signature,
$\sqrt{\dot{x}^2}{}|_{DW-F}=\sqrt{-m^2\dot{x}^2}|_{our})$.}

The action \re{ham00} is {\it covariant} w.r.t. infinite-dimensional
Lie algebra
$\cG$ which has the structure of a semidirect product of all canonical
transformations with parameters $\e(x,p)$ to the
abelian ideal of "hyper Weyl" transformations $a(x,p)$:
\be \label{def}
S_{H+\delta H} [x +\d x, p+\d p, \l +\d \l]=
S_H [x,p,\l] +\mbox{ boundary terms }
\ee
\be\label{lie00}
[\delta_{\e_1,a_1},\delta_{\e_2,a_2}]=\delta_{\{\e_1,\e_2\},
\{\e_1,a_2\}-\{\e_2,a_1\}},\ee
where
\begin{equation}
\label{invar00} \delta{x^m}=\{x^m, \epsilon(x,p)\}, \; \delta{p_m}=
\{p_m, \epsilon(x,p)\}, \;\delta \l = -2\,a(x,p)\, H(x,p)
\end{equation}
and
\be\label{canon00}
\delta H(x,p) =2 \,a(x,p)\, H(x,p)+ \{\epsilon, H(x,p)\}. \ee
($\{ , \}$ stands for
the canonical Poisson bracket, $\{x^m, p_n\}
=\delta^m_n\,,\,\{x^m, x^n\}=\{p_m, p_n\}=0$).
$\e$ and $a$ are supposed to have the same structure of
power series in momenta as $H$ has,
\be\ba
\e(x,p)=\sum \limits_{k=0}^{\infty} \e^{m_1...m_k}(x) p_{m_1}...p_{m_k}=\sum \limits_{k=0}^{\infty} \e_k ,\\ \\
a(x,p)=\sum \limits_{k=0}^{\infty} a^{m_1...m_k}(x) p_{m_1}...p_{m_k} = \sum \limits_{k=0}^{\infty} a_k.\\ \\
\ea\ee
Then the transformations from $\cG$
can be rewritten in terms of tensor fields being the coefficients of power
series \re{hclass00} and corresponding parameters $\e^{m(k)}(x),
a^{m(k)}(x)$. The lowest components of
$\e$ and $a$, $\e_0, \e_1, a_0$ parameterize the "low-spin" algebra
of $U(1)$ "phase" transformations, $x$-diffeomorphisms and
Weyl dilations, correspondingly.

According to the definitions   of the previous subsection,
our consideration shows that  there exists the geometry with the source
$O=$``point particle'', geometric structure $H(x,p)$ \re{hclass00}
constituting the collection of symmetric tensors of all ranks,
coupling  $S_H$ being the action of the point particle in general background fields \re{ham00}, and the covariance algebra $\cG$
(\ref{invar00}, \ref{canon00}) being the semidirect product of all canonical transformations by an abelian ideal of
hyper-Weyl transformations.

The set of variables $o$ is provided, in the Hamiltonian picture,
by the particle's coordinates $x^m$, momenta $p_m$ and the Lagrange multiplier
$\l$.

According to the general strategy, described in the previous subsection,
one should look for an "induced"
action functional for the background fields $H(x,p)$,
which is invariant w.r.t. covariance transformations \re{canon00}.
For our purposes it is better to postpone this procedure to the next subsection,
where the quantum particle is considered.
Nevertheless, it is instructive to have a glimpse of what kind of theory
the invariant induced action could describe.
To this end one performs the {\it linearization}
of the gauge transformations \re{canon00} around a particular "vacuum" configuration of
background fields of the form
\be\label{vac17}
H=-p^2-m^2=-\eta^{mn} p_m p_n-m^2, m^2=Const
\ee
($\eta^{mn}$ is the inverse Minkowski metric), which sets the dynamics of
massless particle in flat Minkowski space. The linearization means introducing
fluctuations $h(x,p)$ by representing general Hamiltonian as
\be
H(x,p)=-p^2-m^2+h(x,p),
\ee
rewriting gauge transformations \re{canon00} in terms of the fluctuation $h(x,p)$
and keeping in gauge transformations only the terms of zero order
in fluctuation.
In so doing, the linearized gauge transformations are always given by the
variation of the "vacuum".  One gets
\be\label{bsqinf54}
\d h(x,p)=\d (-p^2-m^2)=-2a(x,p) (p^2+m^2) -\{\e(x,p),p^2\}=-a(x,p)(p^2+m^2)
-2p\pa_x \e(x,p),
\ee
which is nothing but the {\it gauge laws of the deformed conformal higher spin
theories \re{baseq102} in the case $\m(\Box)=m=Const$}. These theories were shown
in Sec. \re{dthst}
to be
equivalent to ordinary conformal higher spin theories with conformal spins
$s=0,1,2,3,...$, formulated in terms of {\it traceless} tensors of rank $s$.

Recall the well-known fact (see also App. \re{ApA}):
if an action is gauge-invariant w.r.t.
full gauge transformations, then the quadratic part of its expansion around
any vacuum is gauge invariant w.r.t. linearized transformation.
Therefore, if one could have constructed an invariant induced action which possesses
$H$ of the form \re{vac17} as a solution of the equations of motion, then
the quadratic part of the action around this vacuum is a sum
of conformal higher spin theories with conformal spins $s=0,1,2,3,...$ described in terms of
symmetric traceless tensors of rank $s$.

Note that the origin of this infinite number of {\it traceless} tensors from the
Hamiltonian of the particle has a clear interpretation.
Indeed, the fluctuations of the Hamiltonian are parameterized by traceless tensors
as their traces are gauged away by the
linearized  hyper-Weyl transformations parameterized by $a(x,p)$. But
the meaning of hyper-Weyl transformations is quite transparent.
Namely, the classical dynamics of the particle is being developed on the
{\it constraint surface} $H(x(\t),p(\t))=0$, and
the hyper-Weyl-covariance expresses the fact that the {\it
the particle fills only the shape of the constraint surface $H(x,p)=0$,
not the form of the Hamiltonian as a function of phase space variables
outside the constraint surface. In this sense, any two hamiltonians,
which differ by  multiplication by a function which never comes to zero,
are equivalent}. At the linearized level, fluctuations of the Hamiltonian
around the vacuum $H=-p^2-m^2$ are equivalent if they differ by a function
$-(p^2+m^2) a(x,p)$, so the fluctuations of the constraint surface
$H=0$ are represented by the equivalence classes of functions
$h(x,p)$ modulo the equivalence relation $h(x,p) \sim h(x,p)-(p^2+m^2) a(x,p)$.
This equivalence does not affect only the special {\it traceless} parts of $h(x,p)$,
described in App. \re{ApC}. (For $m=0$ these special traceless parts become
ordinary traceless parts of $h^{m(s)}, s=0,1,2,3,...$).
Therefore, the traceless parts of the fluctuation
of the Hamiltonian around the vacuum $H=-p^2-m^2$ are in one-to-one correspondence with
fluctuations of the shape of the constraint surface $H(x,p)=0$.

The equivalence of linearized covariance transformations and of gauge transformations
of $\m=m$ deformed conformal higher spin theories allows to state that
{\it infinite collection of the
conformal higher spin fields is in one-to-one correspondence with fluctuations of the
shape of the constraint surface $H(x,p)=0$ around the vacuum $H=-p^2-m^2$.}

Therefore, we have found a physical origin of conformal higher spin fields
as background fields of the point particle. A gauge-invariant action of
conformal higher spin fields, once constructed,
will play a role of "induced" action \re{inducedac}.
However, it appears to get the physically relevant results one has to quantize
the particle. After all, it will be seen why the classical geometry
is useful only as a motivation for, and as a limit
of the full quantum treatment.

\subsection{Quantum geometry}\label{subs42}

The covariant quantization of the model is performed easily,
since there is the single first-class constraint and hence
the algebra of constraints is trivial.

The particle wave functions $|\psi>$ are realized in
a Hermitian representation space of $d$-dimensional Heisenberg-Weyl algebra,
spanned by canonical coordinates and momenta $\hat{x}^m,\hat{p}_n$, with
commutation relations
\be
[\hat{x}^m,\hat{p}_n]=i\hbar \d^m_n.
\ee
One may choose the coordinate
representation, then $|\psi>$ is represented by
a complex field $\psi(x)$ over ${\cal M}^d$,  $<\psi|$ -- by a complex
conjugated $\bar{\psi} (x)$ and the scalar product $<\psi_1|\psi_2>$ -- by
the integral $\int d^n x \bar{\psi_1}(x) \psi_2(x)$.

The covariant quantization is performed by
imposing the constraint
\be \label{hpsi}
\hat{H} (\hat{x}, \hat{p}) |\psi> =0
\ee
where some ordering of $\hat{x}$ and $\hat{p}$ is implied which is
consistent with the hermiticity of the quantum Hamiltonian,
$\hat{H}=\hat{H}^{\dag}$. The role of the constraint \re{hpsi} is to
extract the physical states subspace\footnote{For general Hamiltonian, the issue
of existence of a positive inner product in the physical subspace is
a subtle question which we do not consider in this paper}.
The classical analogue of this
equation is the constraint equation (\ref{mshell}).
After specifying appropriate
boundary conditions, this equation
determines the quantum dynamics completely (for example, for
hamiltonians with $H_k=0, k>2$, it becomes the Klein-Gordon equation in external low spin
fields).

The equation \re{hpsi} follows via the variational principle from
the action \be \ba\label{php} S[\psi] = \frac{1}{2}\left( <\psi|\hat{H}|\psi>
+ (<\psi|\hat{H}|\psi>)^*\right)=\\ \\
 <\psi|\hat{H}|\psi> +\mbox{boundary terms}.\ea \ee The quantum analogue of the
covariance transformations \re{def} is the group of the most
natural transformations preserving the action
\re{php}: \be\label{bsinv55}
\label{qequiv} S_{H'}[\psi']= S_H[\psi]+\mbox{boundary terms}, \ee
\be \label{bsq}
\hat{H}' = {\hat \Omega}^{\dag}\, \hat{H} \,{\hat \Omega}, \ee \be
\label{bsqp} |\psi'> = {\hat \Omega}^{-1} \,|\psi>, \ee where
$\hat{\Omega} (\hat{x}, \hat{p})$ is some invertible operator. The
transformations (\ref{qequiv},\ref{bsq},\ref{bsqp}) define the
{\it quantum covariance transformations}. The infinitesimal form
of (\ref{bsq},\ref{bsqp}) is extracted easily via the substitution
$\hat{\Omega} = e^{\omega}$ \be \label{bsqinf} \delta \hat{H} =
{\hat \omega}^{\dag} \,\hat{H} +\hat{H} \,{\hat \omega} , \ee \be
\label{bsqpinf} \delta |\psi> = -{\hat \omega}\, |\psi>, \ee where
$\hat{\omega} (\hat{x}, \hat{p})$ is arbitrary.

Thus, {\it one has the geometry with the source $O$=quantized point particle;
$o=$complex scalar field, geometric structure $\hat{H}(\hat{x},\hat{p})$,
coupling $S_H[\psi]$ being the action of the complex scalar field $\psi(x)$, and
the covariance algebra \re{bsqinf}}.

The quantum covariance transformations are interpreted as
gauge transformations of the tensor fields which form the Hamiltonian
as a power series in momenta.
To formulate these transformations in a closed
form it is useful to employ the well known "symbols of operators"
technique (see e.g. \cite{berezin}), which is equivalent to considering
the operators $\hat{f}(\hat{x}^m,\hat{p}_n)$ as functions on the classical
phase space $f(x^m,p_n)$ so that there exists (in a wide class of
functions and operators) a standard invertible
map $\hat{f}=Q f,\;f=Q^{-1} \hat{f}$.
In this technique, the associative product of operators
$\hat{f_1} \hat{f_2}$
becomes an associative $*$-product of functions in the phase space
\be
f_1 * f_2 = Q^{-1} \hat{f_1} \hat{f_2}
\ee
Specifying $Q$ is said as choosing the symbol. Hereafter we choose
the Weyl symbol \cite{berezin}. Then the star product takes the standard
"Weyl-Moyal"\cite{berezin}) form
\be \label{star}
(f_1 *f_2)(x,p)=f_1 \exp (\frac{i\hbar}{2}  {\stackrel{\leftrightarrow}{\nabla}}  ) f_2,
\ee
where
\be  \label{nab}
{\stackrel{\leftrightarrow}{\nabla}} =\frac{\pa}{\pa \stackrel{\leftarrow}{x^n}}
\frac{\pa}{\pa \stackrel{\rightarrow}{p_n}} -
\frac{\pa}{\pa \stackrel{\leftarrow}{p_n}}
\frac{\pa}{\pa \stackrel{\rightarrow}{x^n}} .
\ee
This product have a structure of "semiclassical"
expansion by the powers of $\hbar$. Equivalently, the star product may be
written down in integral form,
\be \ba\label{starintegral}
(f_1 *f_2)(x,p)= (\pi \hbar)^{-2d} \int d^d x_1 d^d p_1
d^d x_2 d^d p_2 \\ \\ \exp^{\left\{ \frac{2i}{\hbar}
\left((x-x_2)p_1+(x_1-x)p_2+(x_2-x_1)p\right) \right\}  } f_1(x_1,p_1) f_2(x_2,p_2),
\ea \ee
Besides associativity,
the star product possesses the standard properties of classical limit
\be
\lim\limits_{\hbar \rightarrow  0}
(f_1 *f_2)(x,p)=f_1 f_2 \;\;,\;\;
\lim\limits_{\hbar \rightarrow  0} \frac{1}{i \hbar}
(f_1 *f_2 -f_2*f_1)=\{f_1 , f_2\} = f_1 \na f_2
\ee
In the Weyl symbol, the Hermitian conjugation
\be
f^{\dag}= Q^{-1} \hat{f}^{\dag}
\ee
is an ordinary complex conjugation:
\be
f^{\dag}=f^*
\ee
Below we will nevertheless write $"f^{\dag}"$ instead of $"\bar{f}"$.
The action of the operator with the Weyl symbol $f(x,p)$ on the wave function
is described in App. \re{ApH}.

In the subsequent calculations, we use operators and their symbols
on equal footing, to distinguish between two
pictures we mark the operators by hat and write
$"*"$ for their symbols product.

Now, we may rewrite \re{bsqinf} as
\be \label{bsqinf1}
\delta H =  \omega^{\dag} * H +H * \omega.
\ee
Introducing the real and imaginary parts of $\omega$
\be
a = Re\; \omega =\fr (\omega+\omega^{\dag})\;\;,\;\; \frac{i}{\hbar}\,\e=
i \;Im\; \omega= \fr
(\omega-\omega^{\dag}) ,
\ee
one rewrites \re{bsqinf} as
\be  \label{bsqinf2}
\delta H =  [a , H]_{+} - \frac{i}{\hbar}[\e , H]
\ee
where $[$ $]_{+}$ and $[$ $]$ are the anticommutator and commutator
w.r.t.  $*$-product.
Applying the realization of $*$ product \re{star}, one gets
\be  \label{bsqinf3}
\delta H = 2 a  H +\{ \e , H \} +o(\hbar^2)
\ee
In this formula one recognizes
the quantum deformation of classical gauge
symmetry \re{canon00}.
The corrections
$o(\hbar^2)$ contain the higher derivatives of gauge parameters $a$ and
$\e$.

One concludes that the gauge laws \re{canon00} may acquire $o(\hbar^2)$
higher derivative corrections. The whole Lie algebra of infinitesimal
gauge transformations \re{lie00}
deforms to
\be \label{qtrs}
\ba
[\delta(\omega_1) ,
\delta(\omega_2) ] H =  \delta(\omega_3)  H\\ \\ \omega_3= - [\omega_1,\omega_2]
\ea
\ee
which reads as follows in terms of real and imaginary parts of $\omega$:
\be \label{qtrs2}
\ba
[\delta(\e_1,a_1) ,
\delta(\e_2,a_2) ] H =  \delta(\e_3,a_3)  H\\ \\ \e_3= -\frac{i}{\hbar}[\e_1,\e_2]
+i\hbar [a_1,a_2]\;\;,\;\; a_3=-\frac{i}{\hbar}([\e_1,a_2] -[\e_2,a_1]),
\ea
\ee
$\e$ parameterize the  quantized canonical
transformations which we shall call infinitesimal unitary transformations.
In contrast with the classical transformations \re{lie00}, $a$ do not form
a subalgebra. These commutation relations are typical for
the Lie algebras possessing Cartan decomposition, when $\e$ play the role
of real generators while $a$ correspond to imaginary ones, so one may call,
in analogy with $so(p,q)$ Lie algebras,
$\e$ the canonical rotations and $a$ the canonical boosts. On the other hand,
the whole algebra exhibits itself as a complexification of the algebra of
infinitesimal unitary transformations.

Of our interest is the subalgebra of power series in momenta
\be \label{hclass2}
\o = \sum\limits_{k=0}^{\infty} \o^{m_1...m_k} (x) p_{m_1}...p_{m_k}
\ee
in the associative algebra of all phase space functions (they do form
a subalgebra w.r.t. $*$-product \re{star}, at least in the framework of the
$\hbar$-expansion).

In the classical case, the zero-order and linear in momenta $\e'$s,
$\e=\ve+\xi^m p_m$,
lead to $U(1)$ and general coordinate transformations. On the other hand,
Weyl product has a property that the commutator of the functions $\e_{1,2}$,
which are linear in momenta (or independent on momenta) reduces to their Poisson bracket:
\be\ba  \label{gencoor}
\e_{1,2}=\ve_{1,2}(x)+\xi_{1,2}(x)^m p_m \Rightarrow
[\e_{1},\e_{2}]=i \hbar\{ \e_{1},\e_{2}\}
\ea\ee
therefore, in the quantum case one may associate the
$U(1)$ and general coordinate transformations with first order
$\e$'s like in the classical case,
as their algebra is not deformed by the quantum corrections.
At the same time, their representation in the space of phase space functions does
deform,
\be\label{xdifcor}
\d H =-\frac{i}{\hbar}[\e,H]=\{ \e,H\}+o(\hbar).
\ee
On the other hand, this deformation is a trivial one in the sense there exists
change of variables from the coefficients $H^{m(s)}$ of the symbol $H(x,p)$ to
ordinary tensor fields ${\bf H}^{m(s)}$, see Sec. \re{ctsdwt} for more detail.

\subsection{Semiclassics. Generalized $W$-geometry}\label{wgeom}

In this subsection we once again make a little digression and show that the coupling
$"<\psi|\hat{H}|\psi>"$ may be naturally decomposed in semiclassical series in $\hbar$ in such a way that
the first, "classical" term of decomposition presents a kind of sigma-model which we call {\it Hamilton-Jacobi sigma-model}
or {\it a generalized $W$-sigma model}, by the following reasons.

Let us write the wave function $\psi(x)$ as
\be\label{quans7}
\psi(x)=\r(x)  e^{\frac{i}{\hbar} \vp(x)},
\ee
where $\r(x),\vp(x)$ are arbitrary real fields and  substitute it to the coupling \re{php}.
Note that an operator $\hat{H}$ with the Weyl symbol $H(x,p)=\sum \limits_{s=0}^{\infty} H^{m(s)}
(x) p_{m_1}...p_{m_s}$ acts on wave functions as
\be
\hat{H}\psi(x)=\sum \limits_{s=0}^{\infty} [(-i\hbar)^s H^{m(s)
}(x) \;\pa_{m_1}...\pa_{m_s} + ... ] \psi(x),\ee
where $"+..."$ denotes terms with less derivatives (for the exact expression, see \re{weyl988}).
It is easily seen the coupling reads
\be\label{hjsm}
S[\r,\vp]= \sum \limits_{s=0}^{\infty}\int d^d x \r^2  H^{m(s)
}(x) \;\vp_{,m_1}(x)...\vp_{,m_s}(x) + o(\hbar)=S_{cl}+o(\hbar)
.
\ee
Note that the modulus of the wave function $\r(x)$ enters the Lagrangian as an overall multiplier while
the phase $\vp$ provides derivative terms.

The variation w.r.t. $\r$ leads to the equation of motion
\be\label{hjeqn}
\frac{\d S_{cl}}{\d \r(x)}=2 \r H(x^m,\vp_{,m} (x))=0,
\ee
which, if $\r\neq 0$
is nothing but {\it Hamilton-Jacobi equation for the action function $\vp(x)$}. The variation of
\re{hjsm} w.r.t. $\vp$ leads to the equation
\be\label{treqn}
\frac{\d S_{cl}}{\d \vp(x)}=- \pa_n \left(\r^2 \frac{\pa}{\pa p_n} H(x,\vp_{,m} (x))\right)=0,
\ee
which is nothing but the generalized {\it transport equation} for the density $\r^2$.

This interpretation presents no surprise as the equations \re{hjeqn} and \re{treqn} are nothing but
first terms of the semiclassical expansion of the full quantum equations of motion
$\hat{H}\psi(x)=0$, while the of Hamilton-Jacobi equation and transport equation originate
exactly from the substitution of the ansatz \re{quans7} to the "generalized Klein-Gordon" equation $\hat{H}\psi(x)=0$.

A less trivial  fact is
that the quantum covariance transformations \re{bsqpinf} have their semiclassical counterparts,
obtained by rewriting \re{bsqpinf},
\be
\d \psi(x) = - \hat{\o} \psi(x)
\ee
in terms of $\r$ and $\vp$, to get
\be\label{hullstrs}
\d \vp(x)= - \e(x,\vp_{,m} (x))+o(\hbar),
\ee
\be\label{newtrs}
\;\d \r^2(x) = - \pa_m \left(\r^2 \frac{\pa}{\pa p_m} \e (x,\vp_{,m} (x))\right) -
2 a(x,\vp_{,m}) \r^2( x)+ o(\hbar)
\ee
and taking limit $\hbar \to 0$, i.e. throwing away $o(\hbar)$ terms.
The resulting transformations, parameterized by arbitrary power series in momenta $\e(x,p),a(x,p)$,
form exactly the same algebra as classical covariance transformations \re{lie00} and
provide {\it the covariance algebra of the Hamilton-Jacobi sigma-model \re{hjsm}},
\be
{S_{cl}}_{H+\d H}\,[\r+\d \r,\vp +\d \vp]={{S_{cl}}_{H}}\,[\r,\vp]
+\mbox{boundary terms}
\ee
The gauge transformations of the
Hamiltonian are given by the classical terms of the quantum gauge transformations \re{bsqinf} and thereby
coincide with classical gauge transformations \re{canon00}.

It is notable that one has such a realization \re{hullstrs} of canonical transformations of $2d$-dimensional phase space
in terms of just one scalar field $\vf(x)$ in $d$ dimensions.
This fact has many clear explanations but we do not dwell on this point here
(perhaps, the simplest explanation is the derivation just delivered).

It appears the transformations of such a kind were already discussed in literature.
In fact, the transformation law \re{hullstrs} is nothing but the basic input of {\it W-geometry} introduced by Hull
\cite{Hull:1992vj}.
Hull did not introduce the second field $\r(x)$ and
was looking for the coupling $S_{\tilde{F}}[\vp]=\int d^d x \tilde{F}(x^m,\pa_m \vp)$ which would possess covariance property
$S_{\tilde{F}+\d \tilde{F}}[\vp +\d \vp]=S_{\tilde{F}}[\vp]$. Here $\tilde{F}(x,p)$ called according to Hull's terminology as
"cometric density" is considered as a power series in momenta whose coefficients should transform via themselves w.r.t. canonical transformations.
Hull has shown the cometric density does not exist if $d>2$, while
in $d=2$ the coupling exists if the canonical transformations parameters as well as coefficients of the
power series $\tilde{F} (x,p)$ are subject to some constraints (while in $d=1$, the cometric density does exist).
This constrained geometry is what was named $"W-geometry"$ as it was shown to underly two-dimensional $W$-gravity models
which gauge global $w_\infty$ symmetries of free massless scalar.

What we just have shown here is that if one introduces, in addition to $\vp$, an additional field
$\r$ with transformation laws
\re{newtrs} then, in any dimension $d$ it is possible to construct covariant action of the form \re{hjsm},
and this action is nothing but
the "classical limit" of the simple quadratic action $"<\psi|\hat{H}|\psi>"$ of the quantized particle. In view of the  above it
is natural to call the "Hamilton-Jacobi sigma-model" \re{hjsm} also as "generalized $W$-sigma model".

We will return to this model in Sec. \re{spec2} where we will argue {\it the action \re{hjsm} has interpretation of
world volume theory of a tensionless $d-1$-brane in $d+1$ dimensions}.

\subsection{Linearization and the theorem on structure of invariant induced action around a flat vacuum.}\label{corper}

Having defined the quantum geometry relations \re{php}-\re{bsqpinf} one has to look
for the invariant induced action $\cA[H]$ which is  invariant w.r.t. covariance transformations
\re{bsqinf}. This is done in the next sections. But before doing this a useful information may be extracted,
as in previous subsection, by studying linearization of gauge transformations around a natural vacuum
\be\label{vac951}
{\bar H}=-(p^2+m^2),
\ee
where $m^2$ is a real constant and $p^2$ is built with flat metric $g^{mn}=\eta^{mn}$.
As it is recalled in App. \re{ApA}, if
${\bar H}$ is a solution of the equation of motion of the action $\cA[H]$ then the quadratic part of the action expanded
around this solution is gauge invariant w.r.t. linearized gauge transformations.
As in Sec. \re{classgeo}, introduce fluctuation $h(x,p)$ around the vacuum \re{vac951} as
\be
H(x,p)={\bar H}+h(x,p),
\ee
rewrite the gauge transformations (\ref{bsqinf},\ref{bsqinf2})
in terms of $h$ and extract the zero order in $h$, then the linearized gauge transformations
are given by the gauge variation of the vacuum,
\be\label{bsqinf37}
\d h =\d {\bar H} = a * {\bar H} +{\bar H} *a - \frac{i}{\hbar} (\e * {\bar H} - {\bar H} *\e)=
-2(p^2 -\frac{\hbar^2}{4}\Box +m^2) a(x,p)- 2p\pa_x \e(x,p).
\ee
As compared to the analogous "classical" gauge transformations \re{bsqinf54} of the previous subsection,
one gets the quantum correction $-\frac{\hbar^2}{2} \Box\; a(x,p)$. The gauge transformations \re{bsqinf37}
are the ones of {\it deformed conformal higher spin theories with}
 \be\label{171819}\m^2(\Box)=m^2-\frac{\hbar^2}{4}\Box,\ee
which, as we know from the results of Sec. \re{dthst}, are equivalent to the sum of ordinary conformal higher spin theories.
More precisely, the conformal higher spin theories arise after deformed models are rewritten in terms of the "undressed"
variables \re{baseq1024}. For general $\m=\m(\Box)$ the undressing map is nonlocal as it is manifested by the inverse square root
of $\Box$ in the argument of the undressing operator, this nonlocality may lead to subtleties in
identifying the deformed theories with undeformed ones. It is remarkable, however, that {\it in the case \re{171819}
with $m=0$,
the undressing map is local}, as in this case
\be
\m^2=-\frac{\hbar^2}{4}\Box \Rightarrow \frac{\m}{\sqrt{\Box}}=\pm \frac{i\hbar}{2}
\ee
The $\pm$-arbitrariness is inessential as the undressing operator is an even function of it's argument.

As a result, one gets the following theorem.
\begin{teo}\label{linearize}
If one constructs an action $\cA[H]$ being invariant w.r.t.
infinitesimal gauge transformations
\be
\d H(x,p)=\o^\dag (x,p) * H(x,p)+H(x,p)*\o(x,p),
\ee
that possesses a vacuum of the form
\be\label{vac4897}
H=-p^2,
\ee
then the quadratic part of the perturbative expansion of $\cA[H]$ around this vacuum
is a sum of deformed conformal higher spin theories (\ref{dthst3}-\re{baseq1024})
with $\m^2=-\frac{\hbar^2}{4}\Box$, which is equivalent
to a sum of conformal higher spin theories (\ref{defac}-\ref{f0})
with conformal spin $s=0,1,2,3,...$, the equivalence is provided by local undressing and reconstruction maps
\re{undr44} and \re{812}, respectively.
\end{teo}

It will be shown in the next sections $\cA[H]$ exists indeed, and possesses the vacuum \re{vac4897}. Therefore,
$\cA[H]$ will present a consistent conformal higher spin theory action. In the subsequent treatment, we not only
construct the action $\cA[H]$ (Sec.\re{action111}), and prove $H=-p^2$ is a vacuum (Sec.\re{scfv}),
but explicitly calculate the quadratic action
around this vacuum and  get the sum of conformal higher spin theories with $s=0,1,2,3...$ (Sec.\re{corollary}).

Concluding the description of the geometry let us mention that a similar treatment of point particle in background fields
has been undertaken in two-time approach \cite{Bars:2001xv}.
In this approach, any $d$-dimensional relativistic system may be reformulated
as a particular gauge of a $2+d$-dimensional one, that usually allows to highlight conformal properties of the model.
On the other hand, as a consequence of two additional dimensions introduced,
the analysis of dynamical content of the theory may appear to be more involved.
In particular, the linearization of the equations proposed in \cite{Bars:2001xv} has not been shown yet to lead to
some known free gauge fields in $d$-dimensions. To make contact to our treatment it is desirable to
have a proposition analogous to Prop. \re{linearize}, which would unambiguously indicate dynamical content of fluctuations
around a vacuum. At the moment it seems
unlikely that the theory proposed in \cite{Bars:2001xv} describes the same dynamics as
our one, one of the reasons is that the action of the theory of \cite{Bars:2001xv}
is of cubic order in dynamical fields that is not easy to relate to our non-polynomial action
\re{actionteta1} constructed in the next section.

\section{The Action}\label{secaction}\label{action111}

\subsection{Formal operator approach}\label{foapp}
According to our strategy one has to find an action $\cA[H]$ being gauge-invariant w.r.t. covariance transformations
\be \label{bsqinf4321}
\d H(x,p)=\o^\dag (x,p) * H(x,p)+H(x,p)*\o(x,p),
\ee
Below we will use what we call "the formal operator approach".
This means writing all formulae in terms of operators rather
than their symbols, and exploiting the standard machinery of quantum mechanics like decomposition
of any operator $\hat{f}$ in terms of a full set of orthogonal states $|i>$ and their duals $<j|$ like
\be
\hat{f}=\sum \limits_{i,j} |i><i|\hat{f}|j><j|= \sum \limits_{i,j} f_{ij} \;|i><j|
\ee
In fact, this formalism is equivalent to treating operators as large matrices.
This formalism is very useful for elucidating the structure of the theory and for checking the algebraic
properties of diverse quantities in the model. After a quantity is formulated in the formal operator approach
it may be rewritten in terms of symbols. After that, one should check whether the formal properties valid for
quantities built from large matrices are still valid for the quantities built from operators whose symbols are power
series in momenta \re{hclass00}.

In terms of operators, the gauge transformations \re{bsqinf} read
\be\ba\label{bsqinf561}
\d \hat{H}(x,p)=\hat{\o}^\dag \hat{H}+\hat{H}\hat{\o}\leftrightarrow \\ \\
\d H_{ij}= \bar{\o}_{ki} H_{kj} +H_{ik}\o_{kj}
\ea\ee
Representing  $\o$ as $\o=a+\frac{i}{\hbar}\e$ one gets
\be
\d H_{ij}= a_{ik} H_{kj} +H_{ik} a_{kj} +\frac{i}{\hbar}
(-\e_{ik} H_{kj} +H_{ik}\e_{kj}).
\ee
$\e$ parameterize infinitesimal unitary transformations, as $H$ is hermitian, it may be
put into diagonal form $H_{ij}=H_i \d_{ij}$, where $\d_{ij}$ is Kronecker's
$\d$-symbol.
Then transformation laws \re{bsqinf561} are rewritten as
\be
\d H_{i}=2 a_{ii} H_i,
\ee
In the finite form, one has
\be \label{bs77}
H'_{i}=e^{2a_{ii}} H_i
\ee
These transformations do
not affect only three quantities, namely, the numbers of states $|i>$ with $H_i>0,H_i=0$ or
$H_i<0$.
Let us denote
\be\ba
\sharp_+=\mbox{dimension of the subspace with} \;H_i > 0\\ \\
\sharp_0=\mbox{dimension of the subspace with} \;H_i = 0\\ \\
\sharp_-=\mbox{dimension of the subspace with} \;H_i < 0.
\ea\ee
To write down these invariants without referring to any particular basis of states one introduces three orthogonal
projectors
$\pi_+(\hat{H}), \pi_0(\hat{H}), \pi_-(\hat{H})$ to the subspaces in which $H_i>0, H_i=0$ or $H_i<0$, respectively,
\be
\pi_+ +\pi_0 +\pi_- =\hat{1}, \; \pi_+\pi_0=\pi_+\pi_-=\pi_0 \pi_-=0,\\ \\
\pi_+^2=\pi_+,\; \pi_0^2=\pi_0,\; \pi_-^2=\pi_-.
\ee
Then
\be\ba
\sharp_+=Tr \pi_+(\hat{H});\;\sharp_0=Tr \pi_0(\hat{H});\;\sharp_-=Tr \pi_-(\hat{H}),
\ea\ee
where trace of an operator is defined as
\be\label{trace000}
Tr \hat{f} = \sum \limits_i f_{ii}.
\ee
Thus, the only functional being
invariant w.r.t. gauge transformations \re{bs77} is
\be\label{actionform}
\cA[\hat{H}]=\a_+  \sharp_+[\hat{H}] + \a_0 \sharp_0[\hat{H}] + \a_- \sharp_-[\hat{H}],
\ee
where $\a_+, \a_0, \a_-$ are some constants.
As the sum of all dimensions $\sharp_+ +\sharp_0+\sharp_-$ does not depend
on $H$, only two constants are independent.

The projectors $\pi_+,\pi_0,\pi_-$ may be formally written down if one uses the notion of a function of operator.
Given a function of a real variable $F(\s)$ and the decomposition of the Hilbert space into a direct sum of eigenspaces
$|H_i>$ of a hermitian operator $\hat{H}$,
\be
\hat{H}|H_i>=H_i|H_i>
\ee
the function of operator $\hat{H}$ is defined as the operator which acts
in each eigenspace as
\be
F(\hat{H})|H_i>=F(H_i)|H_i>.
\ee
Let us define three functions of a real variable $\s$ by the rule
\be\ba
t_+(\s)=0, s\leq 0,\;t_+(\s)=1, s>0\\ \\
t_0(\s)=0, s \neq 0,\;t_0(\s)=1, s=0\\ \\
t_-(\s)=0, s\geq 0,\;t_-(\s)=1, s<0\\ \\
t_+ + t_0 +t_-=1, t_+^2=t_+,\; t_0^2=t_0,\; t_-^2=t_-; t_+t_0=t_+t_-=t_- t_0=0
\ea\ee
Then
\be\ba
\pi_+=t_+(\hat{H}),\;\pi_0=t_0(\hat{H}),\;\pi_-=t_-(\hat{H}),
\ea\ee
and general invariant functional \re{actionform} takes the form
\be
\cA[H]=Tr ( \a_+ t_+[\hat{H}] +\a_0 t_0[\hat{H}] +\a_- t_-[\hat{H}]).
\ee
As $t_+ + t_0 + t_-=1$ the action is rewritten as
\be
\cA[H]=Tr ((\a_+ - \a_-) t_+ [\hat{H}] +(\a_0-\a_-)t_0[\hat{H}] +\a_-),
\ee
that is, up to an infinite constant  $Tr \a_-$ is
\be
\cA[H]=Tr (\g_+ t_+ [\hat{H}] + \g_0 t_0[\hat{H}]),
\ee
To actually calculate this expression for general $H(x,p)$ one has to rewrite the last expression in the langauge of symbols.
This is done in the next two subsections for the case $\g_+=1,\g_0=\fr$, while
generalization to arbitrary $\g$'s does not produce new results.

\subsection{General class of functionals Tr F(H*).\\Semiclassical expansion as a low-energy expansion.}

First of all we show how to rewrite general functional of the form
\be\label{faction0}
A_F[H]=Tr F(\hat{H})
\ee
in terms of symbols. Let $\hat{H}$ be an operator, which acts on the wave functions by the
rule
\be
(\hat{H}\psi)(x)=\int d^d x' \tilde{H}(x,x') \psi(x'),
\ee
with the integral kernel $\tilde{H}(x,x')$. The trace of the operator is defined as
\be
Tr \hat{H} = \int d^d x \tilde{H}(x,x),
\ee
and for a wide class of operators this definition coincides with \re{trace000}.

On the other hand, if $H(x,p)$ is the Weyl symbol of $\hat{H}$,
connected with the kernel by the formula \re{weyl98} of Appendix \re{ApH},
the same expression is identically
rewritten as
\be
Tr \hat{H} =(2\pi\hbar)^{-d} \int d^dx d^dp\;H(x,p)
\ee
In the symbol picture, the notion of function of operator is transformed into the notion of
$*$-function. Given an ordinary function $F(\s)$ and a function $H(x,p)$,
one defines {\it $*$-functions}
of $H$ by the following rules \cite{Bayen:1977hb}.
$*$-polynomials are defined as
\begin{equation}  \label{starpols}
(H*)^k = H*...*H \;\;(\mbox{k times}).
\end{equation}
Then, if a function is decomposed in Tailor series, $*$-function is
defined by the same Tailor series with the (polynomials $\leftrightarrow$
$*$-polynomials) substitution. This way one can define $*$-exponent:
\begin{equation}  \label{zvexp}
 \exp(\lambda H*)= \sum\limits_{k=0}^\infty \frac{1}{k!} (\lambda
H*)^k\;\;,\lambda \in C.
\end{equation}
Note that in contrast to ordinary exponent series which converges for
arbitrary arguments, the convergence of $*$-exponent is an issue.
Then, let
\begin{equation}  \label{fl} F(\sigma)=\int_{C} d\lambda \tilde{F}
(\lambda) e^{\lambda \sigma}, \end{equation} where the integration is
performed along a contour $C$ in a complex plane, be some Fourier-Laplace
type representation for $F(\sigma)$. Then $F(H*)$ is defined as
\begin{equation}  \label{fl1}
F(H*)=\int_{C} d\lambda \tilde{F} (\lambda)  \exp(\lambda H*)
\end{equation}
For a wide class of symbols and operators the symbol of the function of operator $F(\hat{H})$ is given
by the $*$-function $F(H*)$, and the expression \re{faction0} is mapped to
\be\label{action101}
A_F[H]=Tr F(H*)=(2\pi\hbar)^{-d}\int d^d x d^dp\; F(H*)
\ee
Now discuss the invariance properties of the trace operation. In the formal operator formalism,
$Tr(F(\hat{H}))$ is invariant w.r.t. infinitesimal unitary transformations $\d \hat{H} =-\frac{i}{\hbar} (\hat{\e} \hat{H}
-\hat{H} \hat{\e})$ as every eigenvalue $H_i$ is invariant. In the symbol formalism, this invariance is
controlled by the property
\be\label{totder16}
(A*B - B*A) =  \mbox{(total derivative)},
\ee
and therefore (needless to say, w.r.t. infinitesimal unitary transformations $F(H*)$ transforms by the same rule as $H$,
$\delta F(H*) = -\frac{i}{\hbar}[\epsilon , F(H*)]$)
\be
\d Tr F(H*) =-\frac{i}{\hbar}\int d^d x d^d p\; [\epsilon , F(H*)] =\int d^d x d^d p \;\mbox{(total derivative)}
\ee
and equals zero if total derivative vanishes at infinity (we suppose there are no bursts in the total derivative so
integral of the total derivative resides at infinity, it will always be the case
for the expressions we deal with in the paper). In our treatment, the vanishing of the total derivatives at infinity
will take place due to reasons to be discussed a little bit later (see discussion after \re{totder8}).

Note also the ``simple multiplication''property of the Weyl symbol
\be\label{simmul}
Tr (A*B) = Tr (AB)+ \int d^dx d^dp\; \mbox{(total derivative)}
\ee
useful in diverse calculations.

Our main tool for calculating expressions like \re{action101} is as follows.

Every functional (\ref{action101}) may be rewritten in a standard field theoretic form
providing a kind of low-energy expansion, like the $\alpha^{\prime}$ expansion in string
theory where, if one neglects the overall $\frac{1}{(2\pi\hbar)^d}$ prefactor,
$\hbar^2$ plays the same role of parameter counting
the number of space-time derivatives as $\alpha^{\prime}$ does. The
$\hbar^0$-term
contains no space-time derivatives and exhibits itself as a cosmological term.
Indeed, every $*$-function $F(H*)$ turns out to possess a semiclassical expansion
in powers of $\hbar$
\begin{equation}  \label{quasi}
F(H*) = \sum \limits_{k=0}^{\infty}\hbar^{2k} F_{2k}(H)\;\;,\;\; F_0(H)=F(H).
\end{equation}
To ground this statement, consider $*$-exponent {(\ref{zvexp})}
\begin{equation}  \label{deu}
u(\lambda H | \hbar) \equiv  \exp (\lambda H*).
\end{equation}
It satisfies the differential equation
\begin{equation}  \label{deu1}
\frac{\partial}{\partial \lambda} u(\lambda H | \hbar) = H\,*\,u(\lambda H |
\hbar)= H  \exp (\frac{i\hbar}{2} {\stackrel{\leftrightarrow}{\nabla}} )
u(\lambda H | \hbar).
\end{equation}
Writing down $u(\lambda H | \hbar)$ in a form of the semiclassical
expansion
\begin{equation}  \label{deu2}
u(\lambda H | \hbar)= \sum \limits_{k=0}^{\infty} (\frac{i\hbar}{2})^k
u_k(\lambda H)\;\;
\end{equation}
one gets the simple recurrent differential equations for $u_k(\lambda H)$:
\begin{equation}  \label{deu3}
\begin{array}{c}
\frac{\partial}{\partial \lambda} u_0 = H u_0 \\
\\
\frac{\partial}{\partial \lambda} u_1 = H {\stackrel{\leftrightarrow}{\nabla}%
} u_0 +H u_1 \\
\\
\frac{\partial}{\partial \lambda} u_2 = \frac{1}{2} H ( {\stackrel{%
\leftrightarrow}{\nabla}} )^2 u_0 + H {\stackrel{\leftrightarrow}{\nabla}}
u_1 +H u_2 \\
\\
... \\
\\
\frac{\partial}{\partial \lambda} u_k = \sum \limits_{l=0}^{k} \frac{1}{l!}
H ( {\stackrel{\leftrightarrow}{\nabla}} )^l u_{k-l}
\end{array}
\end{equation}
etc., wherefrom accounting the initial condition $u(0) =1$ one gets the
solution\footnote{The expressions like $(A {\stackrel{%
\leftrightarrow}{\nabla}} B {\stackrel{\leftrightarrow}{\nabla}}
C)$ are to be understood in a sense that
${\stackrel{\leftrightarrow}{\nabla}} $ acts only on its
neighbors, i.e. the left ${\stackrel{\leftrightarrow}{\nabla}} $
acts on $A$ and $B$ while the right one acts on $B$ and $C$.}
\begin{equation}  \label{deu4}
\begin{array}{c}
u_0 = e^{\lambda H} \\
\\
u_1 = e^{\lambda H} \frac{\lambda^2}{2} (H {\stackrel{\leftrightarrow}{\nabla
}} H) =0\\
\\
u_2 = e^{\lambda H} \left(\frac{\lambda^2}{4} (H {\stackrel{\leftrightarrow}{
\nabla}} ^2 H) - \frac{\lambda^3}{6} (H {\stackrel{\leftrightarrow}{
\nabla}} H {\stackrel{\leftrightarrow}{\nabla}} H) ) \right) \\
\\
... \\
\\
u_{2k+1}=0,\;\;\; u_{2k} = e^{\lambda H} v_{2k} (\l H, \l \partial H, ..., \l(\partial)^{2k} H) \\
\end{array}
\end{equation}
where $v_{2k}$ is a polynomial of the highest degree $3k$ in $\l
H$ and its derivatives w.r.t. phase space variables $(x,p)$. The
total order of derivatives in each $\hbar^k$ term equals $2k$.
Note that in the Weyl symbol {(\ref{nab})} $H
{\stackrel{\leftrightarrow}{\nabla}} H =0 $, therefore the terms
including this combination vanish.

Substituting the decomposition (\ref{deu1}-\ref{deu4}) in a Fourier-Laplace
representation {(\ref{fl1})} one gets the semiclassical decomposition for
an arbitrary $*$-function
\begin{equation}  \label{fquasi}
\begin{array}{c}
F(H*) = F(H)
+\\ \\
- \frac{\hbar^2}{4} \left( \frac{1}{4} F^{\prime\prime}(H) (H {\stackrel{
\leftrightarrow}{\nabla}} ^2 H) - \frac{1}{6}F^{\prime\prime\prime}(H)(H {\stackrel{\leftrightarrow}{\nabla}} H {\stackrel{\leftrightarrow}{%
\nabla}} H )\right)+ \\
\\
+... + (\frac{\hbar}{2})^{2k} \sum \limits_{n=2}^{3k} F^{(n)}(H)
Pol_{2k,n}(H, \partial H, ... \partial^{4k} H) + ...
\end{array}
\end{equation}
where $F^{\prime\prime}, F^{\prime\prime\prime},..., F^{k}$ are
the second, third and highest  derivatives of $F$ by it's
argument, $Pol_{2k,n}$ are some monomials of degree $n$ in $H$ and
its derivatives w.r.t. phase space variables $(x,p)$ up to the
order $4k$.

It is clear from the structure of $*$ product and
${\stackrel{\leftrightarrow }{\nabla}} $ operator that the numbers
of $x$ and $p$-derivatives in each term are equal to each other
and equal $2k$. Thus, every $\hbar^{2k}$-order of the
semiclassical expansion $F(H*)$ is nothing but the $2k$-th order
in $x$ derivatives. Therefore, the semiclassical expansion takes a
form of low energy expansion for $F(H*)$ where $\hbar$ is an
expansion parameter counting number of $x$-derivatives.
Calculating the trace (\ref{action101}), one finds the low-energy
decomposition of the action
\begin{equation}  \label{inva5}
A_F [H] = A_0  + \hbar^2 A_2 +\hbar^4 A_4 + ... =(2\pi \hbar)^{-d}\int
d^d x \,d^d p \;F (H(x,p))  +\hbar^2 A_2 +\hbar^4 A_4 + ...
\end{equation}
The classical term $A_0$ is invariant w.r.t. classical canonical transformations,
$\d H =\{\e,H\}$ up to a total derivative. Here it represents
the "cosmological term". Therefore, the cosmological term is intimately
related to the classical limit of the quantized particle geometry\footnote{
This, very natural, appearance of the cosmological term conforms with
Fradkin - Vasiliev results on the inevitable appearance of cosmological
backgrounds in higher spin models\cite{Fradkin:ah}.}.

To make contact with standard field theoretic actions one has to
integrate in {(\ref{inva5})} over momenta, then one gets the standard
actions for the infinite collection of symmetric tensor fields on the
$x$-space.

There do exist such "Lagrangians" $F$ that the integration
over momenta is well-defined in the class of power series in momenta
(\ref{hclass00}), at least within the framework of some perturbative procedure.
At least with our choice for $F$ (see the next section, Eq. \re{fteta}),
all integrations  over $p_m$ are finite
and easily calculated at any order in $\hbar$.

\subsection{The Action}

Among general functionals of the form \re{action101}, there exists the unique one,
where the gauge invariance $\d H=-\frac{i}{\hbar}[H,\e]$ is extended to the full algebra of
similarity transformations of the form
\be\label{bsqinf5432}
\d H =\omega^\dag * H + H*\omega,
\ee
which is nothing but  the covariance algebra of our geometry.
Gauge parameters $\o$ are required to have the form of power series in momenta, with coefficients being
tensor fields with compact support in $x$-space. Note that this restriction defines a subalgebra of the total algebra
of smooth phase-space functions \re{hclass2}, at least perturbatively in $\hbar$.

We recall that in the trace
operation the order of product factors is inessential up to  a total derivative,
then the variation of $A_F [H]$ under  \re{bsqinf5432}
\begin{equation}
\delta A_F [H] =2 \int d^d x d^d p\;  F^{\prime}(H*) \,*\, H * \,a +\mbox{total derivative}.
\end{equation}
The only solution providing vanishing of this variation (modulo total derivative) for arbitrary $H$
and arbitrary $a$ is
\begin{equation} \label{fteta} F(H)=\n_1 \theta(\n_2 \sigma)+\n_3;\;
F^{\prime}(\sigma) =\nu_1 \nu_2\delta (\nu_2 \sigma).  \end{equation}
here $\n_{1,2,3}$ are real constants, which may be set, without loss of
generality, to $\n_1=\n_2=1, \n_3=0$, and $\theta(\s)$ is the ``step''
$\theta$-function:
\be \ba
\theta(\sigma)=0\;\;,\;\; \sigma < 0 \\ \\
\theta(\sigma)=\fr  \;\;,\;\; \sigma = 0\\ \\
\theta(\sigma)=1\;\;,\;\; \sigma > 0,
\ea
\ee
while $\d(\s)=\frac{\pa}{\pa \s} \theta(\s)$ is an ordinary $\d$-function.
The variation vanishes due to the identity \be
\delta(H) * H \equiv 0.\ee
This identity is not a formal one, in fact being decomposed in
semiclassical series like \re{fquasi} it generates an infinite number of
relations expressed in terms of derivatives of the equation $\d(\s)\s\equiv0$
and valid as identities for distributions.

The choice of $F(H)$ we just have made conforms with considerations based on the formal operator approach
of subsec. \re{foapp}. and corresponds to the option
\be\label{choice500}
\g_+=1, \; \g_0=\fr.
\ee Thus, the invariance w.r.t. covariance transformations {(\ref{bsq}
)}, {(\ref{bsqinf2})}  fixes the form of action {\it unambiguously}, up to
the choice of the constant $\theta(0)$
which, as it will be seen below, is inessential so our choice \re{choice500} does not
spoil generality.
Thus,
\be\label{actionteta1}
\cA[H]=Tr \;\theta(H*).
\ee
The action can be formally represented as an "invariant induced action"
\be
\cA[H]=\int \cD x \cD p  \frac{\cD \l}{\l} \d(\dot{\l})
 \exp \{\frac{i}{\hbar} \int \limits_{0}^{1} d\t\left( p\dot{x}-\l H(x,p)\right)\},
\ee
where the $\d(\dot{\l})$-insertion implies the natural gauge $\l=const$ \re{gaugecond}, and
the remaining global $\t$-reparametrizations \re{globmodes} are used to set the
limits of $\t$-integration to 0 and 1.
After this gauge is implied the
functional integral by $\l$ reduces to an ordinary one by $\l$'s zero mode.
Choosing  a special integration contour leads to our action \re{actionteta1}. Indeed,
recalling the functional integral representation of the trace of the evolution operator \cite{berezin}
\be
Tr  \exp^{i\l\hat{H}}=\int \cD x \cD p  \;\exp \{\frac{i}{\hbar}\int \limits_{0}^{\l} d\t \left(p\dot{x}-H(x,p)\right)\},
\ee
and the integral representation for $\theta$-function
\be
\theta(\s)=\frac{1}{2\pi i} \int \limits_{-\infty}^{\infty} \frac{d \o}{\o-i\ve} e^{i\o\s},
\ee
one gets \re{actionteta1} after specifying integration contour in $\l$-space as
\begin{equation}  \label{optionCC}
C_{\theta}: \lambda(\tau) = \tau -i\varepsilon ;\;\tau \in (-\infty,\infty),\;
\varepsilon \rightarrow 0^+.
\end{equation}
Note that this is {\it not} the contour that provides the one-loop effective action $Tr Ln (\hat{H})$,
the latter would arise after choosing the contour
\be
C_{\mbox{one-loop}}: \l(\t) =\t, \t \in [\ve, \infty)\;,\; \ve \rightarrow 0^+
\ee
which is ultraviolet divergent in general as $\ve$ tends to zero.

Thus, the action has the form
\begin{equation}  \label{fquasi111}
\begin{array}{c}
\cA[H]=Tr \;\theta(H*) = (2\pi\hbar)^{-d} \int d^dx d^dp\; \left\{\theta (H) +
\right.\\ \\
- \frac{\hbar^2}{4} \left( \frac{1}{4} \d^{\prime}(H) (H {\stackrel{
\leftrightarrow}{\nabla}} ^2 H) - \frac{1}{6}\d^{\prime\prime}(H) (
H {\stackrel{\leftrightarrow}{\nabla}} H {\stackrel{\leftrightarrow}{
\nabla}} H ) \right) +\\
\\ \left.
+... + (\frac{\hbar}{2})^{2k} \sum \limits_{n=2}^{3k} \d^{(n-1)}(H)
Pol_{2k,n}(H, \partial H, ... \partial^{l} H) + ...\right\}
\end{array}
\end{equation}
Below, we will refer the $(2\pi\hbar)^{-d} \hbar^{2k}$- terms  as $\hbar^{2k}$-ones, in so doing
$2k$ appear to be equal to the total number of $x$-derivatives of dynamical fields.

\subsection{Classes of hamiltonians. $\cC$, $\cC_p$ and $\tilde{\cC}_p$}\label{classesh}

It is important to find a reasonable class of hamiltonians $H(x,p)$ for which
the action is well-defined. One example of such a class, which we denote $\cC$ is provided by  hamiltonians
for which the constraint surface $H(x,p)=0$ is a boundary of a {\it compact} "inside" domain $In$ in the phase space
\be
\{In| (x,p) \in In \Leftrightarrow H(x,p)<0 \} ,\;\{\cC| H\in \cC \Leftrightarrow In \;\mbox{is compact}\}.
\ee
Note that this class of hamiltonians is invariant, at least perturbatively in $\hbar$, w.r.t. gauge transformations
\re{bsqinf5432}. \footnote{As $H=0$ is a compact surface, for general $H \in \cC$ this is equivalent
to requiring that the surface $H=E$ is also compact, for $E$ being small enough.
This implies that $\hat{H}\in \cC$ possesses a discrete spectrum which starts from $E=0$.}

To show $H\in \cC$ is a good option for the action \re{actionteta1},
consider first the classical, $\hbar^0$ term, proportional to $\theta(H)$.
The constraint surface $H=0$ is a boundary of two domains
$In$ and $Out$ in the phase space, such that the phase space is split into two regions:
$In$ ``inside'', with $H>0$, and $Out$``outside'',  with $H<0$.
As $\theta(H)=1$
``inside'' and $\theta(H)=0$ ``outside'', the classical term equals
(up to the multiplier) the volume of $In$. Note that the value $\theta(0)$
is inessential as the surface $H=0$ has zero measure in the phase space.

The higher $\hbar^{2k}$-terms, interpreted as {\it quantum corrections to the classical term},
are proportional to derivatives of $\d$-function. Therefore, they are also
finite as due to the presence of $\d^{(k)}(H)$ the integrals
reside on the compact constraint surface $H=0$.

Thus, one has the following statement.
\begin{teo}
In the case constraint surface $H\in \cC$, i.e. if the constraint surface
$H(x,p)=0$ is a boundary of a compact domain,
the full action is the ``quantized volume'' of the ``inside'' region
$H(x,p)>0$ in the phase space.
After the expansion into the semiclassical series \re{fquasi111},  the classical term
is the volume of the ``inside'' region. The quantum corrections of order $\hbar^{2k}, k=1,2,...$
are given by integrals of densities localized on the constraint surface $H(x,p)=0$.\end{teo}

So far we have described the case when $H \in \cC$ and the action is finite (at least semiclassically, i.e.
in each order in $\hbar$).
Note this case is not precisely what we need. Actually, we need the situation when 1) $H(x,p)$ is
a power series in momenta with coefficients $H^{m(k)}(x)$ being arbitrary smooth tensor fields in $x$-space 2) the result
of integration
by momenta in \re{actionteta1} is finite, while the action need not be finite. In this case, the action is interpreted as
a conventional one represented by an integral by $x$ of a Lagrangian being a function of tensor fields $H^{m(k)}(x)$
and their $x$-derivatives. Therefore, it is vital to find a reasonable class
of $H(x,p)$ for which the integrals by momenta converge and can be calculated
explicitly. A good option is the analog of the class $\cC$ described above, the class $\cC_p$
of hamiltonians $H(x,p)$
for which the constraint surface $H(x,p)=0$ is, for each $x$, a boundary of a compact "inside" domain
$\{In_x | p \in In_x \;\mbox{if}\; H(x,p) > 0\}$.

Indeed, then already before $x$-integration
each term of the semiclassical expansion \re{inva5} is an integral of a regular expression
by $In_x$ (classical term) or by the constraint surface (quantum corrections). As $In_x$ and $H=0$ are compact
each term of the $\hbar-$expansion is well-defined.

Thus, provided $H\in \cC_p$ the action \re{actionteta1} has a form of an $x$-integral of the Lagrangian, depending
on components of $H(x,p)$ and their $x$-derivatives.

The first, classical term of the semiclassical expansion of the Lagrangian, corresponding
to the action \re{actionteta1} has the
interpretation of the {\it volume of the domain bounded by the constraint
surface}. This is clear due to the same reasoning as for the $H\in \cC$ calculations.
Namely, $\theta(H)=1$
``inside'' and $\theta(H)=0$ ``outside'', therefore, the classical term in the Lagrangian equals
(up to the multiplier$(2\pi\hbar)^{-d}$) the volume of $In_x$. This
classical term does not contain $x$-derivatives of the components of $H$ and is
interpreted as a {\it cosmological term}. Again the value $\theta(0)$
is inessential as the surface $H=0$ has zero measure in the phase space.

The higher $\hbar^{2k}$-terms, proportional to derivatives of $\d$-function,
contain $2k$ space-time derivatives of the
dynamical fields and are interpreted as {\it quantum corrections} to the
classical(=cosmological) term. Thus the Lagrangian of the theory may be spelled as
follows.\begin{teo} In the case constraint surface $H\in \cC_p$, i.e. if for each $x$ the constraint surface
$H(x,p)=0$ is a boundary of a compact domain,
the full Lagrangian is the ``quantized volume'' of the ``inside'' region
$H(x,p)>0$ in the phase space.
After the expansion into the semiclassical series \re{fquasi111},  the classical term
is the volume of the ``inside'' region, it does not contain $x$-
derivatives of the dynamical fields and therefore plays the role of a
cosmological term. The quantum corrections of order $\hbar^{2k}, k=1,2,...$
contain $x$-derivatives of the dynamical fields of order $2k$ and are given by
integrals of densities localized on the constraint surface $H(x,p)=0$.\end{teo}

Note that of our main interest in this paper are the hamiltonians $H(x,p)$ which are the power series in momenta \re{hclass00}.
Therefore, we are interested if it is possible to have an overlap between $\cC_p$ and the space of power series in momenta.
It is clear that these two classes have wide overlap which we denote $\tilde{\cC}_p$.
In order a power series in momenta $H(x,p)$ to belong to $\cC_p$ the coefficients of this power series should satisfy
some inequalities which do not impose algebraic constraints on the coefficients $H^{m(k)}$ of $H(x,p)$ and
therefore do not reduce the dimension of the space of power series. We refer these inequalities
as {\it $\cC_p$-restrictions}.
Although we do not find these restrictions explicitly we suppose they are satisfied so that
\be
H(x,p)=\sum \limits_{s=0}^{\infty} H^{m_1...m_s}(x)\;p_{m_1}...p_{m_s} \in \cC_p
\ee
A good test of the validity of the last formula would be the convergence of all
integrals by momenta in all calculations we are going to perform, while the appearance of
divergencies would be interpreted as spoiling the $\cC_p$-restrictions.

For practical calculations, it is convenient
considering general hamiltonians as a sum of
the ``low spin'' and ``higher spin'' parts,
\be \ba \label{wdecomp}
H(x,p)=H_2(x,p)+w\Delta H(x,p)\\ \\
H_2(x,p)=\sum \limits_{s=0}^{2} H^{m_1...m_s}(x)\; p_{m_1}... p_{m_s}=
D(x)-g^{mn}(x) \Pi_m \Pi_n,\;\; \Pi_m=p_m-A_m(x)\\ \\
\Delta H(x,p)=\sum \limits_{s=3}^{\infty} H^{m_1...m_s}(x) p_{m_1}... p_{m_s},
\ea\ee
where $D(x)\geq 0$, the metric $g_{mn}$ is Euclidean. Then, if $\D H=0$ then $H\in \cC_p$ as, for each $x$,
the constraint surface $H=H_2=0$ is a generalized sphere of radius $\sqrt{D \det(g)}$.
If to treat general $H$ of the form \re{wdecomp} perturbatively by considering $\D H$ as a fluctuation with the
expansion parameter $w$, then all the expressions in \re{fquasi111}
are to be calculated perturbatively in powers of $w$.
Notably, given a fixed order of the semiclassical expansion \re{fquasi111},
each term of its $w$-expansion is {\it finite}, because it is represented by integrals of the form
$\d^{(k)}(H_2)L(H_2,\D H, \pa H_2, \pa \D H, ...)$ localized on the generalized sphere $H_2=0$.
Thus, the Lagrangian is finite in the framework of the perturbative expansion in $\hbar$ and $w$ in the sense
each term of this expansion is finite for arbitrary hamiltonians of the form \re{wdecomp}, even if
$\cC_p$-restrictions are not obeyed.
In the case $H\in \cC_p$, the series in $w$ converge to exact expressions written down in terms of entire $H$ without its decomposition
into low spin and higher spin parts. In general, summing the power series in
$w$ may lead to divergencies, we interpret these divergencies as arising because $H(x,p)$ ceases to belong to $\cC_p$.

An important remark concerns the difference between even- and odd- dimensional case.
All the integrals to be calculated are given by successive
derivatives by $D$ of the expressions like $\int d^dp\; \theta(D-p^2)
p_{m_1}...p_{m_{2s}} \sim D^{\hd+s}\sg (g_{m(2)})^s$.
Therefore,
in even dimensions all expressions contain only {\it non-negative} powers of $D$ while in odd
ones arbitrary negative powers of $D$ are possible. One more point is that in even dimensions
one has a huge variety of $D^0$-terms, i.e. there exists a well-defined and nontrivial
limit $D\rightarrow 0$.

After integrals by momenta are evaluated either in terms of $w$-expansion for $H$ of the form \re{wdecomp} or explicitly
for $H \in \cC_p$ one gets the Lagrangian that depends on the tensor fields $H^{m(k)}$ and their $x$-derivatives.
Then one takes away the $\cC_p$-restrictions on the coefficients of the power series
and considers $H^{m(s)}$ of a general form. The Lagrangian remains finite as in terms of
$H^{m(s)}$ it is represented by the same expression. Thus one obtains
the analytic continuation of the Lagrangian to the cases
when it does not have the interpretation of the "quantized volume".
This includes the cases when the metric in \re{wdecomp} is not Euclidean, since the surface $H_2=0$ is clearly non-compact
in this case. To illustrate what happens we  deliver a simple example. Consider
the integrals
\be\ba
I_1[A_0]=\int d^d p\; \d(m^2-\eta^{mn}p_m p_n) A_0(p)\\ \\
I_2[A_1,A_2]=\int d^d p\; \d(m^2-\eta^{mn}p_m p_n) A_1(p)A_2(p),
\ea\ee
where $A_{0,1,2}(p)$ are power series in momenta,
\be
A_{0,1,2}(p)=\sum \limits_{s=0}^{\infty} A_{0,1,2}^{m_1...m_s} p_{m_1}...p_{m_s},
\ee
and the metric $\eta^{mn}$ is Euclidean. The integration runs over $d-1$-sphere of radius $m$ and are clearly
well-defined. The integrals are
\be\ba
I_[A_0]= \sqrt{\det{\eta_{mn}}}\sum \limits_{s=0}^{\infty} b_s m^{d-1+2s} A^{m_1...m_{2s}}\eta_{m_1 m_2}...\eta_{m_{2s-1}m_{2s}}\\ \\
I[A_1,A_2]=\sqrt{\det{\eta_{mn}}}\sum \limits_{s=0}^{\infty} c_s m^{d-1+2s} A_{1}^{m_1...m_s} A_{2\;m_1...m_s},
\ea\ee
where $b_s, c_s$ are some easily calculable numbers, and the contractions are by
the Euclidean metric $\eta^{mn}$.
The analytic continuation of these expression is achieved just by
ascribing another signature to the metric, after this continuation, the expression $I[A],I[A_1, A_2]$ are still
well-defined but become pure imaginary, as it is standard in passing from Euclidean to Minkowski space.
The imaginary unit pre-factor is to be absorbed by the same pre-factor which comes from the
analytic continuation of the integration measure in $x$-space.

Contrary, if one tries to change signature of the metric {\it before} integration by momenta in \re{fquasi111}
one gets divergent expression as the surface $p^2=m^2$ is not a sphere but a noncompact hyperboloid or a cone.
To summarize, after integration by momenta and analytic continuation to the cases when $H \not \in \cC_p$ the Lagrangian
is still well-defined but lacks the interpretation of the "quantized volume".

Note that the analytic continuation of the quadratic Hamiltonian $H_2 \in \cC_p$ \re{wdecomp} to Minkowski space,
achieved by changing the sign of $g^{00}(x)$, lead to the Hamiltonian of {\it tachyon}.
Clearly, the Minkowski version of all $H_2 \in \cC_p$
in the framework of the $w$-expansion \re{wdecomp} lead to hamiltonians that describe tachyon in
low and higher spin fields. Thus one can name the backgrounds $H\in \cC_p$ analytically continued to Minkowski space
as {\it tachyon backgrounds}.

\subsection{Actual gauge invariance of the action}\label{actual23}

Now let us discuss the gauge invariance of the action (\ref{actionteta1}, \ref{fquasi111}).
First of all, by construction it is
"formally invariant", that means its variation w.r.t. gauge transformations \re{bsqinf5432} is the integral
of a total derivative.
It appears this formal invariance may be proved to be an actual gauge invariance,
{\it if $H \in \cC_p$}, at least in the framework of the semiclassical expansion
by powers of $\hbar$.

Instead of finding this total derivative explicitly (although it may be done after some algebra)
we note that as the action \re{fquasi111} and gauge transformations \re{bsqinf5432}
are expanded in power series in $\hbar$
\be
\cA=\cA_0+\hbar^2\cA_2+...,\;\d H=\d_0 H +\hbar \d_1 H +\hbar^2 \d_2 H +...
\ee
the conditions of gauge invariance of the action may be expanded in powers of $\hbar$ either,
\be\label{totder7}
\d \cA =\d_0 \cA_0 +\hbar (\d_1 \cA_0) +\hbar^2 (\d_o \cA_1 +\d_2 \cA_0) +...=\int d^dx d^dp\;
\{ [\mbox{t.d.}]_0+\hbar [\mbox{t.d.}]_1 +\hbar^2 [\mbox{t.d.}]_2 +...\}
\ee
where $\d_k \cA_l$ stands for the variation of $\cA_l$ due to
$\d_k$-terms in the variation of $H$, and "t.d.$_n$"
are the coefficients of the expansion of the total derivative \re{totder7} in the power series in $\hbar$.
Explicitly, one gets
\be\label{totder8}
[\mbox{t.d.}]_0 = 0,\; [\mbox{t.d.}]_1= \e \stackrel{\leftrightarrow}{\nabla} \theta(H),\;[\mbox{t.d.}]_2=\frac{2}{3} H \stackrel{\leftrightarrow}{\nabla} \left(\d'(H)
(a \stackrel{\leftrightarrow}{\nabla} H)\right),...
\ee
each $[$t.d.$]_k$  is an expression of the form $A \stackrel{\leftrightarrow}{\nabla} B$, i.e. a total derivative indeed.
All the total derivative terms vanish at infinity by the following reasons.
General structure of these terms is $[$t.d.$]=\frac{\pa}{\pa x^m} f^m +\frac{\pa}{\pa p_m} g_m$. After integrating by $(x,p)$ these terms
are transformed in integrals by $x$-infinity ($f$-term) and by $p$-infinity ($g$-term).
As $x \to \infty$, the $f$-term vanishes as it
contains gauge parameters which have compact support in $x$ space and  vanish, together with all their derivatives,
outside the support domain. As $p\to\infty$, the total derivative vanishes as $g_m$
has the form $ \d^{k}(H) \tilde{g}_m, k=0,1,2,...$
localized at the constraint surface $H=0$ which does not contain $p$-infinity, and therefore $g^m(p\to\infty)=0$.
This reasoning applies in the framework of the semiclassical expansion.
Thus, one has \begin{teo}\label{teort}
If $H\in \cC_p$ then
the "quantized volume" action (\ref{actionteta1},\ref{fquasi111})
is gauge invariant, at least perturbatively in $\hbar$, w.r.t. gauge transformations
\re{bsqinf5432} with gauge parameters being power series in momenta with coefficients being smooth tensor fields with
a compact support in $x$-space.\end{teo}

The conditions of this theorem may be relaxed but we do not consider this task here.
For our needs the theorem \re{teort} is sufficient. Indeed, as we saw above there exists a natural choice of
$H$ where the constraint surface is compact, namely, $H=H_2$ where the metric inside $H_2$ is Euclidean and the dilaton
is non-negative. General hamiltonians are treated perturbatively like in \re{wdecomp} where it is
supposed $\cC_p$-restrictions are obeyed.
Thus, one has \begin{teo}
The "quantized volume" action (\ref{actionteta1},\ref{fquasi111}) is
gauge invariant (perturbatively in $\hbar$) for $H$ being the power series of the form \re{wdecomp}
with Euclidean metric and
$D\geq 0$, provided that tensor fields
entering the power series $\D H$ are subject to restrictions (which do not reduce the dimension of the space of fields)
insuring that the constraint surface $H_2+\D H=0$ is compact.\end{teo}

After the integration by momenta is performed in the Euclidean regime (i.e. when $H_2$ contains Euclidean metric)
the result of integration may be analytically continued, along the lines explained above,
to the cases when the metric has another signature.

In so doing, the gauge invariance is not spoiled as after integration by momenta in Euclidean regime,
because check of gauge invariance does not involve the $\cC_p$-restrictions. Indeed,
these restrictions do not impose algebraic constraints on $H^{m(k)}$ and therefore could not insure vanishing
of any expression built from $H^{m(k)}$. On the other hand,
the variation of the action is zero in the Euclidean regime, therefore, after integration by momenta
its vanishing is controlled by the very structure of the Lagrangian expressed in terms of the component fields,
this structure of course remains the same whether $H\in \tilde{\cC}_p$ or not.

\section{Restoration of the low spin data. Low spin ansatz: Weyl-invariant dilaton gravity}\label{srest}

In this section, we show that the low-spin truncation of the "quantized volume" action \re{fquasi111}  naturally leads to a
Weyl-invariant dilaton  gravity.

\subsection{Calculation}\label{calcttt}

The action has the form
\be \label{acti}
\cA[H]=Tr \;\theta(H*)=\frac{1}{(2\pi\hbar)^d} \int d^dx d^dp\; \theta(H*).
\ee
The formula \re{fquasi111} is rewritten as
\be\label{semicl3}
\theta(H*)=\theta(H)- {\frac{\hbar^2}{8}}\left[ \theta''(H) C_1
+\frac{1}{3} \theta'''(H) C_2 \right] +o(\hbar^4),
\ee
where
\be\ba
C_1 =H_{mn} H^{mn} - H^m_n H^n_m \\ \\
C_2= H_m H_n H^{mn} +H^m H^n  H_{mn} - 2 H^m H_n H^m_n,
\ea\ee
where the lower and the upper indices
denote the derivatives w.r.t. coordinates $x^m$ and momenta $p_m$,
respectively.
In this section we perform the calculation of this action for the particular
"low spin" Hamiltonian of the second order
\begin{equation}  \label{quadr111}
H_2=D(x)-g^{m n }(x)p_m p_n,
\end{equation}
representing general gravity+Maxwell+dilaton background.
To simplify the calculation we set zero the Maxwell field $A_m(x)$.
As we argued above, in the case $D \geq 0$ and if the metric is Euclidean what we suppose henceforth
in this section, the Hamiltonian \re{quadr111} is a simplest example when $H\in \cC_p$, therefore all
integrals will converge.

In the case under consideration, the last formula turns to
\be\ba
C_1=( D_{,mn} -  g^{kl}_{,mn} p_k p_l)(-2 g^{mn}) -4
g^{nk}_{,m} p_k g^{ml}_{,n} p_l \\ \\
C_2= (D_m -  g^{kl}_{,m} p_k p_l)( D_{,n}-  g^{k'l'}_{,n} p_{k'}p_{l'})
(-2 g^{mn}) +\\ \\+4(D_{,mn} - g^{kl}_{,mn} p_k p_l) g^{mr} g^{ns} p_r p_s -
8 g^{mk} p_k (D_n -  g^{ef}_{,n} p_e p_f) g^{nr}_m p_r.
\ea\ee
It is seen one has to calculate integrals like
\be\label{inttk}
I^{(k)}_{m_1...m_s}(x)=\int d^dp\; \theta^{(k)}(H_2)p_{m_1}...p_{m_s}.
\ee
In fact only even-$s$ integrals are to be calculated, because
odd-$s$ integrals equal zero.
The simplest integral, $I^{(0)}$, which reproduces the classical part
of the semiclassical expansion \re{fquasi111} is
\be\label{intt0}
I^{(0)}(x)=\int d^dp\; \theta (D - p^2).
\ee
Repeating the reasoning from Sec. \re{classesh} let us
note that the surface $H_2=D- p^2=0$ splits the phase space into two
domains: ``inside'', $H_2>0 \leftrightarrow p^2
< 2D$ and ``outside'' $H_2<0 \leftrightarrow p^2 > D$. By definition of the $\theta$-function,
the integrand is zero outside. Therefore, {\it all the integrals \re{inttk} are well-defined}
as the actual integration runs over a compact domain.

The integral \re{intt0} is
equal to the {\it volume of the domain bounded by the constraint surface
$H_2=0$}.
$I^{(k)}$ are obtained via $I^{(0)}$ by the successive differentiation
by dilaton $D$,
\be
I^{(k)}_{m_1...m_s} (x)=(\frac{\pa}{\pa D})^k I^{(0)}_{m_1...m_s}(x),
\ee
while $I^{(0)}_{m_1...m_s}(x)$ are encoded by the generating function
\be
G_d (l) =\sum \limits_{s=0}^{\infty} \frac{i^s}{s!} I^{(0)}_{m_1...m_s}(x) \;l^{m_1}
... l^{m_s}=\int d^dp\; \theta (H_2) e^{ip_m l^m}.
\ee
$G_d$ is easily calculated (see App. \re{ApD}) to be
\be \label{intt1}
G_d(l)=2^{\hd}\pi^{\hd} l^{-\hd} \sqrt{g} D^{\frac{d}{4}} J_{\hd}\left(l\sqrt{D}\right),
\ee
where $l=\sqrt{l^2}$,
$g = \det (g_{mn})$. $J_{\hd}$ is the Bessel's function of the first kind (App.
\re{ApE}),
\be
J_{k}(z)=(\frac{z}{2})^p \sum \limits_{k=0}^{\infty}
\frac{(-)^k}{k! \G(p+1+k)} (\frac{z}{2})^{2k}
\ee
Note that the generating function \re{intt1} is a power series
in even degrees of $l$, i.e., a power series in $l^2$.
Expanding both sides of \re{intt1} in $l^m$ one gets the expressions of
integrals \re{inttk} in terms of tensor powers of the metric $g_{mn}$.

Employing \re{intt1} and its derivatives by dilaton $D$ (see \re{genf4}), one can easily
get the answer for any integral \re{inttk}.
In this subsection we need only integrals \be
\ba \int d^dp\; \theta (D- p^2)=\sqrt{g} \;v_d \;D^{\hd}\\ \\ \int d^dp\;
\theta (D- p^2) \,p_m p_n=\sqrt{g} \;v_d \;D^{\hd +1}\frac{1}{d+2}
g_{mn}\\ \\ \int d^dp\; \theta (D- p^2) \,p_k p_l p_{k'} p_{l'}= \sqrt{g}\;
v_d \;D^{\hd+2}\frac{1}{(d+4)(d+2)} G_{klk'l'}, \ea\ee where \be
G_{klk'l'}=g_{kl} g_{k'l'} +g_{kk'} g_{l'l}+g_{kl'} g_{lk'},
\ee
where
\be
v_d=\frac{\pi^{\hd}}{\G(\hd+1)}
\ee
is the volume of a unit ball in $d$ dimensions,
and the consequences of these integrals obtained by a few
differentiations w.r.t. $D$. This is enough for easily writing down the
result of integration by momenta for the expression \re{acti}:
\be  \ba\label{actires}
\cA[H_2] \equiv \cA[D(x), g_{mn}(x)] = \\ \\
\frac{v_d}{(2\pi\hbar)^d}
\int d^dx \sg\left\{ D^{\hd} -\frac{\hbar^2}{12}
\hd(\hd-1)(\hd-2)D^{\hd -3} g^{mn} D_{,m} D_{,n}
-\frac{\hbar^2 d}{8}  D^{\hd-1} Y \right. +\\ \\
+\left.\frac{\hbar^2}{12} \frac{d}{\sg}  \pa_n (\sg g^{mn} \pa_n
D^{\hd-1}) \right\} +o(\hbar^4), \ea \ee
where $Y$ is an expression composed purely from the metric and its
first and second derivatives,
\be\label{yterm}
Y=\fr g^{kl}_{,mn} g^{mn} g_{kl} - g^{nk}_{,m} g^{ml}_{,n} g_{kl}
+\frac{1}{3}G_{klk'l'}(-\frac{1}{4} g^{kl}_{,m} g^{k'l'}_{,n} g^{mn} - \fr
g^{kl}_{, mn} g^{mk'} g^{nl'})
\ee
Now we have to discuss the gauge invariance of this action. By
construction, general action is gauge-invariant w.r.t.
gauge transformations \re{bsqinf5432}. In the case of the low spin ansatz
\re{quadr111} one has to restrict to the subalgebra of gauge transformations which
preserves the second-order ansatz. This subalgebra, $\cG_0$  is spanned
by $\e=\ve (x)+\xi^m (x) p_m$ and
$a=\alpha(x)$, parameterizing infinitesimal $U(1)$ (phase)
transformations, $x$-diffeomorphisms and Weyl dilations, respectively. The
commutation relations are quite the same as in the classical case due to
special properties of the Weyl symbol. Nevertheless, the action of $\cG_0$
on the components of the Hamiltonian $H_2$ acquires quantum corrections.
One has
\begin{equation}  \label{qanstr1}
\delta _{\varepsilon,a} H_2=[\a,H_2]_+ +\frac 1{i\hbar }[\varepsilon, H_2]
\end{equation}
For the case at hand the $\hbar$ series terminate at $\hbar^2$
terms. One has
\be\ba  \label{qtrans1}
\delta _\varepsilon D =-\xi ^m \partial _m D-\frac{\hbar ^2}4(\partial _m
g^{a b })\;\partial _a \partial _b \xi ^m +2\a D
+\frac{\hbar^2}{2} g^{mn} \;\alpha_{,mn} \\
\\
\delta _\varepsilon A_m =-\xi ^k \partial _k A_m +A_k \partial _m \xi ^k
-\partial _m \varphi   \\ \\
\delta _\varepsilon g^{m n } =-\xi ^k \partial _k g^{m n }+g^{m k
}\partial _k \xi ^n + g^{n k }\partial _k \xi ^m +2\a g^{mn}, \ea\ee
where $A_m=g_{mn}A^m$.
Here $\hbar^2$-independent terms represent
standard $U(1)$ transformations, $x$-diffeomorphisms and Weyl dilations,
while $\hbar^2$-correction to the dilaton's gauge transformations
prohibits its interpretation of a scalar field.
Note that these gauge transformations are already exact expressions
for the gauge transformations. They form the closed algebra, being
isomorphic to its classical limit.
It appears there exists the redefinition of variables such that the
new variables transform in a standard way, i.e. as follows
\be\ba  \label{qtrans2}
\delta _\varepsilon D' =-\xi ^m \partial _m D'
 +2\a D'
\\
\\
\delta _\varepsilon A'_m =-\xi ^k \partial _k A'_m +A'_k \partial _m \xi ^k
-\partial _m \varphi   \\ \\
\delta _\varepsilon g'^{m n } =-\xi ^k \partial _k g'^{m n }+g'^{m k
}\partial _k \xi ^n + g'^{n k }\partial _k \xi ^m +2\a g^{mn}, \ea\ee
The redefinition reads (see Appendix \re{ApF}) \be \ba\label{redef75}
D'= D-\frac{\hbar^2}{4}(Y-\frac{d-4}{3(d-1)} R),\\ \\
g'^{mn}=g^{mn}\;\;,\;\; A'^m=A^m,
\ea\ee
where $Y$ is from \re{yterm} while $R$  is a scalar
curvature constructed of the Riemann tensor $R^m_{nkl}$ of the
metric $g_{mn}$ by the rule $R=g^{mn} R^k_{mkn}$.

The action \re{actires} is to be expressed in terms of new "standard" variables. Then it should
be a action of Weyl-invariant dilaton gravity, described in Sec. \re{weylg}, because the gauge laws
for $D', g'^{mn}$ just coincide with those of \re{qtrans25}.
The result is
\be  \ba\label{weylg33}
\cA[H_2] \equiv \cA[D'(x), g_{mn}(x)] =
\frac{v_d}{(2\pi\hbar)^d} \int d^dx \sg\left\{ D'^{\hd} +\right.\\ \\
-\frac{\hbar^2 d(d-4)}{24} \left(
\frac{d-2}4 D'^{\hd -3} g^{mn} D'_{,m} D'_{,n}
+ \frac{1}{d-1}  D'^{\hd-1} R\right)
+\\ \\
+\left.\frac{\hbar^2}{12} \frac{d}{\sg}  \pa_n (\sg g^{mn} \pa_n
D'^{\hd-1}) \right\} +o(\hbar^4), \ea \ee
In the last formula, the first row is the classical, $\hbar^0$, term, which does
not contain space-time derivatives of the dynamical fields. It
has clear meaning of the cosmological term. The dependence
of the cosmological term on the dilaton exhibits the fact that,
by the very construction, it represents the volume of the $d$-ball of
radius $\sqrt{D}$ bounded by the surface $H_2=0$ in the momentum space.

According to our predictions, the derived action is nothing but the action of Weyl invariant dilaton gravity
\re{weylgrav}, described
in the beginning of the paper in Sec. \re{weylg}. Evidently, the inclusion of a nonzero Maxwell field
$A_m(x)$ easily comes along the same lines.
The higher orders in $\hbar$ may be computed along the same lines.
As the gauge transformations of the modified dilaton $D'$ and the metric
$g^{mn}$ are just the standard diffeomorphisms while the action is gauge
invariant by construction, each $\hbar^{2k}$ term of the semiclassical expansion
will be given (modulo total derivatives) by Weyl-invariant combinations of the
scalar $D'$, Riemann curvature and their covariant derivatives, with the
total number of derivatives of $D'$ and $g^{mn}$ being equal $2k$.
Thus we have recovered in our approach the low spin data of  Sec. \re{weylg} and
have shown that
\begin{teo}
The low-spin truncation of the "quantized volume" action \re{fquasi111} provides, in the
Euclidean regime, Weyl-invariant dilaton
gravity with cosmological term given by dilaton potential.\end{teo}

\subsection{Comparison to Schwinger-Dewitt technique}\label{ctsdwt}

There exists another interpretation of the redefinition \re{redef75}, that
leads us naturally to conjecture about the form  of the invariant induced action in
even dimensions.

Namely, let us recall that the standard tensor fields, dilaton and metric,
of Weyl-invariant gravity are the components of the second-order operator
\be
{\bf H}=g^{mn}\hat{\na}_m \hat{\na}_m -\frac{d-2}{4(d-1)} R + D'(x)
\ee
governing the quadratic part of the action of free conformal scalar
\be\label{weyloper1}
S_{\bf H}[\vf]=\int d^dx \sg \c^* {\bf H} \c.
\ee
where $g=\det{g_{mn}}$. ${\bf H}$ scales as
\be
{\bf H}\rightarrow e^{(1+\hd)a(x)}{\bf H}e^{(1-\hd)a(x)}
\ee
w.r.t. Weyl dilations
\be
g^{mn} \rightarrow e^{2a(x)} g^{mn}\;,\;D' \rightarrow e^{2a(x)} D'
\ee
so that the action \re{weyloper1} is Weyl invariant provided $\c\rightarrow e^{(-1+\hd)a(x)}\c$.

Clearly, the wave operator $\bf H$, scalar $\c$ and the action $S_{\bf H}[\c]$ are related
to some Hamiltonian $\hat{H}$, wave function $\psi(x)$ and the coupling \re{php} by the map
\be\ba\label{redef82}
S_{\bf H}[\c]=S_H[\psi]\\ \\
{\bf H}= g^{-\frac{1}{4}}(x) \hat{H} g^{\frac{1}{4}}(x), \; \c (x) =g^{-\frac{1}{4}}(x) \psi(x),
\ea\ee
Given ${\bf H}$, this map gives a second-order operator $\hat{H}$
which starts from $g^{mn}(x) \pa_m \pa_n$. It proves that by the $x$-diffeomorphism covariance and
Weyl covariance, this operator is nothing but the operator with the Weyl symbol (let $\hbar=1$)
\be
H=-g^{mn}(x) p_m p_n + D(x)=
-g^{mn}(x) p_m p_n + D'+\frac{1}{4}(Y-\frac{d-4}{3(d-1)} R),
\ee
where the relation of $D(x)$ and $D'(x)$ is from Eq. \re{redef75}.

Note that the origin of the map \re{redef82} is clear by the following reasons.
In the Weyl symbol, $x$-diffeomorphisms are generated by the generating function
\be
\o=\frac{i}{\hbar}\xi^m (x) p_m,
\ee
which is the Weyl symbol of the operator
\be
\hat{\o}=\fr(\xi^m \pa_m +\pa_m \xi^m)=\xi^m(x) \pa_m +\fr (\pa_m \xi^m(x)).
\ee
The latter operator defines representation of general coordinate transformations characteristic for the
scalar density of weight $\fr$, that is manifested by the $\fr \pa_m \xi^m$ addition to ordinary vector
field $\xi^m \pa_m$. The $g^{\frac{1}{4}}$ factors in \re{redef82} just compensate this addition rendering
infinitesimal general coordinate transformations to be ordinary vector fields.
To summarize,
$x$-diffeomorphisms, which look as  $\d H =-\frac{i}{\hbar} [\xi^m p_m,H]_*$ in the symbol
approach and lead to unusual transformation laws \re{qtrans1} for the dilaton $D$,
in terms of $\bf H$ are just ordinary vector fields which transform $D'(x)$ as a scalar. In so doing,
$D'$ is just a scalar part of the generally covariant differential operator $\bf H$.

Having established the map between the symbol $H(x,p)$ and operator ${\bf H}$ one can
use the Schwinger-Dewitt technique for evaluating the expression
\be
Q(i\t)=Tr  \exp (i\t {\bf H})
\ee
and compare the result with our calculations in symbol formalism.
The benefit of using the Schwinger-Dewitt
approach is that ${\bf H}$ is parameterized by ordinary tensor fields and
$Q(i\t)$ is expressed as an integral of the covariant expression \cite{Barvinsky:an}
built from Riemann curvature, dilaton and their covariant derivatives.
Namely, let
\be\ba
{\bf H}=g^{mn}\hat{\na}_m \hat{\na}_m -\frac{1}{6} R +P(x)+m^2\Rightarrow,
\ea\ee
where
\be\label{identif}
P(x)+m^2=D'(x)+\frac{d-4}{12(1-d)}R
\ee
Then the Schwinger-Dewitt technique provides the expression \cite{Barvinsky:an}
\be\ba
Tr  \exp(i\t{\bf H})=\frac{ e^{i m^2\t}}{(4\pi i\t)^{\hd}}  \sum \limits_{n=0}^{\infty}
 (i\t)^n \int d^d x \sg\; a_n(x,x)
\ea \ee
where $a_n(x,x)$ are built from
$P(x)$, Riemann curvature and their covariant derivatives, \cite{Barvinsky:an}
\be\ba\label{sdwc}
a_0(x,x)=1,\;\a_1(x,x)=P,\;\\ \\a_2(x,x)=\frac{1}{180}(R_{mnkl}R^{mnkl}-R_{mn}R^{mn} +\Box R)+\fr P^2 +\frac{1}{6}\Box P,
\ea\ee
and so on.

Our action is given by the formula
\be\ba
\cA[H]=Tr \; \theta(\hat{H}) = Tr \frac{1}{2\pi i} \int
\limits_{-\infty}^{\infty} \frac{d \o}{\o-i\ve} e^{i\o \hat{H}}
\ea\ee
Suppose changing the order of integration by $\o$ with $"Tr"$ operations is a legal operation
\footnote{In general such a changing is hard to justify. We nevertheless omit the detailed
discussion of this issue supposing the changing of contours is meaningful at least for
a wide class of hamiltonians.}
one gets
\be\ba\label{sdw546}
\frac{1}{2\pi i} \int
\limits_{-\infty}^{\infty} \frac{d \o}{\o-i\ve} Q(i\o)=
\frac{1}{2\pi i} \int
\limits_{-\infty}^{\infty} \frac{d \o}{\o-i\ve}
\frac{e^{i m^2\o}}{(4\pi i\o)^{\hd}}   \sum \limits_{n=0}^{\infty} (i\o)^n \int d^d x \sg \;a_n(x,x).
\ea\ee
Let $d$ be even. For $m^2 >0$ the integrals by $\o$ are calculated easily as integration contour
may be continued to the closed contour in a full complex plane by adding the point $Im \; \o =+\infty$,
\be\label{weylan2}
\cA[H]=\frac{1}{(4\pi)^{\hd}} \sum \limits_{n=0}^{\hd} \frac{(m^2)^{\hd-n}}{(\hd-n)!}
\int d^d x \sg \;a_n(x,x).
\ee
Using the identification \re{identif} one can see that this expression coincides, modulo boundary terms,
with that obtained in the Weyl symbol formalism \re{weylg33}.
But the expression \re{weylan2} is nothing but the time-independent term of the decomposition of the
trace of the evolution operator,
\be\label{sdw765}
\cA[H]=Tr\;\theta(\hat{H})=\lim \limits_{\t \rightarrow 0^+}^{} Tr \exp (i\t {\bf H})
\ee
We expect analogous phenomenon takes place for general Hamiltonians which start from small positive dilaton.

Let us comment on general coordinate invariance in our model. Of course, the action
is general coordinate invariant as $x$-diffeomorphisms form a subalgebra of the gauge algebra of the theory.
On the other hand, the coefficients of the decomposition of the Weyl symbol $H(x,p)$ in power series in
momenta are not ordinary tensors but acquire quantum corrections to ordinary tensor transformation laws
\re{xdifcor}.
Luckily, there exists the parameterization of general differential operator in terms
of ordinary tensor fields. It is achieved by transferring to the new operator
${\bf H}(x,\na)$ being a power series in covariant derivatives.
Namely, given general Hamiltonian with Weyl symbol
\be H(x,p) =\sum \limits_{s=0}^{\infty} H^{m(s)}(x) p_{m_1}...p_{m_s} \ee
one has differential operators
\be\hat{H}=\sum \limits_{s=0}^{\infty} (-i\hbar)^s H^{m(s)}(x) \pa_{m_1}...\pa_{m_s} + ...
\ee
(see (\ref{weyl98},\ref{weyl988}) for exact formula),
and
\be\label{genhna}
{\bf g}^{-\frac{1}{4}}(x) \hat{H} {\bf g}^{\frac{1}{4}}(x)={\bf H}
=\sum \limits_{s=0}^{\infty} (-i\hbar)^s {\bf H}^{m(s)}(x) \na_{m_1}...\na_{m_s}
\ee
where the inverse metric is nothing but ${\bf H}^{m(2)}(x)$ and ${\bf g}=\sqrt{\bf \det{g_{mn}}}$.
Thus one gets
the map from the set of tensor fields ${\bf H}^{m(s)}$ comprising power series
$\bf{H}(x,\na)$ to the set of coefficients $H^{m(s)}$ of decomposition of $H(x,p)$ in momenta, the latter
transforms w.r.t. $x$-diffeomorphisms according to \re{xdifcor}.
Thus any $x$-diffeomorphism invariant expression built from $H^{m(s)}$ may be rewritten as
invariant one in terms of
${\bf H}^{m(s)}$ i.e. as an integral $$\int d^d x \sqrt{\bf g} \,F(\na_{r_1}...\na_{r_k} R_{efgh},
 \na_{q_1} ... \na_{q_l} {\bf H}^{m(s)}, ...)$$
of a function of Riemann curvature, ${\bf H}^{m(s)}$ and their covariant derivatives.
Rewriting of the action in terms of tensor fields ${\bf H}^{m(s)}$ and their covariant derivatives
is what is likely more simply performed in the Schwinger-Dewitt approach.

Despite the Schwinger-Dewitt technique is less elaborated in the case of general
operators \re{genhna} one can make perturbative calculations along
the lines of $w$-decomposition of Sec. \re{classesh}, i.e. taking the higher spin part as a
small fluctuation. In so doing, one will always get $Q(i\t)$ as a power series of the
form
\be
Q(i\t) = (4\pi i \t)^{-\hd} e^{im^2\t} \sum \limits_{n=0}^{\infty} \int d^d x \,\sg\, a_n,
\ee
where $a_n$ are covariant expressions built from Riemann curvature, tensor fields ${\bf  H}^{m(s)}$
and their covariant derivatives which reduce to ordinary Schwinger-Dewitt coefficients
\re{sdwc} when higher rank fields are switched off. Then as $m^2 >0$ the integration by $\o$ in \re{sdw546} may be always
continued to run over the closed contour around origin in the complex plane and thereby gives
the simple residue of the expression $Q(z)$ at the origin or just $Q(0)$. Thus, at least in
the framework of the $w$-expansion one has the formula \re{sdw765} for an arbitrary Hamiltonian
(in even dimensions). This provides an alternative definition of the action of the conformal higher spin theory.

But the same expression provides the logarithmically divergent term
of the one-loop effective action,
\be\ba
-\fr Tr Ln \hat{H} =\fr \int \limits_{\l=\m}^{\infty} \frac{d\l}{\l} Tr\; e^{\l \hat{H}}=
-\fr (\ln \m) \,Q(0) + \\ \\\mbox{+ power divergencies in $\m$ + regular terms in $\m$}.
\ea\ee
In this sense, in even dimensions, the action of the model of the paper is nothing but the
logarithmically divergent term in the one-loop effective action of complex scalar field for an arbitrary
background wave operator $\hat{H}$. This means the {\it invariant induced action} is a logarithmically
divergent part of true {\it induced action} $-\fr Tr Ln \hat{H}$ of the quantized particle.

\subsection{Conformally flat vacuum}
According to the derivation of Sec. \re{weylg}, in $d>2$, the theory (\ref{acti},\ref{weylg33})
possesses the vacuum of the form
\be
D'=0, A_m=0, g^{mn}=\eta^{mn}\ee with $\eta^{mn}$ being flat metric, or, equivalently, of the form
\be
H=-p^2.
\ee
As the theory is Weyl-invariant, the Weyl-dilated configuration
\be\ba \label{vaccf}
H_{a(x)}=e^{a(x)}*(-p^2)*e^{a(x)}=-e^{2a(x)}(p^2-\frac{\hbar^2}{4}\Box a(x)) \Leftrightarrow \\ \\
D'=0, A_m=0, g^{mn}=e^{2a(x)}\eta^{mn}
\ea\ee
is a vacuum either. Therefore, the Hamiltonian
$H_{a(x)}$ represents a family of {\it conformally flat vacua} of the low spin truncated theory.
Note that among the vacua \re{vaccf} there are anti-de Sitter and de Sitter spaces corresponding to the
following choices of the conformal factor
\be\label{dsads}
e^{2a_{AdS}}=(1- R^{-2} x^2 )^{2}\;,\;e^{2a_{dS}}=(1+R^{-2} x^2 )^{2},
\ee
where $R$ is "radius" of AdS or dS space.
It is natural to wonder if the vacuum \re{vaccf} can be a vacuum of the whole non-truncated theory. The
answer is yes, the proof is the subject of the next section.

\section{Conformally flat vacuum.}\label{scfv}

Consider the perturbative expansion of the "quantized volume" action \re{fquasi111} around the configuration
\be
\bar{H}=D-p^2,
\ee
where $D\geq 0$ and $p^2$ is built with flat Euclidean metric.
To this end one represents general $H$ as
\be
H(x,p)=\bar{H}+eh(x,p),
\ee
where $h(x,p)$ is a fluctuation,
and expands the action \re{fquasi111} in powers of $eh$,
\be
\cA[\bar{H}+h]=\cA[\bar{H}]+e\cA_{1}[h]+e^2 \cA_{2}[h] +...
\ee
The zero order term is easily calculated as any $*$-function of $\bar{H}$ amounts to ordinary function,
\be
F(\bar{H}*)=F(\bar{H})
\ee
Therefore, the action is given by the classical term while the quantum corrections vanish.
The result is
\be
\cA[\bar{H}]=\we \int d^d x d^dp\; \theta (D-p^2)=\frac{v_d}{(2\pi\hbar)^d} D^{\hd} \int d^d x
\ee
Of course this expression diverges but this is not dangerous for our purposes.

Of our main interest in this section is the $eh$-linear term. This term equals
the variation of the
action \re{fquasi111}, under the variation of the Hamiltonian of the form $\d H=e h$, around  the configuration $H=\bar{H}$.
Taking into account that the
order of product factors under the sign of trace operation is inessential (up to total derivative terms) one gets
\be\ba
\cA[\bar{H}+h]-\cA[\bar{H}]=\we \int d^dx d^dp\; \left\{
\theta((\bar{H}+e\,h)*)-\theta(\bar{H}*)\right\}=\\ \\
\we \int d^dx d^dp\; \delta(\bar{H}*)* e h+\mbox{tot. derivative}=\we\int d^dx d^dp\; \delta(\bar{H})\,e \,h +\mbox{tot. derivative},
\ea\ee
where in the last equality two $*$-products turn to ordinary products because of the
Eq. \re{totder16} and "simple multiplication" rule
\re{simmul}.
The total derivatives by $x$ are transformed into the integrals by $x$-infinity and vanish as $h$ is required to vanish at
$x$-infinity. The total derivatives by $p$ may be shown to vanish either, as they are proportional to derivatives
of delta-function $\d(\bar{H})$ (that may be shown along the same lines as in Sec. \re{actual23}).

The main part of the variation, the expression
\be
\we\int d^dx d^dp\; \delta(\bar{H}) h(x,p)
\ee
presents the integral by the constraint surface $\bar{H}=0$ being $d-1$-sphere. This expression
is easy calculable to be
\be
\we\int d^dx d^dp\; \delta(\bar{H}) h(x,p)=\we  \sum \limits_{s=0}^{\infty} D^{\hd-1+s}
d_s  \int d^d x  \;{h^{m_1...m_s}}_{m_1...m_s}(x),
\ee
where $d_s$ are non-zero coefficients which may be extracted e.g. from generating functions \re{genf4}.

Now consider the case
\be
D=0.
\ee
Then the variation vanishes provided $d>2$ and thus the configuration $H=-p^2$ is a solution of the classical equations of motion of the "quantized volume"
action, i.e. it is a vacuum. This vanishing of the variation has a transparent meaning as $\sqrt{D}$ is the radius of the sphere
$\bar{H}=0$ and therefore the limit $D\to 0$ corresponds to the case when the sphere shrinks to a point. As a consequence,
the "quantized volume" as well as its variation, vanishes in even dimensions (in $d>2$). At the same
time, in odd dimensions the $w$-decomposition of the invariant induced action in the sense of Sec. \re{classesh}
contains all negative powers of the dilaton, therefore, the fact that the linear variation cancels
is a less trivial phenomena which manifests that all terms in the linear variation of $\cA[H]$
with negative powers of $D$ become total derivatives as $D\rightarrow 0$.
Thus, we get the \begin{teo}
In $d>2$, the Hamiltonian $H=-p^2$ is a solution of the classical equations of motion for the "quantized volume" action \re{fquasi111}.
\end{teo} Now one recalls Proposition \re{linearize} to derive
\begin{teo}
In $d>2$, the quadratic in $h$ terms of the perturbative expansion of the "quantum volume"
are described by a sum of deformed conformal higher spin theories with $\m^2=-\frac{\hbar^2}{4}\Box$.
\end{teo}
In the next section, by direct calculation we show this is the case indeed. Thereby, the "quantized volume"
is a consistent action for the (Euclidean phase of) conformal higher spin theory.

As the whole theory is invariant w.r.t. Weyl dilations, one gets the family of vacua
\be
H_{a(x,p)}=-e^{a(x,p)*}*p^2*e^{a(x,p)*},
\ee
We call any vacuum of the form a {\it conformally flat vacuum}. Below we perform all the calculations
in the case $a=0$, but one has to keep in mind that the results of calculations are related by the gauge transformation
to another conformally flat vacua, e.g. $d$-dimensional de Sitter or anti-de Sitter spaces
\re{dsads}.

\section{Restoration of the linearized data.  Perturbative expansion around the conformally flat vacuum}
\label{corollary}

In this section, we consider perturbative
expansion and linearization of the theories \re{fquasi111}, firstly around an arbitrary vacuum solution,
and then around the
conformally flat vacuum. We uncover, at the quadratic order, a sum of
conformal higher spin theories with $s=0,1,2,3,...$ that presents our main result.

\subsection{General perturbative expansion.}

Consider the functional
\be\label{onceag1}
A_F[H*]=Tr F(H*)
\ee
and perform its expansion in perturbative series around a particular point
$H=A$. This means representing general Hamiltonian as
\be
H={\bar H}+eh,
\ee
and expand the action \re{onceag1} in Tailor series in $h$. This is achieved as
follows. First, we recall the representation of the function $F(H*)$ in
Fourier-Laplace type series \re{fl}, which gives
\be
Tr F(H*)=\int \limits_C d \l \tilde{F}(\l)Tr  \exp (-\l H*),
\ee
where $\l\in C$ while $C$ is a contour in a complex plane.

One expands the expression $Tr u(-\l)= Tr  \exp (-\l H*)$ and then
integrates by $\l$. The expansion of the exponent is performed by
making use of the formula
\be
 \exp(-\l ({\bar H}+eh)*)= \exp(-\l {\bar H}*) * T  \exp (-\int \limits_{0}^{\l} e^{\l {\bar H}*}*eh*
e^{-\l {\bar H}*})
\ee
which is proved easily by the differentiation by $\l$. $T$ denotes the
$T$-ordering,
\be
T(B_1(\l_1) B_2(\l_2))=\theta(\l_1 -\l_2) B_1(\l_1) B_2(\l_2) +
\theta(\l_2 -\l_1) B_2(\l_2) B_1(\l_1).
\ee
According to this formula, the functional \re{onceag1} is equal, up to $h^3$ terms,
\be \ba \label{expan}
\cA_F[H*]=Tr F(({\bar H}+eh)*)=
\\ \\
\int \limits_C d\l f(\l) Tr e^{-\l ({\bar H}+eh)*}=\int \limits_C d\l f(\l) e^{-\l {\bar H}*}
* T  \exp (-\int \limits_{0}^{\l} e^{\l {\bar H}*}*eh*
e^{-\l {\bar H}*}) =\\ \\= \cA_0[{\bar H}]+e\cA_1[{\bar H},h]+e^2\cA_2[{\bar H},h]+
e^3\cA_3[{\bar H},h]+o(h^4)\\ \\
\cA_0[{\bar H}]=Tr F(H*)\\ \\
\cA_1[{\bar H},h]=Tr F'(H*)*h\\ \\
\cA_2[{\bar H},h]=\int \limits_C d\l f(\l)Tr e^{-\l {\bar H}*} * \int \limits_{0}^{\l} d \t_1 \int \limits_{0}^{\t_1}
d \t_2 e^{\t_1 {\bar H}*} *h* e^{-\t_1 {\bar H}*} * e^{\t_2 {\bar H}*} *h* e^{-\t_2 {\bar H}*}\\ \\
\cA_3[{\bar H},h]=-\fr \int \limits_C d\l f(\l) Tr e^{-\l {\bar H}*}*\\ \\
{}* \int \limits_{0}^{\l} d \t_1 {\int \limits_{0}^{\t_1}} d \t_2 \int \limits_{0}^{\t_2} d\t_3
e^{\t_1 {\bar H}*} *h* e^{-\t_1 {\bar H}* }* e^{\t_2 {\bar H}*} *h* e^{-\t_2 {\bar H}*} * e^{\t_3 {\bar H}*} *h* e^{-\t_3 {\bar H}*}
\ea
\ee
In what follows, we apply these formulae to the case
\be
F(\s)=\theta(\s),
\ee
what is equivalent to choosing the function $\tilde{F}(\l)$ and the integration
contour as follows,
\be
\tilde{F}(\l)=\frac{1}{2\pi i \l},\;\l=-\ve+i\t,\;\t\in {\bf R}, \ve\to 0^+
\ee
We also make use of the notation\be a_d=\frac{1}{(2\pi\hbar)^d},\;b_d=(\pi\hbar)^d\ee

It is worth discussing briefly where the leading terms of the conformal higher spin models will come from.
The calculation we perform is in the framework of expansion in two parameters
$\hbar$ and $e$, the former counts the number of $x$-derivatives and the latter
counts "the number of  tails" i.e. the degree of $h$, such that
a term $\sim \hbar^{2k}e^l$ is a homogeneous polynomial in $h$ of degree $l$ with the
total number of $x$-derivatives equal to $2k$. Using integration technique explained in Sec. \re{calcttt}
and App. \re{ApD} it may be readily shown that
after taking all integrals by momenta, the invariant induced action has a form like
\be
\cA[D-p^2+eh]\sim \int d^dx \sum \limits_{r} \sum \limits_{l=0}^{\infty}e^l
D^{-r}(\hbar\pa)^{d-2l+2r +\sum
\limits_{i=0}^{l} s_i}  h_{s_1}...h_{s_l},
\ee
where partial derivatives act in a diverse way on the rank-$s_1...s_l$ components of the fluctuation.
In even dimensions terms with negative powers of $D$ are zero by the reasons discussed in Sec.
\re{classesh} while in odd dimensions arbitrary negative powers of $D$ are available. The terms with
negative degrees of $x$-derivatives are absent by the very construction, as we are
doing semiclassical decomposition in $\hbar\pa$, this implies summations are actually
restricted to the domain $2r\geq -d+2l-\sum \limits_{i=0}^{l} s_i$. Of our main interest $D=0$ case,
\be
\lim \limits_{D\rightarrow 0^+} \cA[D-p^2+eh]
\sim \int d^dx \sum \limits_{l=0}^{\infty}  e^l (\hbar\pa)^{d-2l+\sum \limits_{i=0}^{l} s_i}  h_{s_1}...h_{s_l},
\ee
While in even dimensions the latter expression is a straightforward limit of $D > 0$ one,
in odd dimensions the situation is more subtle, the limit $D \rightarrow 0$
exists in the sense this may be a vacuum but not in the framework of perturbative expansion, just like
the point $\s=0$ is not a good point for expansion of the function $f(\s)=\s^{\frac{k}{2}}, \k=1,2,3,...$.
Our calculations will naturally incorporate odd-dimensional case, despite it is only the even-dimensional
case when the perturbative expansion is well-defined in a usual sense.

Consider the quadratic action, i.e. $l=2$, and let $s_1=s_2=s$. Then one has
\be
\lim \limits_{D\rightarrow 0^+} \cA[D-p^2+eh]
\sim e^2 \int d^dx\; \sum \limits_{s=0}^{\infty} (\hbar\pa)^{d-4+2s}\;  h_{s} h_{s} +...+o(e^3h^3),
\ee
where "..." stands for non-diagonal terms. Terms $\sim (\hbar\pa)^{d-4+2s} h_s h_s$ are
nothing but the  leading terms of free conformal higher spin models of spin $s$. It is seen that the
conformal spin-$s$ model originates from the ${d-4+2s}$-th quantum correction to the classical
volume of the domain bounded by the constraint surface, in the limit $D\rightarrow 0$ when
the surface shrinks to a point in the momentum space.

\subsection{Calculation of the quadratic part of the action}

Calculate the quadratic term in the expansion \re{expan} around the vacuum solution
\be
{\bar H}=-p^2.
\ee
One has
\be \ba
e^{\l p^2 *} = e^{\l p^2} \\ \\
e^{\l p^2*} * h(x,p)* e^{-\l p^2*}=h(x-2i\hbar\l p,p).
\ea \ee
After making use of the trace  property \re{totder16}, the ``simple multiplication'' property
\re{simmul} and making the change of variables
\be
t=\t_1 -\t_2\;,\;\t_1=\t_1
\ee
one rewrites the quadratic action form \re{expan} as
\be\ba
\cA_2=
-a_d \int \limits_C d \l \tilde{F}(\l)  \int dx dp e^{\l p^2}
\int\limits_{0}^{\l} d \t_1 \int \limits_{0}^{\t_1} dt h(x,p) * h(x-2i\hbar t,p)
\ea \ee
It is useful to employ the integral representation for the star-product
\re{starintegral} to rewrite the last
expression as
\be \ba
\cA_2=
-a_d b_d^{-2}\int \limits_C d \l \tilde{F}(\l)  \int dx dp dx_1 dp_1 dx_2 dp_2 e^{\l p^2}
\int\limits_{0}^{\l} d \t_1 \int \limits_{0}^{\t_1} dt\\ \\
 \exp(\frac{2i}{\hbar}[(x-x_2)p_1 +(x_1-x)p_2+(x_2-x_1)p]) h(x_1,p_1) h(x_2-2i\hbar tp_2,p_2),
\ea\ee
or, making shift of the integration variables $x_2\rightarrow x_2+2i\hbar tp_2$,
\be \ba
\cA_2=
-a_d b_d^{-2}\int \limits_C d \l \tilde{F}(\l)  \int dx dp dx_1 dp_1 dx_2 dp_2 e^{\l p^2}
\int\limits_{0}^{\l} d \t_1 \int \limits_{0}^{\t_1} dt\\ \\
 \exp(\frac{2i}{\hbar}[(x-x_2)p_1 +(x_1-x)p_2+(x_2-x_1)p -2i\hbar t p_2(p_1-p)])
h(x_1,p_1) h(x_2,p_2).
\ea\ee
The next step is  rewriting the integral by $\l,\t_1,t$ using the formulae
\be \ba
\int\limits_{0}^{\l} d \t_1 \int \limits_{0}^{\t_1} dt
e^{t\s} =\frac{1}{\s}(\frac{1}{\s}(e^{\l\s}-1) -\l) \\ \\
\int \limits_C d \l \tilde{F}(\l)  e^{\l \r}  \frac{1}{\s}(\frac{1}{\s}(e^{\l\s}-1) -\l)=\\ \\
=\int \limits_{0}^{1} d\a \int \limits_{0}^{1} d\b \a F''(\a\b \s-\r)
\ea\ee
to get
\be \ba
\cA_2=
-a_d b_d^{-2}\int \limits_{0}^{1} d\a \int \limits_{0}^{1} d\b \a\int dx dp dx_1 dp_1 dx_2 dp_2
\d'(\a \b \s-\r )
\\ \\
\exp(\frac{2i}{\hbar}[(x-x_2)p_1 +(x_1-x)p_2+(x_2-x_1)p ])
h(x_1,p_1) h(x_2,p_2),
\ea\ee
where
\be
\s=-4(p-p_1)p_2\;\;,\;\; \r=-p^2.
\ee
The integration over $x,p_2$ is elementary, with the result
\be \ba \label{quadrres}
\cA_2=
-a_d b_d^{-1}
\int \limits_{0}^{1} d\a \int \limits_{0}^{1} d\b \a\int dp dx_1 dp_1 dx_2
\d'(4(p-p_1)p_1 \a \b -p^2 )
\\ \\
 \exp[\frac{2i}{\hbar}(x_1-x_2)(p_1-p)]
h(x_1,p_1) h(x_2,p_1),
\ea\ee
Instead of calculating the quadratic action
\begin{equation}\label{Saction11}
  \cA_P[h]=\sum \limits_{k=0,k'=0}^{\infty} \int d^d x
  h^{m_1...m_k}(x) P_{\{m_1...m_k | n_1...n_{k'}\}} (\partial_l)h^{n_1...n_{k'}} (x),
\end{equation}
itself it is worth calculating the function of two vector variables
\begin{equation}\label{generP11}
P(q,q')=\sum \limits_{k=0,k'=0}^{\infty} \frac{1}{k!k'!}
q^{n_1}...q^{n_{k'}} q'{}^{m_1}...q'{}^{m_{k}} P_{\{m_1...m_k |
n_1...n_{k'}\}} (\partial_l)
\end{equation}
with the coefficients depending on $\pa_l$.

\begin{lem} If $\cA_2$ is given by the expression
\be \label{lem1}
\cA_P[h] =\int d^d x_1 d^d x_2 d^dp\; f(\frac{x_1-x_2}{2},p)\,h(x_1,p)h(x_2,p),
\ee
then
\be\label{lem11}
P(q,q',\pa)=2^d\int d^dr d^dp\; f(r,p) \,e^{(q^m+q'^m)p_m -2 r^m \pa_m}
\ee\end{lem}
The proof is by direct calculation.

Now we are ready to calculate the generating function $P(q,q',\pa)$.
According to \re{quadrres} and (\ref{lem1},\ref{lem11}),
the expression for generating function reads
\be \ba
P(q,q',\pa)=
-2^d a_d b_d^{-1}
\int \limits_{0}^{1} d\a \int \limits_{0}^{1} d\b \a\int dr dp dp_1  \\ \\
\d'(4(p-p_1)p_1 \a \b -p^2 )
 \exp(\frac{4i}{\hbar}[r(p_1-p)]+(q+q')p_1-2 r\pa).
\ea \ee
Before doing integrals over $r$ etc. it is useful to make the shift of the
integration variables
\be
p\rightarrow\P = p -2p_1\a\b
\ee
and change the variables in the $\a,\b$-sector as
\be
\eta = 1-2\a\b\;,\; \a=\a
\ee
to get
\be \ba
P(q,q',\pa)=
-\fr 2^d a_d b_d^{-1}
\int_{\Sigma} d\a d\eta \int dr d\P dp_1  \\ \\
\d'(p_1^2 (\eta^2-1)-\P^2)
 \exp[\frac{4i}{\hbar}(r(p_1 \eta -\P) +(q+q')p_1-2 r\pa],
\ea \ee
where the domain of integration $\Sigma$ in the $(\eta,\a)$ plane is the triangle
with vertices
$(1,0)$,$(1,1)$,\\$(-1,1)$.
Now perform elementary integration by $r,\P$ with the result
\be \ba
P(q,q',\pa)=
-\frac{a_d}{2}
\int_{\Sigma} d\a d\eta \int dp_1  \\ \\
\d'((1-\eta^2)n^2-(p_1-i\eta n)^2 +)
e^{fp_1},
\ea \ee
where
\be
n=-\frac{\hbar}{2} \pa\;,\;f=q+q'.
\ee
After shift of the integration variables $p\rightarrow p+i\eta n$, the last integral takes the form
\be \ba
P(q,q',\pa)=
-\frac{a_d}{2}
\int_{\Sigma} d\a d\eta \int dp_1  \\ \\
\d'((1-\eta^2)n^2-p_1^2)
e^{fp_1}e^{i\eta nf}
\ea \ee
The remaining integral over $p_1$ is calculated easily as
it is expressed via generating function $G^{(2)}_d$ (\ref{genf1},\ref{genf4}),
\be
G^{(2)}_d (l) =\int d^dp\; \d' (M^2-p^2)
e^{ip_m l^m}=
 2^{\hd-2} \pd M^{\hd-2} l^{-\hd+2}
J_{\hd-2}(\m l).
\ee
Combining two last formulae one gets
\be \ba
P(q,q',\pa)=
-\frac{a_d}{2} 2^{\hd-2} \pd
\\ \\\int_{\Sigma} d\a d\eta
\;e^{i\eta nf}
 (n \sqrt{1-\eta^2})^{\hd-2}J_{\hd-2}(\pm i |f||n| \sqrt{1-\eta^2})
\ea \ee
One easily checks that the $\pm$-ambiguity in the separate multipliers of the last formula
cancels for the whole expression.

Now integrate over $\a,\eta$. Note that the $\eta$-odd part of the last expression
is purely imaginary and therefore it represents a total-derivative term
(the associated contribution to the quadratic action gives operator of odd order
in derivatives) which arose just because despite we have started from the purely real
expression \re{onceag1}, after that we did some integrations by parts (we have used
the cyclic property of the trace operation after  the Eq. \re{simmul}).

 In summary, $\eta$-odd term
should be omitted and we are interested only in $\eta$-even contribution.
But then the domain of integration may be supplemented (of course, after
multiplying the total expression by $\fr$) by its image under $\eta
\rightarrow -\eta$ reflection to get the rectangle with vertices $(0,1),(1,1),(1,-1),(0,-1)$.
Then the integration by $\a$ gives the factor $1$ as the integrand does not depend on $\a$, while for
$\eta$ it is worth making the substitution
\be
\eta=sin\;\theta\;\;,\;\; \q \in[-\frac{\pi}{2};\frac{\pi}{2}]
\ee
The integral takes the form (wherein we have accounted that the integrand is even w.r.t.
$\theta \to - \theta$)
\be \ba \label{456}
P(q,q',\pa)=
-\frac{a_d}{2} 2^{\hd-2} \pd \\ \\(\pm i f)^{2-\hd}
\int \limits_{0}^{\frac{\pi}{2}} d\q \;\cos \q\;
e^{i\sin\q \,nf}\;
 (n \cos \q)^{\hd-2}J_{\hd-2}\;(\pm i |f||n| \cos\q)
\ea \ee
Remarkably, the last integral by $\q$ is calculated explicitly due to the existence of the
special family of identities for Bessel's functions \re{bessgood}. Taking into account the identity
\be
J_{-\fr} (z)=\sqrt{\frac{2}{\pi z}} \cos z
\ee
the integral \re{456} fits the identity \re{bessgood} under the choice
$q=\fr, p=\frac{d-3}{2}$. For the validity of the formula \re{bessgood} it is necessary
$p\geq 0$ that excludes the case $d=2$, but this case is already excluded from our
consideration as in $d=2$ $H=-p^2$ is not a vacuum.

The final result is ($d>2$)
\be \ba \label{4567}
P(q,q',\pa)= \k_d
(n^2)^{\hd-2} \left(\,\sqrt{-n^2f^2+ (nf)^2}\,\right)^{-\frac{d-3}{2}} J_{\frac{d-3}{2}}
\left(\,\sqrt{-n^2f^2+ (nf)^2}\,\right)\\ \\
n=-\frac{\hbar}{2}\, \pa\;,\;f=q+q',
\ea\ee
where
\be
\k_d=-2^{-\frac{d+7}{2}} \pi^{\frac{1-d}{2}} \hbar^{-d},
\ee
No ambiguity in sign comes from the square root terms, as
the expression $z^{-\n} J_{\nu}(z)$ contains only even powers of $z$.

Let us comment on the emergence of the non-integer powers of $\Box$ in odd dimensions.
From the first glance it seems contradictory to our previous considerations as we
are just calculating power series in derivatives in the framework of the semiclassical expansion.
These apparent problems have arisen because we have used integral formula for the star-product
and then many times integrated by parts neglecting boundary terms. And the quadratic action
corresponding to generating function
\re{4567} in odd dimensions is just a total derivative, if one defines nonlocal operators
$\Box^{\frac{k}{2}}, k=1,3,5,...$ in such a way they are antisymmetric:
\be
\int d^dx\; A \Box^{\frac{k}{2}} B =-\int d^dx\; B \Box^{\frac{k}{2}} A+\mbox{boundary terms}.
\ee
Hereafter we suppose this choice is made. So in odd dimensions the quadratic action
vanishes.

\subsection{Spin decomposition}

As shown in \cite{Segal:2001di} and in Sec. \re{dthst},
conformal higher spin model of integer spin $s$ is characterized by the
generating function
\begin{equation}\label{baseq10}
P^{(s)}_{\mu}(q,q',\pa)= \Box^{\frac{d}{2}-2}\;
(\rho\rho')^{-\frac{d-3}{2}}\;
J_{s+\frac{d-3}{2}} (\mu\rho)\; J_{s+\frac{d-3}{2}}
(\mu\rho') \;\cC^{\frac{d-3}{2}}_s (\frac{\tau}{\rho\rho'}),
\end{equation}
\be \ba
\rho^2= \Box
\,q^2 -(q\pa)^2\;;\; \rho'^2=\Box \,q'^2_{\bot} =\Box \,q'^2
-(q'\pa)^2
\\ \\ \tau = \Box \,(qq') -(q\pa)(q'\pa).
\ea
\ee
$\mu$ is a real operator, $\m=\m(\Box)$. According to the theorem of
Sec. \re{corper}, just calculated quadratic theory,
characterized by the generating function \re{4567} should break up as a sum
of deformed conformal  higher spin theories\re{baseq102} with all spins
$s=0,1,2,3,...$, with deformation parameter
\be
\m^2=-\frac{\hbar^2}{4} \Rightarrow \frac{\m}{\sqrt
{\Box}}=\pm \frac{i\hbar}{2},
\ee
while direct calculation should
give specific values for the coefficients of the sum.
Let us show by direct calculation this is really the case.
To this end, note that the argument of
the Bessel's function in \re{4567} may be rewritten as
\be \ba
-n^2f^2+ (nf)^2=-\frac{\hbar^2}{4}(\rho^2+\rho'^2+2\t)=-\frac{\hbar^2}{4}(\r^2+\r'^2+2\r \r' cos \varphi)\\ \\
cos \varphi =\frac{\tau}{\rho\rho'}.
\ea \ee
The last row is justified as for Euclidean vectors $q,q',\pa$ one has
\be
\tau=\Box (q_{\bot} q'_{\bot}) \leq \rho\rho' =\Box |q_{\bot}||q'_{\bot}|,
\ee
where $q_{\bot}$ stands for the $\pa$-transverse projection of $q$.
The identity for Bessel's functions \re{Bess4} allows one to decompose
the generating function of the quadratic higher spin action \re{4567} as
follows
\be \ba \label{45678}
P(q,q',\pa)=
\k_d \;2^{\frac{d-3}{2}}
\G (\frac{d-3}{2})\\ \\
(n^2)^{\hd-2}
\sum \limits_{s=0}^{\infty}(-)^s (s+\frac{d-3}{2} ) (\pm\frac{i\hbar}{2}\r)^{\frac{d-3}{2}}
(\pm\frac{i\hbar}{2}\r')^{\frac{d-3}{2}}
J_{s+\frac{d-3}{2}}(\pm\frac{i\hbar}{2}\r)  J_{s+\frac{d-3}{2}}(\pm\frac{i\hbar}{2}\r')
\;\cC_s^{\frac{d-3}{2}} (\frac{\tau}{\rho\rho'})=\\ \\=
2^{\frac{d-3}{2}}
(\pm\frac{i\hbar}{2})^{d-3}(\frac{\hbar}{2})^{d-4} \k_d \sum \limits_{k=0}^{\infty}\;(-)^s \;\G ( s+\frac{d-1}{2} )\;
P^{(s)}_{\pm\frac{i\hbar}{2}}(q,q',\pa)=\\ \\
-2^{-1-d} \pi^{\frac{1-d}{2}} (\pm\frac{i\hbar}{2})^{d-3} \hbar^{-4} \sum \limits_{k=0}^{\infty}\;(-)^s\;
\G (s+\frac{d-1}{2} )\;
P^{(s)}_{\pm\frac{i\hbar}{2}}(q,q',\pa).
\ea
\ee
As above, note that the $\pm$-ambiguity
disappears as all expressions are even either w.r.t. $\r$ or $\r'$.

We have shown that the quadratic action is an infinite sum of conformal higher spin models
of all integer spins $s=0,1,2,3, ...$. More precisely, the quadratic action, corresponding to the generating function
\re{45678} is rewritten in terms of undressed variables $\vf_s$ introduced in Sec. \re{dresundres} as a sum
of conformal higher spin actions described in Sec. \re{fchst}. This is the corollary of the paper.

\subsection{Calculation of the cubic part of the action}

We start with the expression for the cubic part of the action $\cA_3$ \re{expan},
which we rewrite using cyclic property of the trace and the equations
\be\ba
{\bar H}=-p^2\;\;,\;\;
e^{\l p^2 *} = e^{\l p^2}
\ea \ee
in the form
\be\ba
\cA_3=-\fr a_d \int_C d\l \tilde{F}(\l) \int dx dp e^{\l p^2}\\ \\
\int \limits_{0}^{\l} d\t_1 \int \limits_{0}^{\t_1} d\t_2
\int \limits_{0}^{\t_2} d\t_3 h * e^{(\t_1-\t_2)p^2}*h *e^{-(\t_1-\t_2)p^2}
e^{(\t_1-\t_3)p^2} *h*e^{-(\t_1-\t_3)p^2}.
\ea\ee
Making change of variables
\be\ba \label{9876}
t_1=\t_1-\t_3\;\;,\;\;t_2=\t_1-\t_2\\ \\
t_2\leq t_1\leq\t_1\;\;,\;\;0\leq \t_2\leq\t_1
\ea\ee
and using
\be\ba
e^{\l p^2*} * h(x,p) * e^{-\l p^2*}=h(x-2i\hbar\l p,p),
\ea \ee
the cubic action is rewritten as
\be\ba
\cA_3=-\fr a_d \int_C d\l \tilde{F}(\l) \int_\Sigma d\t_1 dt_1 dt_2 \int dx dp e^{\l p^2}\\ \\
\int \limits_{0}^{\l} d\t_1 \int \limits_{0}^{\t_1} dt_2
\int \limits_{t_2}^{\t_1} dt_1 h(x,p) * h(x- 2i\hbar t_2 p,p)*
h(x-2i\hbar t_1 p,p),
\ea\ee
where integration over $\t_1,t_1,t_2$ goes over the domain $\Sigma$
specified by the second row of the Eq. \re{9876} and the condition $\t_1 \in [0,1]$.

Employing integral representation for the
star-product \re{starintegral}
and making some simple redefinition of
integration variables one rewrites the last expression as
\be\ba
\cA_3=-\fr a_d b_d^{-2} \int_C d\l \tilde{F}(\l)\int_\Sigma d\t_1 dt_1 dt_2 \int dx dp dx_1 dp_1 dx_2 dp_2 dx'_1 dp'_1 dx'_2 dp'_2
e^{\l p^2}\\ \\  \exp(\frac{2i}{\hbar}[(x-x'_2)p'_1 +(x'_1-x)p'_2 +(x'_2-x'_1)p +
x'_2-x_2)p_1 +(x_1-x'_2)p'_2+\\ \\ +2i\hbar(p_1p_2(t_2-t_1) +t_1p_2p_2' -t_2p_1p_2')])\;
B(x'_1,p'_1)B(x_1p_1)B(x_2,p_2).
\ea \ee
Doing elementary integrations over $x,p,x'_2,p'_2$ one gets
\be\ba
\cA_3=-\fr a_d \int_C d\l \tilde{F}(\l)\int_\Sigma d\t_1 dt_1 dt_2 \int dx_1 dp_1 dx_2 dp_2 dx'_1 dp'_1
e^{\l (p'_1+p_2-p_1)^2} \\ \\ \exp(\frac{2i}{\hbar}[(p-p_2)x'_1 +p'_1(x_2-x_1) +x_1p_2-x_2p_1 +
2i\hbar( (p'_1-p_1)p_2 t_1 +(p_2-p'_1)p_1 t_2) ])
\\ \\B(x'_1,p'_1)B(x_1p_1)B(x_2,p_2).
\ea \ee
To integrate by $\l$ one has to make use of the equality
\be \ba
\int d\l \tilde{F}(\l)
\int \limits_{0}^{\l} d\t_1 \int \limits_{0}^{\t_1} dt_2
\int \limits_{t_2}^{\t_1} e^{-\l \s_3 +t_1 \s_1 +t_2 \s_2} = \\ \\
-\int \limits_{0}^{1} d\a \a^2 \int \limits_{0}^{1} d\b \int \limits_{\b}^{1} d\gamma
F'''(\s_3 -(\gamma \s_1 +\b \s_2)\a),
\ea \ee
with
\be
\s_1=-(p'_1-p_1+p_2)^2\;\;,\;\;\s_2=-4(p'_1-p_1)p_2\;\;,\;\;\s_3=-4p_2(p_2-p'_1),
\ee
to get \label{453}
\be\ba
\cA_3=-\fr a_d
\int \limits_{0}^{1} d\a \a^2 \int \limits_{0}^{1} d\b \int \limits_{\b}^{1} d\gamma
\int dx_1 dp_1 dx_2 dp_2 dx'_1 dp'_1
\d''(\s_3 -(\gamma \s_1 +\b \s_2)\a)\\ \\ e^{\frac{2i}{\hbar}[(p_1-p_2)x'_1]+(x_2-x_1)p'_1 +x_1p_2-x_2p_1]}
\;B(x'_1,p'_1)B(x_1p_1)B(x_2,p_2).
\ea \ee
To exhibit the structure of the cubic action it is worth calculating not the action itself but
a special generating function which encodes the complete structure of the cubic action.

Let
\be
\cA_3=\int dx \sum \limits_{s_1,s_2,s_3,k_1,k_2,k_3=0}^{\infty}
P^{b_1(k_1)b_2(k_2)b_3(k_3)}_{a_1(s_1)a_2(s_2)a_3(s_3)}
h_{,b_1(k_1)}^{a_1(s_1)}(x) h_{,b_2(k_2)}^{a_2(s_2)}(x)h_{,b_3(k_3)}^{a_3(s_3)}(x)
\ee
be general cubic action (commas denote derivatives w.r.t. $x^m$).
Generating function is the function of six vector variables
$b_1^m,b_2^m,b_3^m, d_{1m}, d_{2m}, d_{3m}$ of the form
\be \ba
P_3=\sum\limits_{s_1,s_2,s_3,k_1,k_2,k_3=0}^{\infty}\frac{1}{s_1 ! s_2! s_3!}
P^{b_1(k_1)b_2(k_2)b_3(k_3)}_{a_1(s_1)a_2(s_2)a_3(s_3)}  \\ \\
b_1^{a_1}...b_1^{{a_1}_{s_1}} b_2^{a_2}...b_2^{{a_2}_{s_2}}
b_3^{a_3}...b_3^{{a_3}_{s_3}} d_{1\;b_1}...d_{1\; {b_1}_{k_1}}
d_{2\;b_2}...d_{2\; {b_2}_{k_2}}d_{3\;b_1}...d_{3\; {b_1}_{k_1}}.
\ea \ee
\begin{lem}
Let the cubic action is given by the formula
\be
\cA_3=\int dx_1 dx_2 dx_3 dp_1 dp_2 dp_3 f(Q_2-Q_1,Q_3-Q_2, Q_1-Q_3)
h(Y+Q_1) h(Y+Q_2)h(Y+Q_3),
\ee
where vector variables $Y, Q_{1,2,3}$ are defined as follows
\be
x_{1,2,3}=Y+Q_{1,2,3}\;\;,\;\; Q_1+Q_2+Q_3=0.
\ee
Then the generating function $P_3$ is
\be
P_3=3^{-d} \int dQ_1 dQ_2 f(Q_2-Q_1,-Q_1-2Q_2, 2Q_1 +Q_2)
e^{Q_1(d_2-d_1) +Q_2(d_3-d_1) +p_1b_1 +p_2b_2 +p_3b_3}.
\ee\end{lem}
The proof is by direct calculation.
Combining this lemma with the expression \re{453} one finds
\be\ba
P_3=-\fr 3^{-d} a_d
\int \limits_{0}^{1} d\a \a^2 \int \limits_{0}^{1} d\b \int \limits_{\b}^{1} d\gamma
\int dQ_1 dQ_2 dp_1 dp_2 dp'_1
\d''(\s_3 -(\gamma \s_1 +\b \s_2)\a)\\ \\
\exp(\frac{2i}{\hbar}[Q_1(-p_1+2p_2-p'_1+\frac{\hbar}{2i}(d_2-d_1)
+Q_2(p_2-2p_1+p'_1 +\frac{\hbar}{2i}(d_3-d_1)] \\ \\+p_1 b_1 +p_2 b_2 +p'_1 b_3)
\ea\ee
Doing elementary integration by $Q_1, Q_2, p_1, p_2$ one gets
\be\ba
P_3=-\fr 3^{-d} a_d b_d^2
\int \limits_{0}^{1} d\a \a^2 \int \limits_{0}^{1} d\b \int \limits_{\b}^{1} d\gamma
\int dp'_1
\d''(\s_3 -(\gamma \s_1 +\b \s_2)\a)\\ \\
\exp[(p'_1+\frac{1}{3}(2n_2-n_1))b_1 +(p'_1 +\frac{1}{3}(n_2-2n_1)) b_2 +p'_1 b_3],
\ea\ee
where
\be
n_1=\frac{\hbar}{2i}(d_2-d_1)\;\;,\;\;n_2=\frac{\hbar}{2i}(d_3-d_1)
\ee
\be \ba
\s_1=\frac{4}{3}(p'_1 +\frac{1}{3}(n_2-2n_1))(2n_2-n_1)\;\;,\;\;
\s_2=-\frac{4}{3}(p'_1 +\frac{1}{3}(2n_2-n_1))(n_2-2n_1)\;\;,\;\;\\ \\
\s_3=-(p'_1-\frac{1}{3}(n_2+n_1))^2.
\ea \ee
The argument of $\d''$ is rewritten as
\be\ba
\s_3 -(\gamma \s_1 +\b \s_2)\a=-(p'_1+\d_1 n_1 +\d_2 n_2)^2 +A^2(n_1,n_2|\d_1,\d_2),
\\ \\
A^2=(\d_1 n_1 +\d_2 n_2
)^2 -\frac{2}{9}(\d_2-\d_1)(n_2-2n_1)(2n_2-n_1) -\frac{1}{9}(n_1 +n_2)^2\\ \\
\d_1=\frac{1}{3}(-1+\a(4\b-2\gamma))\;\;,\;\;\d_2=\frac{1}{3}(-1+\a(4\gamma-2\b)).
\ea\ee
Making change of variables $\Pi=p'_1+\d_1 n_1 +\d_1 n_2$ one gets
\be\ba  \label{974}
P_3=-\fr 3^{-d}a_d b_d^2
\int \limits_{0}^{1} d\a \a^2 \int \limits_{0}^{1} d\b \int \limits_{\b}^{1} d\gamma
\int  d\Pi
\d''(-\Pi^2+A(n_1,n_2|\d_1,\d_2))\\ \\ \exp\left(\Pi(b_1+b_2+b_3) -(\d_1 n_1 +\d_1 n_2)(b_1+b_1+b_3)+
\frac{1}{3}(2n_2-n_1)b_1 +\frac{1}{3}(n_2-2n_1)b_2\right)
\ea\ee
Just like in the calculation of the quadratic action, the integration over momenta is performed
easily as it is given by the generating function $G^{(3)}_d$ (see \ref{genf1},\ref{genf4}):
\be
G^{(3)}_d (l) =\int d^dp\; \d'' (M^2-p^2)
e^{ip_m l^m}=
2^{\hd-3}\pd M^{\hd-3} l^{-\hd+3} J_{\hd-3} (\m l)
\ee
Combining the last formula with the expression for the generating function \re{974}
one arrives at the formula (wherein $f=b_1+b_2+b_3$)
\be\ba  \label{9745}
P_3=-2^{-\hd-4}\pi^{\frac{3d}{2}} 3^{-d} \hbar^d
 e^{\frac{1}{3}(2n_2-n_1)b_1 +\frac{1}{3}(n_2-2n_1)b_2}
(f)^{3-\frac{d}{2}} \int_\Sigma d\a d\delta_1 d \delta_2 \\ \\e^{-(\d_1 n_1 +\d_2 n_2)f}
A(n_1,n_2|\d_1,\d_2)^{\frac{d}{2}-3} J_{\frac{d}{2}-3} (f A(n_1,n_2|\d_1,\d_2)),
\ea\ee
where the integration by $\a, \d_1, \d_2$ goes over the domain $\Sigma$ in the
three-dimensional space with coordinates $\a, \d_1+ \d_2+\frac{2}{3}, \d_2-\d_1$, where
each plane with fixed $\a \in [0,1]$ is a triangle with vertices
$\a(1,0,0),\a(1,1,1),\a(1,2,1)$. The integration may be performed by decomposing
analytic function $z^{\hd-3}J_{\frac{d}{2}-3}(z)$ in power series in $z$
then each term of the decomposition of the integrand is a polynomial in $A(n_1,n_2|\d_1,\d_2)$
and the integral runs over the compact domain $\Sigma$.

Let us briefly discuss the structure of cubic vertices. It reproduces general structure anticipated
in the beginning of this section,
\be\label{cubscheme}
A_3[h]\sim
\int d^dx\; \sum \limits_{s_1,s_2,s_3=0}^{\infty} \pa^{d- 6 + s_1 + s_2 + s_3}  h_{s_1}...h_{s_3},
\ee
and the exact expression \re{9745} just gives the special values to the coefficients of separate terms in the
last formula.

Let us comment of the procedure of finding interactions of original conformal higher spin fields,
described by traceless symmetric tensors. We saw in Sec. \re{dresundres} that the quadratic action of deformed
conformal higher spin theories  is expressed via "undressed fields $\phi_s(x),\, s=0,1,2,3,...$
\re{baseq1023},\re{baseq1024}
related to $h_s$ by the invertible field redefinition \re{undr44},\re{812} and
that only traceless  parts of undressed fields $\vf_s$ contribute the action. Thus, at the quadratic level
the action is entirely expressed in terms of $\vf_s$. The cubic action we just have calculated depends,
however, on $h_s$ or, what is the same, on all $\phi_s$, i.e. on $\vf_s$ and on trace parts of $\phi_s$.
This means that in order to formulate the cubic action one has to add all the trace parts of $\phi_s$ to
original traceless fields $\vf_s$ the quadratic action is build of. Thus from a first glance it
seems as if at nonlinear level one has to introduce additional degrees of freedom
that is dangerous, because variation w.r.t.
these additional fields would give new equations of motion and make dynamics inconsistent in general.
However this is of course not the case. First of all let us note that the decoupling of the traces of
$\phi_s$ is controlled at the quadratic level by the hyper-Weyl gauge invariance, which
is being smoothly deformed to nonlinear level, the latter property is in the core of our construction.
Thus one makes sure analogous decoupling phenomena will take place at the nonlinear level, and
one just has to get which form this decoupling takes at the cubic order.

In fact it is easy to
describe the decoupling of the traces at the cubic level, as according to the well-known statement
(App. \re{ApA}),
the cubic action is gauge-invariant w.r.t. linearized gauge transformations provided the fields obey
linearized equations of motion. Therefore, if the fields satisfy the linearized equations of motion then
traces of $\phi_s$ do decouple. This implies the traces of $\phi_s$ enter the cubic action only
as a part of linearized equations of motion.
This phenomenon, and analogous ones which are nothing but the consequences of the full
gauge invariance $ \d H = \o^\dag * H +H* \o $,
make it consistent the procedure of solving full nonlinear equations perturbatively
by representing dynamical fields as a formal power series
in $e$, $h=h_0+eh_1+e^2 h_2+...$ and solving the nonlinear equations order by order in $e$.
For instance, up to the cubic order one has, schematically,
\be
L_2 h_0=0\;; L_2 h_1=L_3 h_0 h_0,
\ee
where $L_2$ and $L_3$ are structures governing the quadratic and the cubic actions. From the
first equation one gets that $h_0$ obeys free equations of motion, the latter are gauge invariant
w.r.t. linearized gauge transformations that implies decoupling of the traces. By the same reasons, the traces
of $h_1$ decouple in the second equation, while the "source" $L_3 h_0 h_0$ is gauge invariant w.r.t.
linearized gauge transformations as $h_0$ is on-shell.
To summarize, the traces of the fluctuation $h_0+eh_1$ do not contribute equations of motion up to cubic order,
therefore, no dangerous phenomena arise. Due to analogous decoupling phenomena, the procedure of perturbative
solution of the equations of motion will be always consistent and will give interactions of the infinite number
of {\it traceless} tensors of all spins, each spin will enter the theory just once.

One more important issue is $d=3,4$ case and comparison of our
results with results by Fradkin and Linetsky \cite{linetsky}.
Although we do not pursue explicit comparison in this paper it is
worth making a few comments. First of all note that in $d=3$ the
action by Fradkin and Linetsky is just a Chern-Simons action which
does not set a dynamical three-dimensional theory, in this sense
our result that the quadratic part of the action of conformal
higher spin theory in odd dimensions is a boundary term conforms
with results of the authors of \cite{linetsky}. In $d=4$ the
authors investigated construction of the cubic interaction in a
supersymmetric theory of conformal higher spin fields with spins
$s=1,\frac{3}{2},2,\frac{5}{2},3,\frac{7}{2},4...$ and concluded
there is no consistent interaction beyond cubic order until each
spin enters the theory with infinite multiplicity. This looks
different from our consideration as we have the theory of all
spins, from zero to infinity, with multiplicity one. Leaving aside
difference of approaches and difference of supersymmetric and
purely boson treatment one notes that perhaps the main difference
is the presence in our formulation of spin-$0$ field that is
purely auxiliary\footnote{Besides, in the purely boson version of
Fradkin and Linetsky's model there is no spin-$1$ Maxwell field.}.
At the quadratic level, it is zero on-shell while at higher levels
$s=0$ field is expressed algebraically via the higher spin fields.
Namely, employing schematic representation of the cubic
interaction \re{cubscheme} one gets at the cubic order \be
h_{0,0}(x)=0;\;h_{1,0}(x)\sim \sum \limits_{s_1,s_2=1}^{\infty}
\pa^{-2+s_1+s_2} h_{0, s_1} h_{0, s_2}, \ee where $h_{0,s},
h_{1,s}$ stand for the zero and the first order of the
perturbative expansion for the component of rank $s$. It will give
additional, "dynamical" vertices for higher spin fields in
addition to those explicitly encoded in generating function
\re{9745}. To summarize, we observe the problems with infinite
multiplicities of all spins (and, of course, problems with
all-orders interaction) encountered by Fradkin and Linetsky are
avoided in our approach, which differs by the presence of the
auxiliary spin-$0$ scalar.

\section{Free particles and conformal higher spin fields}\label{sequat}

It is clear one may consider the united action
\be\ba\label{unitedac}
\cA[H,\psi]=\cA[H]+S_H[\psi]=\\ \\
Tr\theta(\hat{H})+\left(
\fr <\psi|\hat{H}|\psi> +\fr (<\psi|\hat{H}|\psi>)^*\right),
\ea \ee
where $e$ is a coupling constant. In the classical limit, interaction with classical particle is achieved by adding the
coupling
\be  \label{co91}
S_H[x,p,\l]=\int d\tau \{p_m \dot{x}^m -\lambda H (x,p)\}.
\ee
Then the equations of motion read
\be\ba\label{eom256}
\d_H \cA[H,\psi]=\d(\hat{H})+e|\psi><\psi|=0\\ \\
\d_{<\psi|} \cA[H, \psi]= \hat{H}|\psi>=0.
\ea
\ee
The interpretation of the second equation is that $|\psi>$ is a zero mode for $H$, while from the first equation
one gets that $|\psi><\psi|$ is a projector to zero modes subspace. In fact, the second equation is
the consequence of the first one. All this means that the zero modes subspace is one-dimensional. Thus, in the formal operator approach,
the classical dynamics, described by the united action \re{unitedac}, just describes, for $e\neq 0$
the hamiltonians with one-dimensional normalizable zero modes subspace.
In the case $e=0$ one gets the formal equation $\d(\hat{H})=0$ which is can be interpreted as
a condition to $\hat{H}$ not to admit normalizable zero modes in ${\bf L}^2$, this is discussed
in more detail in App. \re{ApB}.

Here it is also worth
remarking that this interpretation is just a formal picture that may have a little to do with actual dynamics of conformal
higher spin fields. On the other hand, in some cases this picture may provide an adequate description, therefore we
find it useful to keep it in mind.

In $d>2$, the equations \re{eom256} possess Poincar\'e-invariant solution
\be
H=-p^2, \psi=0
\ee
The perturbative decomposition of the action \re{unitedac} around this vacuum describes interaction of
massless complex scalar and the tower of real conformal higher spin fields of all integer spins. Note that in
$d=4$ the conformal spin-$0$ scalar is an auxiliary field, while in $d=6$ it is the familiar real massless
scalar, therefore there are in total three real massless scalars in this case.
In the cubic order in fluctuations one gets the picture of sources built from the
bilinear combinations of wave function interacting by means of exchange by conformal higher spin fields.
Namely, the equations \re{eom256} are solved perturbatively, order by order in the coupling
constant $e$, after representing
the Hamiltonian $H(x,p)$ and the wave function $\psi(x)$ in series in $e$ as
\be
H=-p^2 +\sum \limits_{k=1}^{\infty} e^k h_k(x,p), \;\psi
(x)=\sum \limits_{k=0}^{\infty} e^k \psi_k(x).
\ee
Then in zero and the first order in $e$ one gets the equations of motion
\be\ba
\d(p^2)=0\\ \\
\hat p^2|\psi_0>=0\\ \\L \hat{h}_1 = |\psi_0><\psi_0|,\;
\ea\ee
where $L$ is the wave operator for linearized equations of deformed conformal higher spin fields, calculated above.
In the classical limit, one gets
\be\ba
\d(p^2)=0\\ \\
L h_1 (x,p)=\int d\t \l(\t) \;\d(x-x(\t), p-p(\t)),
\ea\ee
where $x(\t),p(\t)$ is an on-shell trajectory, and $\d(x,p)$ is $2d$-dimensional $\d$-function.
The r.h.s. of the second row is nothing but the (relativistic)
phase space density which thereby plays the role of a source for conformal higher spin fields.

The interpretation of these equations is that there exists a source for conformal higher spin fields
built from the bilinear combination of the on-shell wave function $|\psi>$ (or, in the classical limit,
the source is a functional of classical trajectories).
After analytic continuation to Minkowski space, the equations $L h_1=... $  describe radiation of conformal higher spin fields by the
complex scalar field $\psi(x)$, or, in the classical limit, by the massless point particle, and their propagation
through space-time according to free equations of motion.

After this equation is solved for $h_1$ in terms of
$\psi(x)$ and some initial(boundary) data one can substitute the result back to the action
\re{unitedac} to get the first order interaction of the sources
in the form
\be\label{int227}
S_{int}=e\int d^d x\; \psi_0^*(x)\;\hat{h}_1\; \psi_0(x),
\ee
where $\psi_0(x)$ are on-shell, $\hat{p}^2\psi_0=0$. This action
is gauge invariant w.r.t. linearized gauge transformations
\be\label{bsqinflin1}
\d h(x,p) =\o(x,p)^\dag *p^2+ p^2 * \o(x,p),
\ee
as $\hat{p}^2 \psi(x)=0$. This invariance is nothing but the manifestation of the covariance of the coupling
$S_H[\psi]=e<\psi|\hat{H}|\psi>$ in all orders in $e$. Indeed, expanding $S_H$ and gauge transformations
\re{bsqinf} in $e$,
one gets, in the lowest order in $e$, the invariance of $S_{int}$
w.r.t. linearized gauge transformations \re{bsqinflin1} provided $\psi$ is on-shell w.r.t. vacuum Hamiltonian.
As we show in Sec. \re{sNint} the interaction \re{int227} is an economic form of Noether interaction of the on-shell
complex scalar field $\psi$ to the external gauge fields via "Noether currents - one-forms" coupling,
with \re{bsqinflin1} being the counterpart of abelian gauge transformations of linearized connections.

Note that in $d=4$ the linearized equation for the scalar fluctuation $h(x)$ is purely algebraic, while the
corresponding current does not contain derivatives, so one gets the quartic interaction of the form
$\sim (\int d^dx\; |\psi(x)|^2)^2$. $s=1$ contribution in $d=4$ is a standard interaction of charged matter sources
via Maxwell fields.

\section{Global Symmetries. Conformal algebra and its infinite-dimensional
extension, conformal higher spin algebra. Higher spin currents}\label{sgl}

This section presents an independent study of global symmetries of the
conformal higher spin model. The global symmetries are gauge transformations that preserve the vacuum.
The emergence of higher spin fields as background fields of the point particle allows one to describe
the algebra of global symmetries in a simple fashion, as algebra of global symmetries of the particle.
This way we describe all known conformal higher spin algebras (in $d>2$) as algebras of observables of massless
particle in an arbitrary dimension, and besides we give definition of boson conformal higher spin
algebras in an arbitrary dimension $d$.

General technique of symbols of operators we use allows us to derive a simple formula for
all Noether currents of free complex scalar field
for any wave operator, provided its global symmetries are known.
In particular, in the case of the wave operator $\hat{H}=\hbar^2 \Box$ our formula reproduces, in another basis,
the set of currents found in \cite{konsh} by direct calculation.

But first of all we will show that the global symmetries of the vacuum $\hat{H}=\hbar^2 \Box$
contain the conformal algebra $so(d-q+1,q+1)$ and that its action on undressed fluctuations $\vf_s$
is diagonal w.r.t. spin $s$. This means we deal with standard conformal theory in any dimension $d>2$.
Then we will pass to the next issues related to the infinite-dimensional
extension of the conformal algebra to the conformal higher spin algebra.

\subsection{Undressing, Reconstruction, and
Conformal invariance.}

Consider the field $f(x)$ taking  values in
a representation of the Lorentz algebra $so(d-q,q)$ with
generators $M_{ab},\;a,b=0,...,d-1$. Then
one has the associated representation of
the conformal algebra $so(d-q+1,q+1)$ with generators
\be\ba\label{cotens}
\tilde{P}_a=-\pa_a, \tilde{J}_{ab}=-(x_a \pa_b-x_b \pa_a) +
M_{ab}\\ \\
\tilde{D}= x\pa_x+ \g,\; \tilde{K}_a=-(x^2 \pa_a-2 x_a  x\pa_x) -2
x^b M_{ab} -2x_a\g,
\ea \ee
where $\g$ is an arbitrary constant, the conformal weight.

Consider the representation of the conformal group on the components of the
Hamiltonian $H(x,p)$. By construction, it is given by the gauge transformations
\re{bsqinf5432} which leave intact the vacuum Hamiltonian ${\bar H}=-p^2$,
\be\ba\label{glco775}
\d H(x,p)=\o^{\dag} * H +H* \o,\;\;\o =a +\frac{i}{\hbar}\e\\ \\
\e= b^a P_a+b^{ab} J_{ab} + c^a K_a +c D\\ \\
P_a=p_a, J_{ab}=x_a p_b-x_b p_a,\;
D=- (x,p),\; K_a=x^2 p_a-2 x_a  (x,p),\\ \\
a= -c+2c^a x_a,
\ea\ee
that is in accord with \re{glob0135}.
These transformations exhaust global symmetries with $\e$ linear in momenta, that is
the subject of our analysis in this section.
Using the formula \re{star} for the star-product, one derives
\be\ba\label{cotens1}
\d H(x,p) =
\left[b^a \tilde{P}_a+b^{ab} \tilde{J}_{ab} + c^a \tilde{K}_a +c \tilde{D}
+\frac{\hbar^2}{4}c^a (\pa_p^2\pa_a-2(\pa_x \pa_p)\d_a)\right] H(x,p)\\ \\
\ea\ee
where  $\d^a=\frac{\pa}{\pa p_a}$ and operators
$\tilde{P}_a,\tilde{J}_{ab},\tilde{K}_a,\tilde{D}$ map each rank-$s$
subspace of $H$ into itself and act in this subspaces as the generators of
the conformal algebra \re{cotens} with
\be
M_{ab}=-(p_a\d_b-p_b\d_a), \;\;\g=2-p\pa_p
\ee
One observes that the conventional tensor transformation laws \re{cotens}
are deformed by the quantum correction, represented by the $\hbar^3$-term in
\re{cotens1}. As a consequence, the conformal transformations mix fields
of different spins. However, at the linearized level
this mixing miraculously cancels after passing to the
``undressed fields'' of Sec. \re{dresundres}. Indeed, consider representation of
the global symmetry in the space of fluctuation near the conformally flat
vacuum $H=-p^2$, $h(x,p)$. It is given (see App. \re{ApA}) by the differential of the full global
symmetry transformations \re{glco775} evaluated at the vacuum point. However, as
the global transformations \re{glco775} are linear, their
differential looks just the same, with $H\rightarrow h$ substitution,
\be\label{globfluct}
\d h(x,p)=
\o^{\dag} * h +h* \o.
\ee
As it is shown in Sec. \re{dresundres}, at the linearized level it is possible to set
the undressed fields $\phi_s$ \re{baseq1024} traceless without loss of generality,
this implies in turn that $h$ is traceless either (that may be easily seen by
inspecting reconstruction map \re{812}),
\be
\pa_p^2\; h(x,p)=0.
\ee
\begin{teo}
The representation of the conformal group on the ``undressed'' fields \re{baseq1024}
\be
\vp_s=[U_{s+\di}(\frac{i\hbar}{2}\stackrel{\rightarrow}{\rho}) h]|_{p^s},\;
\pa_p^2\vp_s=0,\;\stackrel{\rightarrow}{\rho}=-i\pa_x\pa_p
\ee
is diagonal w.r.t. $s$ and has the form \re{cotens} with
\be\label{diag22}
M_{ab}=-(p_a\d_b-p_b\d_a), \;\;\g=2-s.
\ee \end{teo}
{\it Proof.}
For Poincar\'e generators $P_a, J_{ab}$ and dilations $D$ the proof is trivial as
for these generators there is no quantum correction  because they commute
with the undressing operator. The nontrivial part of the proof is to check that
the quantum correction cancels for the special conformal transformations $K_a$.
By direct calculation one proves that, for any function of
$z=\frac{i\hbar}{2}\stackrel{\rightarrow}{\rho}$,
\be
[f(z), K_a] =-i\hbar^2\left(2(p\pa_p+\hd -1)f'(z)+zf''(z)\right)\delta_a,
\ee
where the last equation is understood as valid when l.h.s. and r.h.s. act on
traceless functions (i.e. those with $\pa^2_p=0$).
Then, for the transformation laws of the undressed field $\vp_s$ one has
\be \ba
\d_{K_a} \vp_s=
[U_{s+\di}(\frac{i\hbar}{2}\stackrel{\rightarrow}{\rho}) \d_{K_a} h]|_{p^s}=\\ \\
\tilde{K}_a \vp_s -i\hbar^2 \{
(zU_{s+\di}(z)+2(p\pa_p+\hd-1) U'_{s+\di}(z)+zU''_{s+\di}(z))
\d_a h \} |_{p^s}=\\ \\
\tilde{K}_a \vp_s -i\hbar^2
\{2(s+\hd-1)-(1+2s+d-3)\} [U'_{s+\di}(z)\d_a h \} |_{p^s}]=\\ \\
\tilde{K}_a \vp_s,\;\g=2-s \ea \ee
where in passing from the second to the third row
we have used that $U_\n(z)$ satisfies the equation \re{uuret}.

Note that the value of the conformal dimension $\g=2-s$ is in accord with
the structure of the free conformal spin-$s$ action \re{baseq10} which may be
schematically written down as \be \cA_s[\vp_s]\sim\int d^d x \;\vp_s \;\Box^{\hd-2}\;
\pa^{2s} \;\vp_s.  \ee
Needless to say, this  value of conformal dimension in length units matches with that in mass
units \re{massun} derived from $AdS/CFT$ considerations.

As the gauge transformations \re{glco775} are linear,
the global transformations are linear either, hence they present the symmetry of
each term of the perturbative decomposition of the ``quantized volume''around
the vacuum $H=-p^2$. In particular, the quadratic action $\cA_2$ is invariant
up to boundary terms. As the transformations are diagonal w.r.t. spin, they
present symmetry of each free conformal higher spin-$s$ theory with the action $\cA_s[\vf_s]$
\re{defac},\re{f1}. Our derivation is a somewhat indirect, but the first, proof of the conformal
invariance of the conformal higher spin theories.

\subsection{Algebra of observables of free quantized point particle and
all bilinear conserved currents in general background fields.}
\label{toki}
Here we simultaneously investigate global symmetries of the model and the Noether currents
of the free action \re{php}.
\subsubsection{Algebra of observables}

Consider the
action \re{php}. The {\it global symmetries} are by  definition
the gauge transformations
(\ref{bsqinf},\ref{bsqpinf}) which do not
change the Hamiltonian. They act on the wave functions by the rule
\re{bsqpinf}, \be \label{bsqpinf22}\d|\psi>=-\hat{\o}|\psi>.
\ee
In accordance with \re{bsqinf1} the infinitesimal global symmetries
are characterized by the equation
\be \label{glob1}
\omega^{\dag} * H +H * \omega =0
\ee
and hence, due to \re{bsinv55}, leave the action \re{php} invariant up to boundary terms.
The global symmetries form the Lie algebra ${\cal G}_{H}$ with commutation relations
\be \label{glalg}
\ba
[\delta(\omega_1) ,
\delta(\omega_2) ]  =  \delta(\omega_3)  \\ \\ \omega_3= -[\omega_1,\omega_2].
\ea
\ee
It is seen that
the antihermitian global symmetries $\om =i\e$ commute with $H$ while hermitian ones
$\om = a$ anticommute with $H$.

The Lie algebra \re{glalg} possesses the
subalgebra ${\cal G}_{trivH}$ of {\it trivial }
symmetries
\be  \label{triv}
\om = i \m* H   \;\;,\;\; \mu ={\mu}^{\dag}
\ee
which annihilate every {\it on-shell} wave function, i.e. the one
satisfying the equation \re{hpsi}.
In fact the trivial symmetries form an ideal since given
a trivial symmetry, its commutator with any symmetry
is a trivial symmetry again:
\be
[i\mu*H , \om]= i(-\mu *\om^{\dag}-{\om} * \mu)*H.
\ee
As a consequence, the factor of the Lie algebra $\cG_H$
by the ideal of trivial symmetries,
\be  \label{observ}
{\cal G}_{o}={\cal G}_{H}/{\cal G}_{trivH}
\ee
is a Lie algebra which we call {\it"algebra of observables" }.

The Lie algebra ${\cal G}_H$ admits a bilinear {\it nonassociative }
multiplication
\be  \label{circ}
\om_3\equiv\om_1 \circ \om_2 = i(\om_1 *\om_2 +\om_2 *\om_1),
\ee
which is symmetric w.r.t. exchange of multipliers, and properly respects
the trivial symmetries as the $\circ$-product of a trivial symmetry with
any other  one is a trivial symmetry again:
\be
(i\mu*H) \circ \om= i(i \om *\m- i\m * {\om}^{\dag})*H.
\ee
Therefore, the algebra of observables ${\cal G}_{o}$ acquires $\circ$-product
either: the $\circ$ product of two elements from factor space
${\cal G}_{o}$ is given by the canonical projection of the $\circ$- product
of their prototypes in ${\cal G}_{H}$.
Analogously, any symmetrized product of symmetries gives a new symmetry:
\be  \label{mcirc}
\ba
\om_N\equiv \{\o_1,...,\o_n\}_{\circ} \equiv
i^{N-1}\sum \limits_{permutations\{1...N\}}^{} ( \om_1 *... * \om_N)=\\ \\
2^{-N}\sum \limits_{permutations\{1...N\}}^{}
(\om_1 \circ
(\om_2 \circ ... \circ (\om_{N-1}\circ \om_N) ...))
\ea
\ee
This product which we call {\it multi-$\circ$} product,
also respects properly the trivial symmetries, as
$\om_N$ is trivial if any of the multipliers is trivial.

\subsubsection{Noether currents}
Any global symmetry $\om$ leave the action \re{php} invariant up
to boundary terms, therefore there exists a {\it Noether current}
$J_{\om}$ given by bilinear expressions in $\bar{\psi}(x),\psi(x)$
and its derivatives. Let us recall a simple derivation of Noether
theorem via studying general variation of an action w.r.t.
transformation of the global symmetry  but with a parameter being
an arbitrary function of $x$. Let $S[\phi^i]$ be an action
(depending on a set of fields $\{\phi^i\}$) invariant under the
global symmetry $\delta \phi^i= \delta_{\n} {\phi^i}$, where $\n$
is a constant parameter.This means that the variation of the
Lagrangian w.r.t. this transformation is a total derivative.
Hence, if one makes the same transformation but with $\n$ being an
arbitrary function of $x$ the variation of the Lagrangian has a
form \be\ba \d L(\f^i, \pa \f^i, \pa^2 \f^i, ...)=\n_{,m} J^m
(\f^i, \pa \f^i, \pa^2 \f^i, ...)+\mbox{total derivative}=\\ \\
-\n \pa_m J^m+\mbox{total derivative}. \ea\ee Hence, if the fields
$\f^i$ satisfy classical equations of motion, the {\it Noether
current} $J^m$ is conserved, \be \frac{\d S}{\d \f_i}[\f^i]=0
\Rightarrow \pa_m J^m(\f^i, \pa \f^i, \pa^2 \f^i, ...)=0.\ee

The Noether currents are defined up to trivial ones of the form
$J^m_{triv} = \pa_n k^{[mn]}$ with antisymmetric $k^{[mn]}$. The
trivial symmetries are defined as \be \label{trivgen} \delta
\phi^i = k^{[ij]} \frac{\pa S}{\pa \phi^j}\;\;;\;\;k^{[ij]}=-
k^{[ji]} \ee and are easily seen to give rise to on-shell zero currents.
Note that our definition of trivial symmetries \re{triv} is a
realization of a general one \re{trivgen} applied to the action
\re{php}.

Now we derive Noether currents associated with the global
symmetries \re{glob1}. To this end one has to study the variation
of the action \re{php}, \be \ba\label{co9} S[\psi] = \fr
<\psi|\hat{H}|\psi>
+\fr (<\psi|\hat{H}|\psi>)^*=\\ \\
 <\psi|\hat{H}|\psi> +\mbox{boundary terms},\ea \ee
w.r.t. transformations \be
\label{bsqpinf222}\d|\psi>=-\n(x)\hat{\o}|\psi>, \ee where
$\n(x)$ is an arbitrary function of $x$.

The variation is \be \ba \d S=-\fr \int d^d x \left\{ (\n
\hat{\o}\psi)^* \hat{H} \psi + \psi^* \hat{H} \n
\hat{\o} \psi \right\} +c.c.=\\ \\=-\fr \int d^d x \left\{
\psi^* (\hat{\o}^\dag \n \hat{H} \psi +  \hat{H} \n \hat{\o})
\psi\right\} +c.c. =\\ \\ =-\fr \int d^d x \left\{ \psi^*
([\hat{\o}^\dag \n] \hat{H} \psi +  [\hat{H} \n] \hat{\o})+
\n(\hat{\o}^\dag \hat{H} + \hat{H} \hat{\o}) \psi\right\} + c.c.,
\ea\ee where square brackets denote commutators. The first term
may be neglected as it will give rise only to on-shell vanishing terms in
$J^m$. The $\hat{\o}^\dag \hat{H} + \hat{H} \hat{\o}$ term equals
zero off-shell, due to \re{glob1}. The remaining term in the
variation is \be -\fr  \int d^d x \left\{ \psi^* [\hat{H}
\n ] \hat{\o} \psi \right\} + c.c. \ee Using the
representation \re{weyl988} for an operator $\hat{H}$ with the Weyl symbol
\be
H(x,p)=\sum \limits_{s=0}^{\infty} H^{m(s)}(x) p_{m_1}...p_{m_s},
\ee
one rewrites the last equation as \be\ba \d S=-\fr
\int d^d x \,\psi^*  \sum \limits_{s=0}^{\infty}
\sum \limits_{k=0}^{s}\sum \limits_{l=1}^{s-k} \\ \\
\left\{(-i\hbar)^s (\fr)^k C^k_s C^l_{s-k}
H^{m(k)m(s-k)}_{,m(k)} \n_{,m(l)} (\hat{\o}\psi)_{,m(s-k-l)}\right\}+c.c.,
\ea\ee
where $C^k_s=\frac{s!}{k!(s-k)!}$ are binomial coefficients.
Integrating by parts, one represents the last expression in the form
\be
\d S=\int d^d x \,\n_{,m} J_\o^m +\mbox{total derivative},\ee
where
\be\ba \label{noet23}
J_\o^m=-\fr \sum \limits_{s=0}^{\infty}\sum \limits_{k=0}^{s}\sum \limits_{l=1}^{s-k} \\ \\
\left\{(-i\hbar)^s (\fr)^k (-)^{l-1}C^k_s C^l_{s-k}
 \pa_{m_1}...\pa_{m_{l-1}}\left[\psi^* H^{m(k)m(s-k)}_{,m(k)}  (\hat{\o}\psi)_{,m(s-k-l)}\right]\right\}+c.c.
\ea\ee
is a formula for the Noether current $J_\o^m$ associated with the symmetry $\o$ of the Hamiltonian $H$.
It is also
seen that the trivial symmetries of the form $\hat{\om} =i \hat{\mu}\hat{H}$
give rise to the on-shell vanishing currents.

For the case of our main concern, the Hamiltonian of the form
\be\label{glob345}
H=-p^2,
\ee
the expression for currents simplifies as only $s=2,k=0,\, l=1,2$ terms are nonzero. The result is
\be\label{noet235}
J_\o^m=- \frac{\hbar^2}{2}\left(\psi^* \pa^m (\hat{\o} \psi)-(\pa^m\psi^*)\hat{\o} \psi\right)
+c.c.
\ee
In order to get the complete list of conserved currents in this case one has to solve the global symmetry
equation \re{glob1} in the case $H=-p^2$, which is rewritten as
\be\label{glob11}
(p^2-\frac{\hbar^2}{4}\Box)a(x,p)+ p\pa_x \e(x,p)=0
\ee
and find the algebra of observables \re{observ}.
We are interested only in global symmetries analytic in momenta (otherwise $\hat{\o}$ is
non-local).
Using $\mc$ product one can construct the solutions of higher degree in momenta $p_m$ in terms of
$\mc$ products of solutions of lower degrees in momenta.
First of all, let us find the symmetries with $\e$ linear in momenta and with $a$ independent of $p$,
\be
\e=\xi^m (x) p_m,\;a=\alpha(x)
\ee
Then \re{glob11} implies
\be\ba\label{glob134}
\eta^{ab}\,\a=-\frac{1}{2} (\pa^a \xi^b+\pa^b \xi^a),\\ \\ \Box\, \a=0.
\ea \ee
The first row is the definition of the {\it conformal Killing vector} $\xi^m$, with general solution of the form
\be\ba\label{glob135}
\e= \xi^m(x)p_m= b^a P_a+b^{ab} J_{ab} + c^a K_a +c D\\ \\
P_a=p_a, J_{ab}=x_a p_b-x_b p_a,\;
D=- (x,p),\; K_a=x^2 p_a-2 x_a  (x,p),\\ \\
\a= -c+2c^a x_a,
\ea\ee
while the second row of \re{glob134} is satisfied as a consequence of the last row of \re{glob135}.

The generators $P_a, J_{ab},K_a, D$ form the conformal algebra $so(d-q+1,q+1)$ of $d$-dimensional conformally
flat space with $P_a, J_{ab}$ being Poincar\'e translations and Lorentz transformations and $K_a, D$ being the special conformal transformations and dilations,
respectively. It is well-known the generators of the conformal algebra may be combined in the second-rank antisymmetric tensor
$M_{AB}, \;A,B, =0...d-1, d, d+1$ (where $d$-th and $d+1$-th coordinates are space-like and time-like, respectively), according to the
rule
\be\label{glob1355}
M_{ab}=J_{ab}, M_{+-}=-\fr D, M_{+a}=\fr  P_a, M_{-a}=\fr K_a,
\ee
where $Y^{\pm}=\fr(Y^d \pm Y^{d+1})$.

Pure imaginary $\o$ which belong to the center of the algebra of all phase-space functions,\be\label{u12}
\o=iConst,\; Const\in {\bf R}\ee is a symmetry either (the localization of this symmetry is the conventional
$U(1)$ gauge symmetry).

The $\mc$ products \re{mcirc} of conformal generators
\be\label{glob67}
\o_{A_1...A_k B_1...B_k}=\left\{M_{A_1 B_1}...M_{A_k B_k}\right\}_\circ,\;k=1,2,...
\ee
and $U(1)$-charge $\o$'s of the form \re{u12} span general solution of the global symmetry equation (\ref{glob1},\ref{glob345},\ref{glob11}), modulo trivial generators
\re{triv}. The proof will be delivered elsewhere \cite{ya}\footnote{An independent  proof
of this statement is in fact contained in paper \cite{Eastwood:2002su}, which has appeared when the present paper has been
prepared for publication.}.
Note that a multi-$\circ$ product with $U(1)$ charge does not lead to new
generators.

By construction, $\o_{A_1...A_k B_1...B_k}$ are antisymmetric w.r.t. changes $A_r \leftrightarrow B_r$, and symmetric
w.r.t. simultaneous change $A_r \leftrightarrow A_{r'}, B_r \leftrightarrow B_{r'}$.
It is easy to prove that antisymmetrization of
$\o_{A_1...A_k B_1...B_k}$ w.r.t. any three indices equals zero. The most efficient way to see this is to use
$d+2$-dimensional realization of the massless particle dynamics (see App. \re{ApG}),
where the generators $M_{AB}$ are realized as
$M_{AB}=Y_A \P_B - Y_B \P_A$ with $d+2$-dimensional coordinates $Y^A$ and momenta $\P_B$, then the antisymmetrization
over three indices is zero because any third-rank antisymmetric tensor built from two vectors $Y_A, \P_B$,
and $d+2$-metric $\eta_{AB}$, is zero.
All this means $\o_{A_1...A_k B_1...B_k}$ is characterized by the Young tableaux of the form
\bigskip
\be
\begin{picture}(70,50)\label{young1}
\put(20,45){k} \put(33,35){\circle*{2}}
\put(25,35){\circle*{2}}
\put(17,35){\circle*{2}}
\put(25,25){\circle*{2}}
\put(17,25){\circle*{2}}
\put(33,25){\circle*{2}}
\put(00,40){\line(1,0){70}}
\put(00,30){\line(1,0){70}} \put(50,30){\line(0,1){10}}
\put(60,30){\line(0,1){10}} \put(70,30){\line(0,1){10}}
\put(00,20){\line(1,0){70}} \put(00,20){\line(0,1){20}}
\put(10,20){\line(0,1){20}} \put(40,20){\line(0,1){20}}
\put(60,20){\line(0,1){20}} \put(50,20){\line(0,1){20}}
\put(70,20){\line(0,1){20}}
\end{picture}
\ee
where the first row is occupied by $A$'s and the second row  by $B$'s.
In every $AB$ column one has the antisymmetry, that corresponds to the antisymmetry of
$M_{AB}$ generators.
At $k=0$, when  the Young tableaux is empty, one has $\o=i\,Const, \; Const \in {\bf R}$.

The representation of the conformal group being a direct sum of
irreducible ones characterized by traceless Young tableaux
\re{young1} with $k=0,1,2,3,...$ is what we denote
$\Upsilon_d$.

If one applies  symmetry generators \re{glob67} to the {\it on-shell} wave functions, i.e. those satisfying
\be
-p^2 \psi(x)=\hbar^2 \Box \psi(x)=0,
\ee
according to the rule \re{bsqpinf22}, one can prove the following identity
\be\label{iden2}
0=N_{AB}\psi(x)=\{(\hat{M}_{AB} {\hat{M}^B}{}_C+\hat{M}_{CB} {\hat{M}^B}{}_A) - \hbar^2 \eta_{AC}(d-2)\} \psi(x),
\ee
that is proved most simply in $d+2$-dimensional formalism (see Appendix \re{ApG}).
As a consequence, the contraction of $\hat{\o}_{A_1...A_k B_1...B_k} \psi(x) $ by any pair of $A,B$ indices
reduces to the linear combination of $\o_{A_1...A_{k'} B_1...B_{k'}} \psi(x),\; k'<k $. (Indeed, after the contraction,
one transfers the generators to be contracted to the right (by commuting them with the rest generators),
to act on the wave function $\psi$, and then uses
\re{iden2}, all this results in $\mc$-products of $M_{AB}$ of a lower order). Thus, only {\it traceless} parts of
$\hat{\o}_{A_1...A_k B_1...B_k} \psi(x)$ are linearly independent.
Studying $\o$'s acting on general on-shell
wave function is equivalent to studying the {\it observables} \re{observ} in the sense that
the ideal of trivial symmetries is projected out in either cases.
Therefore, {\it the traceless parts
of $\o_{A_1...A_k B_1...B_k}$,
\be\label{trl58}
\tilde{\o}_{A_1...A_k B_1...B_k} =\mbox{\rm traceless part of} \;\o_{A_1...A_k B_1...B_k}
\ee
are in one-to-one correspondence with the basis in the algebra of observables \re{observ}}.
In accordance with our considerations, one has the Noether currents \re{noet235}
\be\label{curr47}
J^{m}_{{A_1...A_{k},B_1...B_{k}}},
\ee
associated with traceless symmetries \re{trl58}.
The on-shell currents automatically project out the trivial symmetries as
trivial symmetries \re{triv} give rise to the on-shell vanishing currents. Therefore,
{\it the space of on-shell conserved currents \re{noet23} is in one-to one correspondence with the
algebra of observables}. According to the above treatment, {\it basis in the space of all on-shell
conserved currents is provided by the currents \re{noet235}, corresponding to the
traceless Young tableaux \re{young1}, for $k=0,1,2,...$}
The $U(1)$-current corresponds to $k=0$, $\o=i\, Const,\; Const\in {\bf R}$.
The energy-momentum  and orbital momentum currents are $J^{m}_{a+},J^{m}_{ab}$, respectively.

The traceless Young diagrams \re{young1} are exactly those
characterizing higher spin currents of free massless scalar
constructed by Konstein, Vasiliev, and Zaikin by direct
calculation \cite{konsh}. Naturally, all the currents
obtained in \cite{konsh} are the linear combinations of our
currents \re{noet235}, modulo additions of ``improvements'' terms of the
form $\d J^m = \pa_n N^{mn},\;N^{mn}=-N^{nm}$ which do not
contribute to the conserved charges. Our derivation uncovers
simple algebraic structure underlying the construction of
currents, naturally explains the origin of two-row rectangular
Young tableaux \re{young1} and provides the simple unified formula
\re{noet235} for all currents at once. Besides, the formula \re{noet23} provides an
expression for all Noether currents of complex scalar field in
arbitrary background, i.e. for an arbitrary wave operator
$\hat{H}$.

\subsubsection{Classical global symmetries}
Let us recall that  the classical covariance algebra $\cG^{cl}$ consists of
infinitesimal transformations which act on $H(x,p)$ by the rule
\be
\label{canon0001}
\delta H(x,p) = 2 a(x,p) H(x,p)+ \{\epsilon, H(x,p)\}.
\ee
Given Hamiltonian $H(x,p)$, the subalgebra of {\it classical global symmetries}
$\cG^{cl}_H$ is the subalgebra consisting of the elements $(\e,a) \in \cG^{cl}$ that
preserve $H$,
\be \label{glob000}
\delta_{\e,a} H =2 a(x,p) H(x,p)+ \{\epsilon, H(x,p)\} =0.
\ee
These equations are easily seen to be a contraction of the quantum ones \re{bsqinf},
in accordance with discussion of Sec. \re{subs42}.
In fact, all the expressions below are obtained as a classical limit of corresponding quantum counterparts.

Given $\e$, representing a global symmetry, its $a$-part is restored
unambiguously from \re{glob000}, so to specify the symmetry
transformation it is enough to write down only its $\e$-part, as we do
in some places below.
All global symmetries form an associative commutative algebra w.r.t product
\be\label{circ00}
(\e_1,a_1) \circ (\e_2, a_2) =(\e_1\e_2, \e_1 a_2 +\e_2 a_1)
\ee
This associative algebra structure is compatible with the Lie algebra
structure in the sense the commutator \re{lie00} is a derivation of
the product $\circ$. It is easy to see that this product
is nothing but a classical limit of the $\mc$ product \re{circ}, in fact, one
obtains \re{circ00} from \re{circ} if one neglects the order of product factors.

The trivial global symmetries are by definition those whose
$\e$-part vanishes on the constraint surface:  \be \label{trivcl} (\e_t,a_t)=(\tilde{\m} H,
-\fr \{\tilde{\m}, H\}), \ee where $\tilde{\m}(x,p)$ is an arbitrary function,
related to the analogous quantum parameter $\m$ \re{triv} as $\tilde{\m}=\m \hbar$.
The trivial symmetries form the
ideal $\cG_{trivH}^{cl}$ in $\cG^{cl}_H$ considered either as a Lie algebra w.r.t.
commutator \re{lie00} or as an associative algebra w.r.t. $\circ$-product.
The factor algebra \be
\cG^{cl}_o =\cG^{cl}_H/\cG_{trivH}^{cl}\ee is called {\it algebra of observables}.
Each element of $\cG^{cl}_o$, except the central element $(Const,0)$ is a
nontrivial symmetry of the particle's classical action \re{ham00}.

Let $H=-p^2$. Then the classical algebra of observables is a
contraction of the conformal higher spin algebra,
and is defined as follows.
The generators are given by a pair
$(\e,a)$, where
\be \label{obert}
\e=u_{A_1...A_{s-1},B_1...B_{s-1}}=\left\{ M_{A_1 B_1}... M_{A_{s-1}
B_{s-1}} \right\}_{\circ},
\ee
where $M_{AB}$ are generators of the conformal algebra \re{glob135},\re{glob1355}, while the
$\mc$ product is classical \re{circ00}. As in the quantum case, these generators
are characterized by the Young tableaux \re{young1}. The factorization by
the trivial generators \re{trivcl} leads to the equation
\be
N_{AB}^{cl}\equiv M_{AB} {M^B}_C = 0,
\ee
which holds on the constraint surface $p^2=0$.
Just like in the quantum case,
this equation ensures the {\it tracelessness} (after factorization by the ideal of trivial generators)
of the generators \re{obert} w.r.t. contraction of any two indices.

\subsubsection{Noether currents of the classical particle}
Let us show that all the currents \re{noet23} have their
classical analogs derived  by exactly the same procedure as the quantum
currents. Given a classical  global symmetry $\o=(\e,a)$,
one derives, applying the Noether theorem, the currents
$J^{\o}[\psi]$ corresponding to each global symmetry $\hat{u}$.
The answer is
\be\label{noetclass}
J_{\o}^m (x) =\int d\t \l \d(x-x(\t)) \e(x(\t),p(\t)) \pa_{p_m} H(x(\t)p(\t)),
\ee
where $x(\t),p(\t),\l(t)$ is a solution for classical equations of motion.
The currents are conserved as
\be\ba
\pa_m J_{\o}^m =\int d\t \;\l \;\e \;\pa_{p_m} H \pa_m \d(x-x(\t)) =-\int d\t \;\l \;\e \; \{H,\, \d(x-x(\t))\}=\\ \\=
\int d\t \;\e\; \frac{d}{d\t} \d(x-x(\t)) =\int \;d\t\; \frac{d}{d\t} \;\e \;\d(x-x(\t)) =0,
\ea\ee
as \be\frac{d}{d\t}\e(x(\t),p(\t)) =\{\e,H\}=-aH(x(\t),p(\t))=0.\ee

\subsubsection{Noether currents of the Hamilton-Jacobi sigma-model}

Given a classical symmetry $\o=(\e,a)$, the variation of the action of Hamilton-Jacobi sigma model
w.r.t. transformations with parameter $\n(x)\o$ reads
\be
\d S_{cl}[\r,\vp]=\int d^d x \n_{,m} J_\o^m,
\ee
where
\be
J_\o^m=\r^2(x)(\e \frac{\pa}{\pa_{p_m}} H )(x,\pa_m \vf)
\ee
are Noether currents. They are easily seen to conserve as a consequence of the equations of motions
(\ref{hjeqn},\ref{treqn}).
One easily checks these currents are nothing but classical limit of the quantum currents
\re{noet23}, obtained by substituting \re{quans7} and keeping only first terms of the semiclassical expansion.

\section{Noether interaction of the scalar particle to gauge fields, and conformal higher spin fields}\label{sNint}

Given the full set of conserved currents \re{noet23} it is natural to try constructing Noether interaction of the
complex scalar field to the one-forms which take values in the algebra of observables, and try to
formulate a sensible theory of these Yang-Mills fields. In this section we show that from the point
of view of the influence on the particle dynamics the Noether interaction introduced in such a way is nothing
but a very superfluous formulation of the interaction $\int d^dx\; \psi^* \hat{h} \psi$ and the
only sensible part that is contained in gauge fields is that one contributing to $\hat{h}$.

\subsection{Quantum case}
Given the full set of (bilinear in $\psi$) conserved currents of the particle it is natural to construct
first order interaction of the particle to some gauge fields by the rule
\be\label{flinint}
S_{int} =e\sum \limits_\o \int d^dx\; J^m_\o (\psi,\pa \psi,...\psi^*,\pa \psi^*,...)A^\o_m(x),
\ee
where $\psi$ is on-shell, i.e. $\hat{H}\psi=0$, $e$ is a coupling constant and the sum goes over the basis
in the algebra of observables $\o \in \cG_o$, while $A_m^\o(x)$ are gauge fields being subject to the
abelian gauge transformations of the form
\be\label{ginvab}
\d A_m^\o(x)=\pa_m \e^\o(x),
\ee
with gauge parameters $\e^\o(x)$. Needless to say, the action $S_{int}$ is gauge invariant w.r.t. these
transformations. In fact, the gauge fields $A_m^{\o}(x)$ are one-forms on $x$-space
taking values in the linear space of the algebra of observables $\cG_o$. Naturally, one constructs
Yang Mills curvatures
\be
R^\o_{mn}=\pa_m A^\o_n -\pa_n A^\o_m + e \sum_{\o',\o''} f^{\o}_{\o',\o''} A^{\o'}_m A^{\o''}_n,
\ee
where $ f^{\o}_{\o',\o''}$ are structure constants of $\cG_o$, and tries to construct an action in terms of
these curvatures for the newly introduced gauge fields $A^\o_m(x)$.
As the global symmetry of the free massless particle is the conformal higher spin algebra it is natural to anticipate
one can construct a theory of conformal higher spin fields from $A^\o_m(x)$. This kind of reasoning was
employed by Fradkin and Linetsky in a series of papers where the authors used the same set of one-forms
\footnote{The coincidence of the algebra of global symmetries of the free massless particle and $d=4$ conformal higher
spin algebras by Fradkin and Linetsky is the subject of the next section.} and formulated the linearized dynamics
in terms of $A^\o_m (x)$ subject to some constraints and tried to generalize this to get interaction vertices.
On of the reasons no progress had been achieved beyond the cubic order could be the excess in the number of
gauge fields $A^\o_m(x)$ as compared to the conformal higher spin fields described most naturally
in terms of fluctuation of the hamiltonian $\hat{h}$.

To exhibit this excess let us demonstrate that
the set of fields $A^\o_m(x)$ is superfluous to gauge all the global symmetries  of
the scalar particle, and the Noether coupling is expressed in terms of special combinations of $A_m^\o(x)$,
which number equals the number of conformal higher spin fields.
Indeed, let us use general formula \re{noet235} for Noether currents and integrate
by parts to represent the expression \re{flinint} in the form
\be\ba
S_{int}=e \int d^d x \;\psi^*(x) \hat{h} \psi(x) \\   \\
\hat{h}= -\frac{\hbar^2}{2} \sum \limits_{\o} \{ 2 A^\o_m(x) \pa^m +(\pa^m A^\o_m)\} \hat{\o} + \mbox{hermitian conjugate}+\hat{\z}^\dag  \hat{p}^2 +\hat{p}^2 \hat{\z}
\ea\ee
The $\z, \z^{\dag}$-terms account the ambiguity in the definition of $h$ arising because
$\psi$ is on-shell, $\hat{p}^2 \psi(x) =0$. It appears the Weyl symbol of operator $\hat{h}$ reads
\be\label{map48}
h(x,p)=-i\hbar \sum \limits_{\o}\left(  (A^\o_m(x) p^m) * \o(x,p) - \o^\dag(x,p) * (A^\o_m(x) p^m)\right) +
\z^\dag * p^2 +p^2 *\z,
\ee
that may be checked by direct calculation using the formula \re{weyl988}
and its consequence $\widehat{A^m p_m}=-i\hbar( A^m \pa_m+ \fr (A^m_{,m}))$.

As a consequence of the abelian gauge invariance \re{ginvab},
$h(x,p)$ is subject to gauge transformations which may be
easily deduced as follows
\be\ba\label{bsqinfind}
\d A^\o_m p^m=\pa_m \e^\o p^m =-i\hbar (p^2 * \e^\o -\e^\o *p^2) \Rightarrow\\ \\
\d h(x,p) =\frac{2i}{\hbar} \sum \limits_{\o}\left(  \frac{i}{2\hbar} (p^2 * \e^\o -\e^\o *p^2) * \o(x,p) -
\o^\dag(x,p) * \frac{i}{2\hbar} (p^2 * \e^\o -\e^\o *p^2)\right)=\\ \\=
\fr \sum \limits_{\o} \left(   p^2 * [\e^\o,\o]_+  + [\e^\o,\o^\dag]_+ * p^2)\right)=\\ \\
=\tilde{\o}_\e * (-p^2)+(-p^2)*\tilde{\o}_\e,
\ea\ee
where (in passing from the second to the third row) the definition of $\o$,
$\o^\dag p^2+p^2*\o=0$, is used and
\be\label{map49}
\tilde{\o}_\e=- \fr \sum \limits_{\o}( \e^\o * \o +\o * \e^\o)
\ee
But the set of fields $h(x,p)$ subject to gauge transformations \re{bsqinfind} is nothing but the set of deformed
conformal higher spin fields with $\m^2=-\frac{\hbar^2}{4}\Box$ i.e. the set of conformal higher spin fields.
Therefore, one gets the following statement.

\begin{teo} General Noether coupling of the scalar particle to
abelian gauge fields $A^\o_m(x)$ is a superfluous formulation of
the general coupling $S_{int}=e\int d^dx\; \psi^* \hat{h}\psi$ of
the complex scalar field to conformal higher spin fields encoded
in the operator $\hat{h}$ subject to gauge transformations $\d
\hat{h}= \hat{\tilde{\o}} \hat{p}^2+\hat{p}^2 \hat{\tilde{\o}}.$
The map from "superfluous" gauge fields $A_m^\o$ and "superfluous"
gauge parameters $\e^\o$ to the gauge fields $h(x,p)$ and gauge
parameters $\tilde{\o}$ is provided by eqs. \re{map48} and
\re{map49}.\end{teo}

\subsection{Classical and semiclassical case}

Using the currents \re{noetclass} one constructs the Noether interaction of the classical particle to the
abelian gauge fields $A_m^\o$ as
\be\ba
S_{int} =e\sum \limits_\o \int d^dx\; J^m_\o (x)A^\o_m(x)=\\ \\=
e\sum \limits_\o \int d^dx\; A^\o_m(x) \int d\t \l \;\d(x-x(\t)) \;\e_\o(x(\t),p(\t)) \pa_{p_m} H(x(\t)p(\t))=\\ \\=
e \sum \limits_\o  \int d\t \;\l\; [\e_\o A^\o_m \pa_{p_m} H ] (x(\t),p(\t)),
\ea\ee
where
\be
\o=a_\o+\frac{i}{\hbar}\e_\o
\ee
As in the quantum case, this interaction is expressed via $h(x,p)$ as
\be
S_{int}=-e\int d\t \l h(x(\t),p(\t)),
\ee
where
\be\label{map50}
h(x,p)=-\sum \limits_\o \e_\o(x,p) A^\o_m \pa_{p_m} H  +\z H,
\ee
where $\z(x,p)$ is arbitrary. This is the classical version of the map
\re{map48}. The gauge transformations $\d A^\o_m=\pa_m \e^\o$
are mapped to the conformal higher spin fields gauge transformations
\be
\d h =\{\tilde{\e}, H\},
\ee
for $H=-p^2$ and
\be\label{map51}
\tilde{\e}=-\sum \limits_\o \e_\o(x,p)\; \e^\o (x),
\ee
which is nothing but the classical limit of the quantum relation \re{map49}.

If one introduces Noether interaction of the Hamilton-Jacobi sigma-model, one
derives the same relations \re{map50},\re{map51} between the gauge transformations and
gauge fields, while the interaction takes the form
\be
S_{int}=e \int d^d x \r^2 h(x, \pa \vp).
\ee

\subsection{On connection with Fradkin-Linetsky approach.}

As it's obvious from \re{map48}, the gauge field $A^\o_m (x)$ enters all formulae only via combinations
\be
A^{\o}(x,p) =A^\o_m (x) p^m.
\ee
Consider the subset of $\o$'s of the form \be \o_{a(s-1)}=\frac{i}{\hbar} p_{a_1}...p_{a_{s-1}},\; s=1,2,...,\ee
corresponding to all $\mc$-products of Poincar\'e translations.
Then the contribution of these $\o$'s to $h(x,p)$ is
\be
h(x, p) = \sum \limits_{s=1}^{\infty} (A^{a(s-1)m}(x)  p_m) * p_{a_1}...p_{a_{s-1}} + c.c.=
2 \sum \limits_{s=1}^{\infty} A^{a(s-1)m} p_m p_{a_1}...p_{a_{s-1}} + o(\hbar^2)
\ee
Thus one has the relation
\be\label{map54}
h^{a(s)}(x)=2 A^{a(s-1)a}(x)+o(\hbar^2),
\ee
where the right index $a$ in the superscript is the former one-form world index of $A^{a(s-1)}_m$, and
complete symmetrization over all $a$'s holds. The relation of this kind, expressing the true dynamical
conformal higher spin fields via the linearized connections taking values in an infinite-dimensional
"conformal higher spin algebra", are the basic input of "gauge description"
of conformal higher spin fields, used by Fradkin and Linestky in $d=2,3,4$ \cite{linetsky}
as a version of Vasiliev's "gauge description" \cite{Vasiliev:2001ur}
of Fronsdal higher spin gauge fields.

Contributions with another $\o$ produce the analogous result. Other $\o$ of the form \re{glob67}  give rise to
further contributions to $h(x,p)$ which however do not bring nothing new as \re{map54}
already provides $h(x,p)$ of
general form. Thus, in the "gauge description" of conformal higher spin fields, one introduces a huge set
of Yang-Mills fields which is much larger actual description of conformal higher spin fields calls for.

As a consequence of this excess, in the "gauge description" one
bumps with serious difficulties at the nonlinear level. In fact
one has to imply  some constraints to eliminate the superfluous
fields at the nonlinear level, and these constraints are not known
beyond the approximation of cubic interaction. It appears that due
to certain mechanism these constraints are sufficient for
constructing consistent cubic vertices \cite{linetsky}, but
advancement to next orders in interaction is prohibited without
generalizing the constraints to further orders.

To summarize, our description in terms of the Hamiltonian $H(x,p)$, subject to gauge
transformations \re{bsqinf5432} appears to be a more economic language for the description of conformal higher spin
fields. On the other hands, taking in mind the full nonlinear picture provided in this paper it may be possible to find
a proper nonlinear generalization of the Fradkin-Linetsky results, in particular, to find proper generalization of
the constraints on auxiliary fields, and to express our "quantized volume" action \re{fquasi111} in terms of corresponding
Yang-Mills curvatures.

A motivation for introduction of Yang-Mills fields was to gauge an infinite-dimensional global symmetry
that was anticipated to arise in conformal higher spin models. It is clear from our consideration
this is successfully achieved without introducing Yang-Mills fields. The remaining question is
do our global symmetries \re{observ} match with infinite-dimensional algebras
Fradkin and Linetsky started from. The answer is yes. In fact,
the conformal higher spin algebras, whenever constructed,
exactly match the algebra of observables $\cG_o$ of the massless particle.
This match is the subject of the next section.

\section{On oscillator realizations of conformal higher spin algebras in $d=3,4,6$. Twistors.}\label{stwist}

Here we will show the algebra of observables of the massless particle in flat $d$-dimensional Minkowski space
$chs(d-1,1)$ coincides with infinite-dimensional conformal higher spin algebras constructed previously
by several authors.

As we have shown above w.r.t. its conformal subalgebra $so(d,2)$ this algebra is decomposed
as a direct sum of finite-dimensional representations characterized by the infinite set of two-row
rectangular Young tableaux \re{young1} with spins $s=0,1,2,3,...$, the latter representation is what we denote
$\Upsilon_d$.

On the other hand the same algebra may play an important role in the theory of higher spin fields in $AdS_{d+1}$
as zero modes of linearized gauge transformations of the massless higher spin fields in $d+1$-dimensional AdS space
are in one-to-one correspondence with generators of $chs(d-1,1)$ \cite{ya1}.
Anticipating existence of consistent higher
spin theory in $d+1$ dimensions, the zero modes of linearized gauge transformations
are in one-to-one to global symmetries of the free action being a sum of Fronsdal theories of all
integer spins from zero to infinity. Thus, $chs(d-1,1)$ may be a global symmetry of free Fronsdal theories
in $AdS_{d+1}$. This reasoning conforms with results by
Lopatin and Vasiliev \cite{Lopatin:hz}, who constructed free
massless spin-$s$ theories corresponding to completely symmetric
fields on $AdS_{d+1}$, in terms of one-forms and their
linearized curvatures taking values in $\Upsilon_d$.

All this grounds the following definition of the higher
spin algebras:
\newtheorem{defin}{Definition}
\begin{defin}  Higher spin algebra corresponding to totally
symmetric higher spin fields in $D=d+1$
dimensional anti-de Sitter space is any algebra which contains
$so(d,2)$ as a subalgebra and decomposes w.r.t. it as $\Upsilon_d$ i.e. as direct sum
of finite-dimensional representations, characterized by traceless
two-row rectangular Young tableaux \re{young1}.\end{defin}

The uniqueness of the higher spin algebra is by no means
guaranteed, families of solutions may exist.

As $so(d,2)$ is conformal algebra in  $d$ dimensions, the same
definition may be interpreted as the one for an
infinite-dimensional symmetry algebra of conformal systems in $d$
dimensions (of course, we already know at least two of them: the
first is free massless particle and the second is all-orders
conformal higher spin theory of this paper), that justifies the
definition

\begin{defin} The conformal higher
spin algebra in $d$-dimensional Minkowski space $=$ higher spin
algebra in $D=d +1$ dimensional anti-
de Sitter space. \end{defin}
Till now, the boson higher spin algebras were known only for
$d=3,4,6$, where they have emerged as an even part of the
corresponding higher spin superalgebras via ``oscillator''
construction, the latter ascends to the famous family of
isomorphisms $so(2,1)=sl(2,{\bf R})$, $so(3,1)=sl(2,{\bf C})$,
$so(5,1)=su^*(4)$ and $ so(3,2)=sp(4,R)$, $so(4,2)=su(2,2)$,
$so(6,2)=so^*(8)$ being neatly written in terms of division
algebras ${\bf K}_\n={\bf R,C,H,} \n=1,2,4$ as
$so(\n+1,1)=sl(2,{\bf K_\n})$, $so(\n+2,2)=sp(4,{\bf K_\n})
$\cite{Cederwall:2000km}. Let us show that in these particular
cases our definition of the conformal higher spin algebras
\re{observ} reproduces the even subalgebras of the higher spin
superalgebras in $d+1=4,5,7$ (equivalently, of the conformal
higher spin superalgebras in $d=3,4,6$) constructed previously by
Vasiliev \cite{Vasiliev:2001ur}, Fradkin and Linetsky
\cite{linetsky} and Sezgin and Sundell \cite{Sezgin:2001ij}.

These superalgebras are constructed as Lie superalgebras corresponding to the
enveloping algebras of a set of even oscillators
\be
[a^I,a^J]=0,\;\;[b_I, b_J]=0,\;\;[a^{I}, b_J]=\d^I_J;
\ee
supplemented by some reality conditions, and factorized by certain ideals.
The $so(d,2)$ is realized in terms of bilinear combinations of oscillators, while the
oscillators transform w.r.t. $so(d,2)$ as spinors.

The point is that the origin of spinor generating elements is due to {\it twistors representation}
of the massless particle dynamics in $d=3,4,6$ \cite{twistors},\cite{Bars:2000en}.
Indeed, it is well-known that, in these dimensions, the mass-shell constraint
$p^2=0$ may be solved via introducing even {\it unconstrained} spinors as
\be
p_a=\varsigma{\bar \l} \G_a \l \Rightarrow p^2\equiv 0,\;\varsigma=\pm 1
\ee
where $\G_a$ are gamma-matrices, in $d=3$ $\l$ is a Majorana spinor , in $d=4$ $\l$ is a Weyl spinor, in $d=6$ $\l$ is Weyl or $su(2)$ Majorana-Weyl
spinor.
$\varsigma$ accounts for the ambiguity in the choice of the
connected component of the light cone, for one choice of $\varsigma$ one has  $p_0 \geq 0$, for another one $p_0 \leq 0$.

After introduction of the second spinor\footnote{the pair $\l,\o$ is said to form a ``twistor''.}
\be \o=i \varsigma x^a \G_a \l, \ee
the generators of the conformal algebra \re{glob135} appear to be rewritten in terms of bilinear combinations of
$\l$ and $\o$ in the manner
\be\ba\label{hom500}
P_a= \varsigma {\bar \l}\G_a  {\l},\;\; K_a \sim \varsigma {\bar\o} \G_a  \o\\ \\M_{ab}\sim {\bar \l} [\G_a, \G_b] \o +c.c.,\;\;
D \sim  {\bar\l} \o + c.c.
\ea \ee
Thus, any $so(d,2)$ generator is expressed either in terms of $(x^a,p_b)$ or in terms of $(\l,\o)$.
The commutation relations among any generators of $so(d,2)$ are reproduced if one considers
$\l,\o$ as a pair of canonically conjugated variables in a new, ``twistor'', phase space, like
\be
[\l,\l]=[\o,\o]=0,\;[l,\o]=i{\bf 1}
\ee

According to the analysis of Sec. \re{toki}, an arbitrary element of the algebra of observables of the $d$-dimensional
massless particle is represented by the equivalence class of polynomials built from the generators of conformal group $so(d,2)$,
modulo the ideal of trivial observables
in the enveloping algebra of $so(d,2)$ generated by the combination $N_{AB}$ \re{iden2}
\be\label{ilikk}
g_o\in {\cal G}_o,\;\;g_o=\sum \limits_{k=0}^{\infty} K^{A_1...A_k B_1...B_k} \{ M_{A_1 B_1}...M_{A_k B_k}\}_\circ \\ \\
\ee
\be\ba\label{triv41}
g_o \sim g_o + f_1^{AB}* N_{AB}+N_{AB}* f_2^{AB}\\ \\
N_{AB}=\hat{M}_{AB} {\hat{M}^B}{}_C+\hat{M}_{CB} {\hat{M}^B}{}_A  - \hbar^2 \eta_{AC}(d-2),
\ea\ee
where $K^{A_1...A_k B_1...B_k}$ are real $so(d,2)$ tensors characterized by the traceless Young tableaux \re{young1}, and
$f_{1,2}^{AB}$ are arbitrary elements of the form \re{ilikk}.

Any $g_o$ may be expressed in terms of the twistor $(\l,\o)$, where the monomials of $n$-th degree in $M_{AB}$ turn into the ones of
$2n$-th degree in $(\l,\o)$. We refer this representation of ${\cal G}_o$ as {\it twistor picture}.
Note that, in the twistor picture, the elements of the ideal of trivial
observables \re{triv41} are identically zero
\be\label{ili12}
N_{AB}(\l,\o)\equiv 0.
\ee
Therefore, each equivalence class \re{triv41} is unambiguously represented in the twistor picture, and one gets the homomorphism
\re{hom500},\re{ilikk} of the algebra of even power series in $(\l,\o)$ to the algebra of
observables of $d$-dimensional massless particle ${\cal G}_o$.

\subsection{Quaternion twistors. $D=7, d=6,$ higher spin algebra $hs(8*)$}
To show how it works we consider $d=6$ case, the $d=3,4$ ones are obtained then by the dimensional reduction, while
$d=10$ case is more complicated and will be considered elsewhere \cite{ya1}.
We use the $so(5,1)$ spinor formalism in the form given in \cite{Lyakhovich:1996ne},
which we partially reproduce in Appendix \re{ApI} for
the reader's convenience.

The mass-shell constraint $p^2=0$ is solved in terms of the $su(2)$ Majorana-Weyl spinor
\be
\l^{\a k},\;k=1,2,\;\;\a=1,2,3,4,
\ee
which obeys the reality constraint
\be\label{real64}
(\l^{\a k})^*={I^{\dot{a}}}_{\b} \ve_{kl}\l^{\b l},\;\;\ve_{kl}=-\ve_{lk},\;\;\ve_{12}=1, \; {I^{\da}}_{\b}
{{I^*}^{\b}}_{\dot{\g}}=-{\d^{\da}}_{\dot{\g}}
\ee
as
\be\ba\label{translation}
p_a \rightarrow p^{\a\b}=p_a (\tilde{\s}^a)^{\a\b}=\l^{\a}_k \l^{\b k},\\ \\
p_a (\s^a)_{\a\b}=p_{\a\b}=-\fr \ve_{\a\b\g\d} p^{\g\d}
\ea\ee
where the notation like $A_k B^k =A_k B_l \ve^{kl}$ is adopted. The equation $p^2=0$ is obeyed
as
\be
p^2=-\frac{1}{4} (p\s)(p\tilde{\s})=-\frac{1}{4} p_{\a\b} p^{\a\b}=-\frac{1}{4} (-\fr \ve_{\a\b\c\d} p^{\c\d}p^{\ab})=
\\ \\=\frac{1}{8}\ve_{\a\b\g\d}\l^{\a}_k \l^{\b k} \l^{\g}_{k'} \l^{\b k'} =0.
\ee
Introduce the second component of the twistor,
\be
\o_\a^k=i(x\s)_{\a\b}\l^{\b k},
\ee
which obeys the reality constraint
\be\label{real65}
(\o_\a^k)^*=-{I_{\dot{a}}}^{\b} \ve_{kl}\o_{\b}^l .
\ee
Then it is easy to rewrite the generator of special conformal transformations as follows
\be\ba\label{spectranslation}
K_a \rightarrow K_{\a\b}=K_a (\s^a)_{\a\b}=(x^2 p_a -2 x_a (xp) )(\s^a)_{\a\b}=(x\s)_{\a\g}(p\tilde{\s})^{\g\d} (x\s)_{\d\b}=\\ \\=
(x\s)_{\a\g} \l^{\g}_k \l^{\d k}(x\s)_{\d\b}=\o_{\a k} \o_{\b}^k.
\ea\ee
The Lorentz generators are
\be\label{mom}
\ba
M_{ab} \rightarrow {M_\a}^{\b}={M_{ab}(\s^{ab})_\a}^\b =-\frac{1}{4}{\left( (x\s)(p \tilde{\s})-(p\s)(x \tilde{\s})
\right)_{\a}}^\b=\\ \\=
\frac{i}{2}(\o_{\a k}\l^{\b k}-\frac{1}{4} \d^\a_\b \o_{\g k} \l^{\g k}),
\ea\ee
where it is used that
\be\ba\label{dil800}
(p\s)_{\a\b}=p_{\a\b}=-\fr \ve_{\a\b\g\d} p^{\g\d},\;\;(x\tilde{\s})^{\a\b}=x^{\a\b}=-\fr \ve^{\a\b\g\d} x_{\g\d}.
\ea\ee
The dilation generator is
\be D=-\fr(x^a * p_a + p_a * x^a) =\frac{i}{8} (\l^{\a}_k \o_\a^k + \o_\a^k \l^{\a}_k). \ee
Thus, any $so(6,2)$ generator is expressed either in terms of $(x^a,p_b)$ or in terms of $(\l^{\a k},\o^{k'}_\b)$.

The commutation relations of $so(d,2)$ are satisfied if $(\l^{\a k},\o_{\b}^{k'})$ are canonically conjugated variables with
commutation relations
\be
[\l^{\a k},\l^{\b k'}]=[\o^k_{\a},\o^{k'}_{\b}]=0,\;[\l^{\a k},\o_{\b}^{k'}]=-2i\d^{\a}_{\b}\ve^{kk'}
\ee
The variables $(\l^{\a k},\o_{\b}^{k'})$ form the {\it quaternion twistor} that transforms as
a Dirac spinor w.r.t. conformal group $so(6,2)$.

The generators
\be\label{ili3}
u_{k k'}= \fr (\l^{\a}_k \o_{\a k'}+\l^{\a}_{k'} \o_{\a k})
\ee
are not expressed in terms of $(x,p)$, they form $su(2)$ algebra that commutes to all $so(d,2)$ generators,
\be
[u_{k k'}, M_{AB}]=0.
\ee
Vice versa, any function of $(\l^{\a}_{k'}, \o_{\a k})$ that commutes to $u_{k k'}$ is a polynomial in
$M_{AB}$ (at the level of bilinear combinations of $\l,\o$, this is clear as all $su(2)$-invariants are just
the contractions of $(\l^{\a}_{k'}, \o_{\a k})$ w.r.t. $su(2)$ indices, but these combinations are exactly
$M_{AB}$ (\ref{translation},\ref{spectranslation},\ref{mom},\ref{dil800}), quite analogously one proves this for any monomial of $\l, \o$.

According to the analysis of Sec. \re{toki},
an arbitrary element of the algebra of observables of the $6$-dimensional
massless particle is represented by the equivalence class of polynomials built from the generators of conformal group $so(6,2)$,
modulo the ideal of trivial observables in the enveloping algebra of $so(6,2)$ generated by the combination \re{triv41}
\be\label{ili}
g_o\in {\cal G}_o,\;\;g_o=\sum \limits_{k=0}^{\infty} K^{A_1...A_k B_1...B_k} \{ M_{A_1 B_1}...M_{A_k B_k}\}_\circ \\ \\
\ee
\be
g_o \sim g_o + f_1^{AB}* N_{AB}+N_{AB}* f_2^{AB}
\ee
where $K^{A_1...A_k B_1...B_k}$ are real $so(d,2)$ tensors characterized by the traceless Young tableaux \re{young1}, and
$f_{1,2}^{AB}$ are arbitrary elements of the form \re{ili}.

Any $g_o$ may be expressed in terms of the twistor $(\l^{\a k},\o^{k'}_\b)$, where the monomials of $n$-th degree in $M_{AB}$ turn into the ones of
$2n$-th degree in $(\l^{\a k}, \o^{k'}_\b)$. We refer this representation of ${\cal G}_o$ as {\it twistor picture}.
In twistor picture, any $g_o$ is represented by an arbitrary power series built from {\it even} powers of $(\l^{\a k},\o^{k'}_\b)$,
which commutes to $su(2)$ generators \re{ili3}
\be\label{ili333}
g_o=g(\l^{\a k}, \o^{k'}_\b),\;\;g(-\l^{\a k},-\o^{k'}_\b)=g(\l^{\a k},\o^{k'}_\b),\;\;[u_{k k'}, g_o]=0
\ee
Note that, in the twistor picture,the elements of the ideal of trivial
observables \re{triv41} are identically zero
\be\label{ili1}
N_{AB}(\l,\o)\equiv 0,
\ee
that may be checked either by direct calculation or by noting that the last relation is nothing but the
$so(d,2)$-covariant version of the mass-shell constraint $p^2=0$, the latter being the $N_{++}$-component of \re{ili1}.
Indeed, the function $p^2$ is not a scalar w.r.t. $so(d,2)$ generators, but a component of a
finite-dimensional representation, the symmetric traceless tensor $N_{AB}$. Then, as the mass-shell constraint $p^2=0$
is satisfied by the very construction in the twistor picture, the rest components of $N_{AB}$ are zero either by virtue of the
$so(d,2)$-invariance.

Thus, {\it in twistor picture, the algebra of observables of the $6d$ massless particle
$chs(d-1,1)$ is represented by arbitrary
real (w.r.t. conjugation (\ref{real64},\ref{real65}))
even power series of the components of the twistor $(\l^{\a k},\o^{k'}_\b)$, which commutes to $su(2)$ generators
\re{ili3}.}

But this is nothing but the  $hs(8*)$ algebra of ref. \cite{Sezgin:2001ij}.
Indeed, the authors of \cite{Sezgin:2001ij} constructed the $hs(8*)$ in terms of $so(6,1)$ Dirac spinors
$y_{\hat\a}, \bar{y}_{\hat\b}, \hat{\a}, \hat{\b} =1...8$, satisfying commutation relations
\be
[y_{\hat{\a}},\bar{y}_{\hat\b}]=2C_{\hat{\a}\hat{\b}},\;\;C_{\hat{\a}\hat{\b}}=C_{\hat{\b}\hat{\a}}
\ee
These $8$ pairs of oscillators are identified with $8$ ones $\l^{\a k}, \o^{k'}_\b$.
The $su(2)$ generators of \cite{Sezgin:2001ij} are built as $3$ contractions
\be
\frac{i}{4}y_{\hat\a} y_{\hat\b}\, C^{\hat{\a} \hat{\b}},\;\; \frac{i}{4}{\bar y}_{\hat\a} {\bar y}_{\hat\b} \,C^{\hat{\a} \hat{\b}},
\;\;\frac{1}{4} {\bar y}_{\hat\b}  y_{\hat\a}  \,C^{\hat{\a} \hat{\b}},
\ee
which are identified with our $su(2)$ generators \re{ili3}. After these identifications, the construction of
the $hs(8*)$ coincides with our definition \re{ili333} up to a choice of a basis in the algebra.
Thus one has the isomorphism
\be
chs(5,1)=hs(8*).
\ee
\subsection{Complex twistors. $D=5, d=4,$ higher spin algebra as even subalgebra of $hsc^0(4)=hu_0(1,0|8)$.}
Here we use the Weyl spinor formalism adopted in \cite{WessBagger}. As this formalism is well-known we do not dwell on details of
notation. The mass shell-constraint $p^2=0$ is solved in terms of Weyl spinor $\l_\a$, $\a=1,2$
\be
p_a \rightarrow p_{\a\da}=p_a(\s^a)_{\a \da}=\l_{\a} \l_{\da},\;\;\l_{\da}=(\l_\a)^\dag
\ee
After introducing the second component of the twistor,
\be
\o^\a=i x^{\a\da} \l_{\da},\; \o^{\da}=-(\o^\a)^\dag
\ee
the rest $so(4,2)$ generators are rewritten as
\be\ba\label{confosc5}
K_a \rightarrow K_{\a \da}=- \o_\a \o_{\da}\\ \\
M_{ab} \rightarrow M_{\a\b}=(\s^{ab})_{\a\b} M_{ab}=\frac{i}{4}(\l_\a \o_\b+\o_\a \l_\b)
,\;D=-\frac{i}{2} \l^\a \o_\a = \frac{i}{2} \l^{\da} \o_{\da}
\ea\ee
The generators (\ref{confosc5})
form $so(d,2)$ either if they are realized in terms of $(x,p)$ or in terms of
$\l,\o,\bar{\l},\bar{\o}$ with commutation relations of the form
\be
[\l^\a,\o_\b]=-2i\d^\a_\b,\;\;[\l^{\da},\o_{\db}]=2i\d^{\da}_{\db},
\ee
The spinors $\l_\a,\o^\b$ form the {\it complex twistor} which transforms as a
Dirac spinor of the conformal group $so(4,2)$.

The analog of $su(2)$ algebra of the previous subsection is the $u(1)$ algebra generated by
\be
u=i( \l^\a \o_\a - \l^{\da} \o_{\da})-4,\;[u, M_{AB}]=0.
\ee
It easily seen that any function of $\l,\o,\bar{\l},\bar{\o}$ that commutes to $u(1)$ is expressed via
combinations of $M_{AB}$ and, in the contrast with the previous subsection case,  $u(1)$.
In the twistor picture, an arbitrary power series \re{ilikk} is represented by $u(1)$-invariant
even power series in $(\l^{\a},\l^{\da},\o^{\a},\o^{\da})$,
\be\label{alg4}
g_o=g(\l,\o,\bar{\l},\bar{\o}),\;\;g(-\l,-\o,-\bar{\l},-\bar{\o})=g(\l,\o,\bar{\l},\bar{\o}),\;\;[u, g_o]=0
\ee
Vice versa, if one considers arbitrary powers series of the form \re{alg4}
\be\label{ljg}
g(\l,\o,\bar{\l},\bar{\o}),\;\;g(-\l,-\o,-\bar{\l},-\bar{\o})=g(\l,\o,\bar{\l},\bar{\o}),\;\;[u, g_o]=0
\ee
then ${\cal G}_o$ arises as a Lie algebra
constructed via commutators from the factor algebra of the
associative algebra of power series \re{ljg} by the ideal generated by
$u(1)$:
\be\label{ljg1}
g \sim g + f  u
\ee
where $f$ is an  arbitrary power series of the form \re{ljg}.
The equations (\ref{ljg},\ref{ljg1}) determine the Lie algebra ${\cal G}_o$ which is
exactly the algebra $hsc^{0}(4)$
(where ``hsc'' ascends to ``higher spin conformal'') by Fradkin and Linetsky \cite{linetsky}. In fact,
the authors of \cite{linetsky} constructed cubic interaction of the conformal higher spin fields in terms of
$4d$ Yang-Mills curvatures of the algebra $hsc^{\infty}$ which is nothing but the Lie algebra corresponding
to the power series \re{ljg} without the factorization \re{ljg1}. Giving up the factorization results in the infinite
multiplicity of each conformal spin in the spectrum. It was conjectured   that any consistent
all-orders interaction requires infinite multiplicity of spins \cite{linetsky}.
Our treatment shows this is not the case, as
the model of this paper contains each spin with multiplicity one.

The same algebra ${\cal G}_o=hsc^0 (4)$ is used recently in \cite{Vasiliev:2001wa} in constructing,
up to the cubic level, the action for higher spin
massless fields corresponding to the totally symmetric Young tableaux, in $5D$ anti de Sitter space. In
\cite{Vasiliev:2001wa} it is  named $hu_0(1,0|8)$.
The $su(2,2)=so(4,2)$ oscillators $(a_\a, b^\b), \a,\b=1,2,3,4$ in \cite{Vasiliev:2001wa}
are linear combinations of the components of the twistor $(\l^{\a},\l^{\da},\o^{\a},\o^{\da})$.

Thus one has the isomorphism
\be
chs(3,1)=hsc^0 (4)=hu_0(1,0|8)
\ee
The algebra of observables ${\cal G}_o$ has a reduction to the power series, built from monomials of all {\it odd} degrees
in generators $M_{AB}$. We denote this reduction ${\cal G}^{odd}_o$.
In the twistor picture, ${\cal G}^{odd}_o$ is given in terms of
powers series obeying the constraints
\be
g(-\l,\o,\bar{\l},-\bar{\o})=-g(\l,\o,\bar{\l},\bar{\o}),\;\;g(\l,-\o,-\bar{\l},\bar{\o})=-g(\l,\o,\bar{\l},\bar{\o})
\ee
${\cal G}^{odd}_o$ is isomorphic to $ho_0(1,0|8)$ of ref. \cite{Vasiliev:2001wa} and to
$hs(2,2)$ of \cite{sez}
where it was used for the analysis of the linearized dynamics of the
higher spin massless fields of even spins in $5D$ anti-de Sitter space. Needless to say, the $so(4,1)$ Dirac spinor
$y^\a,\; a=1,2,3,4$ of \cite{sez} is related to the components of the twistor by linear combinations.

\subsection{Real twistors. $D=4, d=3,$ higher spin algebra as even subalgebra of the Vasiliev's
higher spin superalgebra.}

This case is most easily obtained from the $D=5, d=4$ one by the reduction. One just has to impose the constraints
\be
\l_\a=\l_{\da},\; \o^\a=\o^{\da}
\ee
then for $so(4,2)$ and $u(1)$ generators one gets
\be
u=0, p_2=0, K_2=0, M_{2a}=0, a=0,1,3,
\ee
and $so(4,2)$ gets reduced to $so(3,2)$. Thus, this case ${\cal G}_o$ is realized in terms of real oscillators
$(\l^\a,\o_\b)$ forming the {\it real twistor} which transforms w.r.t. conformal group $so(3,2)$ as a Dirac spinor.
It is easy to see generators of $so(3,2)$ are in one-to-one correspondence with all bilinear combinations of the components of the twistor.

W.r.t. $so(3,1)\subset so(3,2)$, the twistor transforms as $sl(2,{\bf C})$ Majorana spinor $y^{\a},{\bar y}^{\da}$,
where $y$'s are linear combinations of $\l^\a,\o_\b$, satisfying the commutation relations
\be\label{osc47}
[y^\a,y^\b]=2i\e^{\a\b},\;\;[y^{\da},y^{\db}]=2i\e^{{\da}{\db}},\;\;[y^\a,y^{\da}]=0.
\ee
Therefore, ${\cal G}_o$ is isomorphic to the algebra od even power series built from oscillators \re{osc47}.
But this algebra is nothing but the even part of the superalgebra used by Vasiliev as a starting point for constructing
cubic action for $4D$ higher spin massless fields and for constructing full nonlinear equations of motion
for this system in the "unfolded" form.

\section{Speculations}\label{spec}

Here we speculate on some issues concerning breaking the higher spin symmetries and then on
the extension of the geometry "point particle + conformal higher spin fields in $d$ dimensions" to the one
"tensionless $d-1$-brane + massless higher spin fields in $d+1$ dimensions".

\subsection{Higher spin compensator}\label{spec1}

One may wonder why the strategy we have followed in the paper has led us to {\it conformal higher spin theories} which
possesses such a huge infinite-dimensional gauge symmetry, instead of more conventional
low spin Maxwell and Einstein gravity theories that possess only low spin gauge symmetries, namely,
$U(1)$ transformations and $x$-diffeomorphisms. One of possible answers is that these theories may arise around
the vacuum with a nonzero dilaton $D(x)$ since after the dilaton is gauge-fixed to a nonzero constant by Weyl dilations,
the dilaton-curvature couplings $\int d^d x \sg D^{\hd-1} R, \int d^d x \sg D^{\hd-2}F_{mn} F^{mn}$
will provide the standard Einstein and Maxwell actions. However, we adhere a more radical point of view that
the standard low spin physics arises when, besides Hamiltonian $H(x,p)$, a new "Higgs-like" field is added,
which transforms w.r.t. canonical transformations
and compensates all the higher spin gauge symmetries in the sense
this "Higgs-like" field may be fully or partly eliminated by gauge transformations and after that
the higher spin gauge symmetries get expired and the only remaining gauge transformations are $x$-diffeomorphisms
and perhaps $U(1)$-Maxwell symmetries.
With this new field added, new structures in the action become possible,
and as a result after it is eliminated
the new structures provide something like "mass terms" for conformal higher spin fields.

As conformal higher spin fields in $d$ dimensions arise as boundary values of massless higher
spin fields on $d+1$-dimensional anti-de Sitter space \cite{Tseytlin:2002gz},
introduction of compensator to the theory of conformal higher spin fields
can have the analog in the theory of massless higher spin fields in $AdS_{d+1}$. Thereby
finding of the compensator may shed the light on the
nature of "Higgsing" the theory of massless higher spin fields \cite{witt}.
Below we briefly present a natural candidate on the role of
this field, which we call "higher spin compensator".
We suppose that the gauge group of the theory in the presence of a compensator remains the same. We
describe only what could happen in a classical limit, i.e. when the gauge group is a semidirect product of
canonical transformations and hyper Weyl ones. Moreover, for a moment we forget about hyper Weyl symmetry and
discuss only the compensation of canonical transformations.

The higher spin compensator is provided by the {\it prepotential of the symplectic form} on the phase space.
Consider the symplectic two-form
\be\label{compens0}
\o_2=2 dp_m \wedge dx^m.
\ee
It may be represented as a differential of a one-form $\o_1$,
\be
\o_2=d \o_1,
\ee
where general solution for $\o_1$ is
\be\label{compens1}
\o_1= 2 p_m d x^m + d \b(x,p),
\ee
where $\b(x,p)$ is an arbitrary power series in momenta, defined modulo transformations $\b \sim \b +Const$.
Under infinitesimal canonical transformations with generating function $\e(x,p)$
\be
\d x^m =\{x^m,\e\}, \;\d p_m =\{p_m,\e\}
\ee
the symplectic form is invariant. As a consequence, $\o_1$ transform to another $\o_1$ of the  form \re{compens1},
equivalently, $\b$ transforms like
\be\label{higgs1}
\d_\e \b =\{\e,\b\} +2 (1-p\pa_p) \e
\ee
One checks by direct calculation  this formula defines a realization of canonical transformations algebra,
so that
\be
[\d_{\e_1},\d_{\e_2}]=\d_{\{\e_1,\e_2\}}.
\ee
Note the appearance of inhomogeneous term in \re{higgs1} signalling of the Higgs-like nature of the field $\b$.

Given one-form \re{compens0}, \re{compens1}, one constructs {\it an outer automorphism $\cN$} of the Lie algebra of canonical transformations,
by the rule
\be
\cN \e =\fr \{\b,\e\}+ (p\pa_p -1)\e =-\fr \d_{\e} \b
\ee
One checks by direct calculation that this operation is an automorphism indeed, i.e.
\be
\cN \{\e_1,\e_2\} =\{ \cN \e_1, \e_2\}+\{ \e_1, \cN \e_2\}.
\ee
As a consequence, if $\cN$ is diagonalizable in a linear space of the phase space functions then
the algebra of all canonical transformations may be graded, namely, let
\be\label{grading1}
\cN \e_k =k \;\e_k,
\ee
then
\be
\{\e_k,\e_{k'}\}=\e_{k+k'}
\ee
This implies the phase space functions which satisfy the equation
\be
\cN \e_0 =0,
\ee
{\it form a subalgebra} in the whole algebra of canonical transformations which we call {\it low spin subalgebra corresponding to the
one-form $\o_1$}. An equivalent definition of this subalgebra is $\e_0$ are transformations that leave intact
$\b(x,p)=0$.

To explain the name "low spin subalgebra" let us consider the case
\be
\b=0 \Rightarrow \cN =p\pa_p - 1.
\ee
The decomposition into direct sum of graded subspaces
\re{grading1} corresponds to the expansion in momenta $p_m$, where the homogeneous
polynomials of $k$-th degree in momenta form the subspace with $\cN=k-1$. With our choice of
the class of phase space function as power series in momenta, the degree $k$ is bounded from below and runs
from $-1$ to infinity, that corresponds to the subspaces of homogeneous polynomials in momenta of order
$k+1$. Then the subalgebra of $\e_0$ is nothing but the subalgebra of $x$-diffeomorphisms $\e_0=\xi^m (x) p_m$.

Thus, if one extends set of dynamical variables by adding $\b(x,p)$ which transforms by the law \re{higgs1},
and then uses
these gauge transformations to gauge $\b$ to zero, then the remaining gauge transformations are just $x$-diffeomorphisms.
As a result, the only remaining gauge field is a metric $g_{mn}(x)$,
while all the higher spin fields would acquire mass terms. These mass terms arise from
additional gauge-invariant structures built already not from $H(x,p)$ alone but from $H(x,p)$ together with
$\b(x,p)$. These structures originate from the fact that, in the presence of $\b$, given a scalar
$G(x,p)$, which transforms as
\be
\d G(x,p)=\{\e,G\},
\ee
there exists an additional scalar
of the form \be
\Phi(x,p)=\cN G=\fr \{\b,G\}+(p\pa_p-1)G \ee
Therefore, from $H$ and $\b$ one can construct many scalars
like
\be\ba
\Phi_k(x,p)=\cN^k H,\\ \\
\Phi_{k_1...k_r}=\{ \{ \Phi_{k_1},\Phi_{k_2} \}...\Phi_{k_r}\}
\ea
\ee
Thus one has the possibility of constructing a variety of canonically-invariant  functionals like
\be
\cA[H,\b]=\int d^dx d^dp\; F(H,\Phi_{k_1...k_r} ),
\ee
where $F$ is an arbitrary function of many variables. If $\b=0$, the family of new scalars is reduced to
combinations like
\be
\Phi_{k_1...k_r}=\{ \{ (p\pa_p)^{k_1} H, (p \pa_p)^{k_2}H \}...(p \pa_p)^{k_r} H\}
\ee
and further structures, obtained by  Poisson brackets of these combinations and by applying operator $p\pa_p-1$.
Note that the canonically-invariant equation
\be
\Phi_1= (s-1) H\; \Leftrightarrow \; \cN H=(s-1) H
\ee
provides the {\it spin-$s$ truncation} of the whole theory as for $\b=0$
its solution consists of the homogeneous polynomials in momenta of order $s$.
It is tempting to reformulate ordinary Einstein gravity on this language as it will provide
description of gravity as of a spontaneously broken phase of a theory with large gauge group
and therefore it may uncover some hidden symmetries of Einstein theory.

The formulae we have delivered root in the {\it contact geometry in $2d-1$ dimensions} \cite{arnold},\cite{Schwarz:qn}
which provide them with a clear geometric interpretation, but we do not dwell on details here.  Here we just stress that the higher spin compensator $\b$ (or, what is the same,
prepotential $\o_1$ of the symplectic form $\o_2$) is not a field a {\it single} point particle is able to test. Indeed,
a single point particle can test only conformal higher spin fields as we have demonstrated in this paper.
In fact $\b$ presents an additional structure that can be tested only by an {\it ensemble} of point particles
with {\it different masses}. It appears when one studies interaction of point particles {\it with different masses}
it is obligatory to introduce the higher spin compensator $\b$ and thus one recovers the standard interaction
via low spin massless gauge fields - gravitons and photons being made up from fluctuation of $H$ and $\b$,
while the interaction via conformal gauge fields being made up just from fluctuations of $H$ arises when
$\b$ decouples \cite{ya1}.

\subsection{From conformal higher spin fields to
Fronsdal higher spin fields. Higher spin fields and
tensionless $D-2$-brane. }\label{spec2}

After the spontaneous breakdown of gauge symmetry
which possibility is displayed in the previous section,
the only remaining gauge symmetries are $x$-reparametrizations
and thereby the only surviving massless field is the spin-$2$ graviton
described by the fluctuation of the metric $g_{mn}$.
In this sense, this spontaneous breakdown
may be characterized as an "ultimate" one.

At the same time, after a more delicate breakdown, more massless
gauge fields may occur. In particular, one expects existence of
such a compensator that there will arise
Fronsdal massless
higher spin gauge theories of spins $s=0,1,2,3,...\infty$.

At the linearized level, the structure of this compensator is clear from
the picture displayed
in the beginning of this paper in Sec. \re{hsfields}. Namely, in addition to
the set of symmetric traceless tensors $\vp^{a(s)}(x)$
of ranks $s=0,1,2,3,...$, representing
the conformal higher spin fields and associated with the fluctuations
of the Hamiltonian, one has to introduce another set of symmetric
traceless tensor fields $\c^{a(s)}$ with gauge transformations
\re{gi1}. This set of fields represents fluctuation of some geometric structure
yet to be uncovered \cite{ya1}. We call this geometric structure
{\it hyper Weyl compensator}.

We expect the unified theory that is formulated in terms of
the Hamiltonian and hyper Weyl compensator may be found along
the lines we have constructed the conformal higher spin theory.
Namely, one has to find an adequate {\it geometry with a source} $O$,
in the sense of definitions given in Introduction, where background
fields are represented in terms of Hamiltonian and hyper Weyl compensator,
and the covariance algebra is {\it the same} as in this paper.
In the case hyper Weyl compensator decouples one recovers the
theory of Hamiltonian alone and therefore, conformal higher spin theory.
In the same case, the object $O$ should get reduced to the point particle.

Our preliminary results indicate that, in $D$ dimensions, $O$ is a
{\it tensionless $D-2$-brane}. This conjecture is
based on the interpretation of conformal higher spin fields as of boundary values of massless
higher spin fields in $AdS_{D}$ as it is described in Sec. \re{adscft}.
Namely, in
the $AdS/CFT$ correspondence, a free $AdS_D$ massless field
$\Phi_s$ of spin $s$, corresponding to totally symmetric Young
diagram is solved, in the framework of the Dirichlet-type problem,
in terms of conformal higher spin fields $\vf_s$ on the
boundary of $AdS_D$. Then, by the very results of this
paper, the full set of $\vf_s$ is nothing but a general background
for $d$-dimensional quantum particle with the wave function
$\psi(x)$, and there exists natural first-order interaction in $d$
dimensions \be S_{int}=e\int d^{d}y \;\psi^*(y) \;\hat{h}\;
\psi(y),\;\Box\psi(y)=0, \ee where \be
\hat{h}\psi(y)=\sum \limits_{s=0}^{\infty} (-i\hbar)^s
\vf^{a_1...a_{s}}(y) \pa_{a_1}...\pa_{a_s} \psi(y) +o(\hbar\pa_b
\vf^{a(s)},\hbar^2 \pa_b \pa_c \vf^{a(s)}, ...),\; a=0,1,...,d-1\ee and
$\vf^{a_1...a_{s}}(y)$ are just mentioned symmetric traceless tensor
fields. As far as $\vf^{a_1...a_{s}}(y)$ are boundary values of
$AdS_d$ higher spin massless fields $\Phi_s$, this implies the
full set of $\Phi_s$ interacts with a $d$-dimensional theory
which, being expanded  around some vacuum, exhibits itself as a free
$d$-dimensional quantized particle, or, equivalently, as a free
{\it complex} scalar field in $d$ dimensions.

It appears that certain limit of this world volume theory
is a particular case of a {\it tensionless
$D-2=d-1$-brane in $D$ dimensions}. To show this is the case, first consider general action \re{php},
\be
S=\int d^{d} y\; \psi^*(x)\;\hat{H}\; \psi(x),
\ee
and extract the leading term of the semiclassical expansion of this functional
(see Sec. \re{wgeom}),
provided by the Hamilton-Jacobi sigma model \re{hjsm},
\be
S_{int}[\psi=\r e^{\frac{i}{\hbar} \vp}]=e\int d^{d}y \sum \limits_{s=0}^{\infty}
\r^2 H^{a_1...a_{s}}(y) \;\vp_{a_1}...\vp_{a_s} + o(\hbar).
\ee
In the case $H^{a_1...a_{s}}(y)=0, s \neq 2$ this action does describe tensionless $D-2=d-1$-brane in
a $D$-dimensional gravity background, where $\vp(y)$ describes the transverse motion of the $D-1$-dimensional world
volume in $D$-dimensional space, while $\r$ is a world-volume density, being somewhat
analogous to the inverse Lagrange multiplier $\frac{1}{e(\t)}$
in the massless particle action $S=\int d\t \frac{\dot{x}^2}{2e}$.

Really, consider a tensionless $D-2$-brane in a gravity background
in $D$-dimensional space (see e.g. \cite{Bozhilov:1999aj}),
with the world volume action
\be
S[V^a(y),x^m(y^a)]=\int d^{d}y\; V^{a}(y) V^{b}(y) \;x^m_{,a} x^n_{,b} \;g_{mn}(x(y)).
\ee
where $m=0...d$, $V^a$ is a vector density,
that transforms w.r.t. $y$-diffeomorphisms as $(\det^{\frac{1}{2}}(\frac{\pa y'}{\pa y}))
\frac{\pa y^a}{\pa y'^b} V'^b(y'(y))=V^a(y)$.
Variation w.r.t. $V^a$ gives
\be\label{ttthhh}
V^{b}(y)\, x^m_{,a} x^n_{,b} \,g_{mn}(x(y)) =0,
\ee
and therefore the metric induced on the brane is degenerate.
Variation w.r.t.
$x^m$ gives
\be
-2\pa_a\, (V^a V^b g_{mn}\,X^n_{,b})=0,
\ee
Using $D-1$-reparametrizations one can (locally on the world volume) impose the gauge
\be\label{localgauge}
x^a(y)=y^a,
\ee
which can be substituted back in action without spoiling the dynamics.
Denoting
\be
x^{D-1}(y)=\vp(y)
\ee
one rewrites the equation \re{ttthhh} as
\be\label{ttthhhnnn}
V_a + \vp_{,a} (V^b \vp_{,b}) g_{D-1,D-1}(y^a,\vf) =0,
\ee
where the indices are raised and lowered via $g_{ab}$ and its inverse (we have put for simplicity $g_{a,d-1}=0$).
With the help of this equation $V^a$ is solved in terms of $\vp$ up to a multiplier,
\be
V_a=\r \;\vp_{,a},
\ee
where $\r$ is a variable which could not be fixed from the equation \re{ttthhhnnn}.
Another consequence of \re{ttthhhnnn} is the constraint
\be\label{hjconstr}
g_{D-1,D-1} (y,\vp)\; g^{ab}(y,\vp)\; \vp_a \vp_b+1=0
\ee
Among the remaining equations are those obtained by variation w.r.t. $x^m$.
In the gauge \re{localgauge} the only independent equation is that with $m=D-1$, which takes the form
\be\label{transp}
-\pa_a(\r^2 g^{ab} \vp_{,b}) =0
\ee
(here $g^{ab}$ is the inverse of $g_{ab}$).
Thus, in the gauge \re{localgauge}, the dynamics of the tensionless $D-2$-brane is completely determined by two equations
\re{transp}, \re{hjconstr}.
But if the metric does not depend on $(D-1)$-th coordinate
these equations are nothing but the equations of motion of the action of Hamilton-Jacobi sigma model
\be\label{hjsm99}
S[\r,\vp]=\int d^{d}y \;\r^2 \;H(y^a,\pa_a \vp),
\ee
with the Hamiltonian
\be\label{hambrane}
H(y^a, p_a)=-1 - g_{D-1,D-1}(y)\; g^{ab}(y) \;p_a p_b,
\ee
which is a general quadratic Hamiltonian without linear and  higher degrees in momenta (general nonzero dilaton may be obtained by
re-scaling $\r$.)
Note also the essential difference of the Hamiltonian \re{hambrane} with the Hamiltonian
$H=-p^2$ which presents a vacuum
of the $d$-dimensional conformal higher spin theory and governs dynamics of a massless $d$-dimensional particle.
With flat metric and the nonzero dilaton $H(y,0)\neq 0$, the Hamiltonian \re{hambrane} is no longer a vacuum of the conformal higher
spin theory.

To summarize, we have shown that {\it a particular form of the $D-1$-dimensional
Hamilton-Jacobi sigma model describes a tensionless $D-2$-brane in $D$ dimensions.}

On the other hand, by the results of this paper,
the quantum deformation of the Hamilton-Jacobi sigma model, being just the free action of the
quantized particle \re{php},\re{hjsm} exhibits interaction to conformal higher spin fields. At the same time,
the latter fields may be
viewed in the $AdS/CFT$ framework as boundary values of Fronsdal higher spin fields in $AdS_{D}$.
Thus, {\it massless higher spin fields in $AdS_d$ interact with a $d-1$ world volume theory, which in a certain
limit describes a tensionless $d-2$-brane}.

In this respect,
the decoupling limit, when the hyper Weyl compensator decouples from the theory on the worldvolume, may have
a transparent interpretation of the limit when the tensionless $d-2$-brane contracts to the tensionless $0$-brane, i.e.
to a massless particle.

We are led to conjecture that

{\it Fronsdal higher spin fields in $D$ dimensions arise as
background fields of an object $O$, which we call "tensionless
$D-2$-brane". After neglecting certain higher derivatives terms in
the worldvolume action, the dynamics of $O$ coincides with that of
tensionless $D-2$-brane in $D$-dimensions. In another case, $O$
exhibits itself as a complex scalar field (quantized particle) in
$D-1$ dimensions. There exists a contraction of $O$'s dynamics
when the world-volume of $O$ collapses to a world line of a
$D$-dimensional massless particle, after this contraction, one
half of the off-shell Fronsdal gauge potentials (hyper Weyl
compensator) decouples and one gets the system "point
particle+conformal higher spin fields" described in this paper.}

One can speculate there exists a limit when tensionless
$D-2$-brane gets contracted not to a massless particle but to a
{\it tensionless string} in $D$ dimensions. Then one gets natural
coupling "tensionless string-higher spin fields". This conjecture
is in accord with recent developments in $AdS/CFT$. Namely, in
\cite{sund} and \cite{witt} it was conjectured that a particular
limit of large-$N$ ${\cal N}=4, d=4, SYM$ theory, in which the
t'Hooft coupling $g_{YM}^2 N$ tends to zero is dual to tensionless
superstring in $AdS_5 \times S^5$. The arguments the authors of
these papers deliver are almost the same as ours, they are based,
in fact, on the emergence of the infinite number of conserved
currents in the free limit of SYM model. The essence of those
arguments is that, in accordance with standard $AdS/CFT$
treatment, each current in $d=4$ couples to a gauge field, which
is a boundary value of a massless field in $AdS_5$. At the same
time, the fact that each massless field in $AdS_D$ has a dual
conserved current built from complex massless scalar in $D-1$
\cite{konsh} has got a simple proof in our treatment.

Let us make a few remarks. We see the phase of the wave function $\vf$ is interpreted as the transverse position
of the brane in $AdS_{d+1}$. This does not contradict to the periodicity of the phase as the period is easily seen to be
$T=2\pi\hbar$ and thus it is zero in the semiclassical limit we consider, so there is no
condition like $\vf=\vf+T$ typical for a finite period. Next, the global $U(1)$-invariance
$\psi'=e^{\frac{i}{\hbar}\e}\psi$ of the full quantum action
\re{php} is translated to shift symmetry $\vf \sim \vf + \e$ in terms of the phase, thus the action
of the Hamilton-Jacobi sigma-model depends on $\vf$ only via its derivatives, the latter fact conforms with
independence of the background fields of $d-1$-brane on the transverse coordinate in $AdS_{D}$. Therefore, the
construction of the full action of the brane in general background fields is related with breaking of $U(1)$
invariance either in the full quantum action $\int d^dx\; \psi^* \hat{H} \psi$ or in its classical limit, Hamilton-
Jacobi sigma-model \re{hjsm}.

At the end of this subsection let us remark that the low-spin analogues of this section conjecture are well-known
M2, D3, M5 branes living in backgrounds of M-theory vacua
$AdS_4\times S^7$, $AdS_5\times S^5$, $AdS_7\times S^4$, the models which become, in the limit
brane is disposed on the boundary of $AdS$,  free conformal theories
of {\it real} scalars describing transverse fluctuations of $D-2$ branes into $AdS_D$
and into corresponding spheres \cite{Claus:1998mw},\cite{Bergshoeff:1987dh}.
These cases may be viewed as corresponding to
broken "supergravity phase" of higher spin models, in this phase, the branes are massive and the higher spin gauge symmetries
are broken so the only gauge fields are the ones of supergravity.
In our case, the branes are tensionless and according to our conjecture
there are massless fields of all spins in $AdS_D$. The real field describing
fluctuations of tensionless brane inside $AdS_D$ is the phase $\vf$ of the wave function $\psi=\r \,e^{\frac{i}{\hbar}\vf}$,
it is not free because variation of the action \re{hjsm99} w.r.t. $\r$ produces the constraint \re{hjconstr}.
The dynamics of $\vf$ would match the dynamics of free real scalar if one puts the constraint
\be
\r=Const
\ee
{\it inside} the action \re{hjsm99}, but then the action \re{hjsm99} ceases to be covariant w.r.t. all
canonical transformations, as the last equation breaks canonical covariance.
This is in accord with absence of higher spin symmetries in the "supergravity phase"
(and in accord with Hull's results \cite{Hull:1992vj} on the absence of $W$-geometries in $d>2$ ($\r$ is absent in Hull's
formulation)).
We postpone more detailed analysis of this mechanism to future study \cite{ya1}.

\section{Conclusion}

Let us make a remark on non-scientific matters, mentioning that
the main results of this paper has been anticipated by the author
two years ago as it is manifested by the Eqs. (33) - (35) of Ref. \cite{Segal:2000ke}
and the remark after Eq. (54) of that paper.
However the structure of perturbative expansion remained rather obscure
as off-shell one has fields of ever-increasing spin with infinite multiplicity of each
symmetric traceless tensor. The progress followed after the paper \cite{Segal:2001di}
where it was realized the importance of "undressing" mechanism for rewriting quadratic action as a sum of
standard conformal higher spin models.

To summarize, we have shown that the functional $\cA[\hat{H}]$
which is equal to the trace of the projector to the positive eigenvalues subspace of an arbitrary Hermitian differential operator $\hat{H}$,
is the action of conformal higher spin theory in the sense that, being expanded in perturbative series around
the Euclidean flat vacuum $\hat{H}=\hbar^2\Box$, $\cA$ reproduces in quadratic approximation direct sum
of free conformal higher spin gauge theories, the latter are higher-derivative models of order $\hd-2+2s$ in derivatives
formulated in terms of symmetric traceless tensors
of all integer spins $s=0,1,2,3,...$
The underlying gauge invariance has a very simple form of similarity transformation of
$\hat{H}$ and in fact fixes the form of the action unambiguously. The latter invariance arises as covariance algebra of
the quantized point particle (free complex scalar field) with $\hat{H}$ being its wave operator.

The covariance transformations preserving the vacuum are global symmetries of both the particle action and of
the conformal higher spin theory action. Considered modulo so called trivial global symmetries (those proportional
to Hamiltonian) the global symmetries define the algebra of observables which in the basic case $\hat{H}=\hbar^2\Box$
are shown to form an infinite-dimensional algebra, the latter contains conformal algebra and decomposes w.r.t. its adjoint action
as a direct sum of finite-dimensional irreducible representations characterized by all
traceless two-row rectangular Young tableaux. The latter algebra
probably arises as a global symmetry in the theory of higher spin massless fields
of all integer spins in $AdS_{d+1}$.

Let us make a remark. When this paper has been in preparation, an interesting paper \cite{Eastwood:2002su} has appeared,
which studies global symmetries of the operator $\hat{H}=\hbar^2 \Box$ and uses
almost the same construction of the infinite-dimensional symmetry algebra.

We have performed almost all calculations in the framework of well-known Weyl symbol formalism, which appears to be
well-suited for our purposes. In particular, we have shown how to calculate the action in the form of a
low-energy-like expansion in $\hbar$ counting number of space-time derivatives, in so doing, the zeroth, "classical"
term of this expansion may be interpreted as a volume of the
domain $H(x,p)>0$ in the phase space (where $H$ is a Weyl symbol of $\hat{H}$) while quantum corrections
are represented by integrals of distributions localized on the constraint surface $H(x,p)=0$.
It is $d-4+2s$-th quantum correction to the classical term that provides in the quadratic approximation
the conformal higher spin-$s$ action.

Note that although we have extensively  used the star-product formalism
we didn't introduce any kind of "unfolded formulation" \cite{Vasiliev:2001ur} which in a sense replaces
each field with all its derivatives at a fixed point of space-time. Most probably, if
there exists a consistent reformulation of our model in "unfolded" form the star product
of unfolded formulations will be the image of the star product in the particle phase space.
This way the noncommutativity inherent to unfolded formulations of higher spin dynamics
is induced by ordinary noncommutativity of quantum coordinates and momenta of a point particle.
In particular, we have shown usual higher spin "star-product algebras" in $d=3,4,6$ constructed out
of power series of even spinor oscillators are nothing but ordinary algebras of observables of
the quantized particle, this way the spinor oscillators arise just as components of the twistor
parameterizing phase space of the massless particle.
Also in the framework of the paper
there is no need in introducing unfolded formulations for realizing infinite-dimensional
higher spin symmetries like it is done in \cite{Shaynkman:2001ip}, as in our case
by construction one has the action of conformal higher spin algebra
\re{glob1} on the wave function \re{bsqpinf22} and on fluctuations of gauge fields \re{globfluct}.

The model of this paper may be easily generalized to different
cases of interest. Below we just mention the most immediate ones.

The description of conformal higher spin fields that carry inner indices
of some finite-dimensional group ${\bf G}$ is achieved by considering
geometry with the source being Wong particle \cite{Wong:fu} which is nothing but
an ordinary point particle supplemented by color degrees of freedom
$Q^i,\, i=1...n$. The supersymmetrization of conformal higher spin theories may be achieved
by considering geometry with the test source being Brink-Schwarz
superparticle \cite{Brink:nb}, note that as Brink-Schwarz superparticle is known to be well-defined
only in backgrounds satisfying appropriate constraints \cite{Witten:1985nt},
the extension to supersymmetric case has to
provide higher spin generalization of those constraints. After that extension higher spin superalgebras
should arise as superalgebras of observables of ($N$-extended)
massless Brink-Schwarz superparticle, with
or without central charges. The origin of oscillator realizations of higher spin superalgebras, used
by Vasiliev in the cases $d+1=4,5$ \cite{Vasiliev:2001wa},\cite{Vasiliev:2001ur} should become clear in this
context as these dimensions $d=3,4$ are right those the so called
doubly-supersymmetric formulation of superparticle dynamics becomes
possible, which uses essentially the twistor representation of dynamics
\cite{superpart}.  On the other hand, our definition does provide a wide class of
superalgebras in any dimension, even when there is no a simple twistor
representation  of a massless vector in $d$ dimensions and therefore
starting from any boson spinor oscillators (as it is conventional in
Vasiliev's approach) is not possible.

Considering particles with higher time derivatives may lead to other
interesting theories. In particular, there are backgrounds when higher derivative
particles describe massless spinning particles with an arbitrary helicity
\cite{Plyushchay:1988wx}, by analogy with the case of this paper one expects that
these backgrounds could be vacua of  some consistent gauge theories with
infinite number of gauge fields.

One of the most intriguing developments of this paper would be
constructing consistent theories of higher spin massless fields, which
at the linearized level have twice as many fields as their conformal higher spin cousins
provide. The additional fields form "hyper Weyl compensator" which does not influence dynamics
of a single point particle. We speculate that a proper source that is able to test/emit
massless higher spin fields in $d$ dimensions is an extended object that in some cases exhibits itself as a tensionless
$d-2$-brane. In the limit the brane shrinks to a point the hyper Weyl compensator
decouples and one recovers the geometry of the present paper
"point particle+conformal higher spin fields". On the other hand, this reasoning
conforms with recent conjecture on the "anti-holographic limit" of the $AdS/CFT$,
where it was argued higher spin fields arise, in a tensionless limit, in the spectrum
of IIB superstring on $AdS_5 \times S^5$ with RR-flux. Indeed, suppose our
proposal about tensionless $d-2$-brane is correct, it implies natural coupling
"tensionless $d-2$-brane/massless higher spin fields". Next, suppose in some situation
tensionless brane collapses to a tensionless string, then there may survive coupling
"tensionless string/massless higher spin fields". Due to basic stringy principles, massless higher spin fields
should appear then in the spectrum of the firstly quantized tensionless string that is in accord with the
"anti-holographic" conjecture \cite{sund},\cite{witt},\cite{Tseytlin:2002gz}.

Another way to draw a link to massless higher spin fields (which in fact motivated the conjecture
of the previous paragraph) is to consider conformal higher spin fields on $d$-dimensional conformally flat space
as boundary values of massless higher spin fields in $AdS_{d+1}$. Although low spin ($s=0,1,2$) analogs of this
phenomenon are in the core of $AdS/CFT$, the higher spin ones are less elaborated.
Namely, the results on "anti-holographic" conjecture usually go along the way "free massless conformal matter in
$d$-dimensions $\rightarrow$ infinite number of conserved currents in $d$ $\rightarrow$ emitting of
massless higher spin fields into the bulk of $AdS_{d+1}$"\cite{sund},\cite{witt}.
Perhaps results of our paper manifest it is
worth extending this line by one more item to "free massless conformal matter in
$d$-dimensions $\rightarrow$ infinite number of conserved currents in $d$ $\rightarrow$
emitting of conformal higher spin fields into $d$-dimensional conformally flat space, with
gauge-invariant action given by the "quantized volume" in the phase space of a particle
$\rightarrow$ the latter fields arise as boundary values for Dirichlet-like problem for
massless higher spin fields on $AdS_{d+1}$". The infinite number of conserved currents is provided, in our approach,
by a free complex scalar field in $d$ dimensions, which in a certain (semiclassical) limit is shown to
be interpreted as a tensionless $d-1$-brane in $d+1$ dimensions, where the phase of the complex scalar plays
a role of the transverse position of the brane inside $d+1$-bulk, while the modulus of the field
plays a role of a world volume density not related to the embedding of the brane into the bulk.

Note that our results are in accord with recent reasoning by Tseytlin \cite{Tseytlin:2002gz} who argued that the action
of free  higher spin massless fields on $AdS_{5}$ evaluated on solutions of Dirichlet problem
(with infra-red regulator $\e$ restricting the limits of integration in $AdS_5$) should
be equal to the quadratic approximation of the one-loop effective action
$A=-\fr Tr Ln \,\hat{H}$ of the operator $\hat{H}=-\Box+\vf_s \pa^s P_s$ in four dimensions,
where $\vf_s$ is a set of symmetric {\it traceless} tensors
and $P_s$ are transverse traceless projectors related to the wave operators $L_s$ of
spin-$s$ conformal higher spin theories by the rule $L_s=\Box^s P_s$. The quadratic term of the action $A$
is proportional to $$\sum \limits_{s=0}^{\infty} \int d^4x \;\vf_s L_s \ln (\e^2\Box)\vf_s,$$
where $\e$ is an ultra-violet regulator, and the logarithmically-divergent part of this expression is nothing
but the sum of quadratic conformal higher spin theories. At the same time,
the construction of the present paper provides all-order
generalization of the sum of quadratic conformal higher spin actions.
In our treatment, there are no tracelessness constraints on the fluctuations of $\hat{H}$, instead, one has
hyper Weyl invariance that gauges away the traces at the linearized level and leads to proper decoupling phenomena at
nonlinear level. As we argued in Sec. \re{ctsdwt}, in even dimension the full nonlinear
action of conformal higher spin theory
$\cA[\hat{H}]$ is nothing but the logarithmically divergent term of the one-loop action
$-\fr Tr Ln \,\hat{H}$, thus our statement matches with Tseytlin's result. On the other hand,
recalling the AdS/CFT motivation of this treatment it is tempting to make the conjecture that
{\it in even dimensions $d>2$, the action of conformal higher spin theory is the logarithmically divergent term
of the action of higher spin massless fields in $AdS_{d+1}$, evaluated on the solutions of Dirichlet-like problem}
As the action of higher spin massless fields is known, for $d>2$,  only
at cubic order in $AdS_4, AdS_5$ \cite{Fradkin:ah},\cite{Alkalaev:2002rq},\cite{Vasiliev:2001wa},
the conjecture, if correct, provides important new data on all-orders interaction of higher spin massless fields.

Note that in even dimension our action equals \re{sdw765},
the time-independent term that appear in the asymptotic
expansion of the trace of the evolution operator of $\hat{H}$, the latter term is
a combination of Schwinger-Dewitt coefficients \re{weylan2} typically encountered in calculation of Weyl anomaly, whence
close connection of our treatment to Weyl anomaly considerations, either traditional
\cite{Duff:1993wm} or "holographic" \cite{Henningson:1998gx}.
Indeed, the logarithmically divergent term of the $AdS$ action is known to produce
"holographic" Weyl anomaly along the lines of refs \cite{Henningson:1998gx}. As in our model one has the
higher spin extension of Weyl transformations to hyper Weyl ones it is interesting to
extend the results of \cite{Henningson:1998gx} to the higher spin case by using Weyl anomaly like considerations.

An important issue is the spontaneous breaking of higher spin gauge symmetries. We have
presented a natural candidate on the role of corresponding Higgs-like field, "higher spin compensator" that
appear to be a prepotential of the symplectic $2$-form in  $d+d$-dimensional phase space.
Considering coupling of this structure to conformal higher spin fields and to their
"source", complex scalar field, one gets many new gauge-invariant
structures as compared to the case higher spin compensator is absent.
For complex scalar this would lead to the appearance of the self-interaction vertices
and loss of conservation of infinite number of Noether currents (except $U(1)$ current and energy-momentum tensor)
\cite{witt}. For $AdS_{d+1}$ image of this picture one would get spontaneous breaking of higher spin gauge symmetries
by some mass-like terms, rendering theory of higher spin massless fields into a model with
low spin massless fields (scalar, photon, graviton) and an extensive collection of massive fields, the latter
situation resembles string theory.

\section{Acknowledgements}
The author is indebted to A. Sharapov for fruitful collaboration on the initial stages of this project.
The author is greatful to I. Bars, A. Barvinsky, R. Metsaev, D. Sorokin, B. Voronov, A. Tseytlin and I. Tyutin
for useful discussions. The work is supported in part by grants RFBR No 02-02-16944,
RFBR No 01-02-30024 and INTAS No 00-00262.

\appendix

\section{On perturbative expansion in gauge theories} \label{I3}\label{ApA}
Let $\cA[H]$ be a functional of a set of fields $H$ invariant w.r.t. gauge transformations
$\d H= \d_g H(H)$, $g$ being a set of gauge parameters.
Suppose $H={\bar H}$ is a solution of the equations of motion,
\be
\frac{\d \cA[H]}{\d H} [{\bar H}]=0.
\ee
The global symmetries are those gauge transformations preserving the vacuum $H=\bar{H}$.
Consider the perturbative expansion of the action $\cA[H]$ around this solution.
Representing general  background field $H$ as
\be
H={\bar H}+h
\ee
one expands the action in powers series in $h$ schematically like
\be \ba
\cA[{\bar H}+h]=\cA[{\bar H}] +\frac{1}{2!} \frac{\pa^2 \cA}{\pa H\pa H} [{\bar H}]\; h h +
\frac{1}{3!} \frac{\pa^3 \cA}{\pa H\pa H \pa H} [{\bar H}] \;h h h +...  =\\ \\=
\cA_0 +\cA_2[h] +\cA_3[h] +...
\ea \ee
The $h$-linear term cancels as ${\bar H}$ is a solution of the equations of motion.

The following well-known propositions are used in the paper:

\noindent{\bf 1.} The quadratic action  $\cA_2[h]$ is gauge invariant w.r.t.
linearized, $h$-independent gauge transformations given by the gauge variation of the ``vacuum''
${\bar H}$,
\be \label{lin888}
\d_g h = \d_g {\bar H}.
\ee
{\bf  2.} The equations of motions of the
quadratic action are invariant w.r.t.
all global symmetries. The global symmetries act on $h$ linearly,
as the differential of the full transformation calculated at the vacuum point
\be
\d_{\tilde{g}} h=(\frac{\pa \d_{\tilde{g}} H}{\pa H}[{\bar H}]) h
\ee
{\bf 3.}  The cubic action is
invariant w.r.t. linearized gauge transformations \re{lin888}
if $h$ satisfies the linearized equations of motions.
\be
\frac{\pa^2 \cA}{\pa H\pa H} [{\bar H}] h=0 \Rightarrow \delta_g \cA_3[h]=0
\ee

\section{Equations of motion.}\label{ApB}

Here we undertake the analysis of the equations of motion of our theory in the formal operator approach.
The variation of the action \re{action101} is found easily as the order of
product factors $H$ and $\delta H$ is inessential:
\begin{equation}  \label{inva3}
\delta {\cal A}_F[H] = Tr( F^{\prime}(H*) * \delta H )= Tr (F^{\prime}(H*)
\delta H )
\end{equation}
where it is used that for the Weyl $*$-product $Tr (A*B) = Tr (A B)$.
Equations of motion take the form
\begin{equation}  \label{eom2}
\frac{\delta {\cal A}_F [{{\bar H}}]}{\delta H} = F^{\prime}(H*) =0.
\end{equation}
These equations are written in terms of a single function $F^{\prime}(H*)$
and encode the full nonlinear dynamics of the infinite collection of
symmetric traceless tensor fields.

The equations, corresponding to variation of \re{action101} w.r.t.
particular space-time field $h^{m_1...m_i} (x)$ are obtained from {(\ref
{eom2})} as follows
\begin{equation}  \label{eom3}
\begin{array}{c}
\frac{\delta {\cal A}_F[H]}{\delta H^{m_1...m_i} (x)} = \int d^d p \; \frac{%
\delta {\cal A}_F[H]}{\delta H} \;\frac{\delta H}{\delta H^{m_1...m_i} (x)} =
\\
\\
\int d^d p \;\frac{\delta {\cal A}[H]}{\delta H} \;p_{m_1}... p_{m_i} (x) =
\int d^dp\; F^{\prime}(H*) (x,p) \;p_{m_1...m_i} =0.
\end{array}
\end{equation}
These are the equations which could have been obtained if one first
evaluates the space-time action by integrating over momenta in \ref{action101}
and then varies it w.r.t.
$H^{m_1...m_i} (x)$. For general $F$ these equations are inevitably highly
nonlinear and perhaps nonlocal, as general function $F^{\prime}(H*)$ has a form of
an infinite semiclassical expansion {(\ref{fquasi})} with $\hbar$ counting
the total number of $x$-derivatives. Definitely, finding solutions to the
equations written in this form is a hard task. Nevertheless, it appears that in the formal
operator approach the equations of motion of the action \re{inva3} have a transparent interpretation.

The equation for symbols {(\ref{eom2})} may be rewritten as operator
equation
\begin{equation}  \label{eom4}
F^{\prime}(\hat{H}) =0,
\end{equation}
where $F$ is a function of operator $\hat{H}$.
In terms of the full set of eigenvalues $H_i$ and eigenspaces $|H_i>$ for $\hat{H}$,
\be
\hat{H}|H_i>=H_i|H_i>,
\ee
the equation {(\ref{eom4})} becomes equivalent to the condition
\begin{equation}  \label{eom5}
F^{\prime}(H_i)=0 \;\;\; \forall i
\end{equation}
Thus {\it in the formal operator approach, an operator $\hat{H}$ is a solution of the classical equations of motion for the
action $\cA=Tr F(\hat{H})$ if the spectrum of $\hat{H}$ is a subset of the set of zeroes of the function $F'(\s)$}.
In other words, the function $F^{\prime}(\sigma)$ is an {\it annulator} of the Hamiltonian $\hat{H}$.

Therefore, if $F'(\s)$ has only discrete set of zeros, then the spectrum $\{H_i\}$ should be discrete either.
The operators $\hat{H}$ which have a nontrivial continuous spectrum can be a solution of equations of motion only
if $F'(\s)$ is zero in a continuous subset of ${\bf R}$.

Consider a simple example. Let $H$ be the Hamiltonian of $d$-dimensional
Euclidean harmonic oscillator:
\begin{equation}  \label{harmony}
H=\frac{1}{2} (\hat{p}^2 + \alpha^2 \hat{x}^2 ) - \hbar \a d/2
\end{equation}
Note that $H\in \cC$ in the sense of section \re{classesh}.
It is well known that it has the following discrete spectrum in $L^2({\bf %
R^d)}$:
\begin{equation}  \label{harmony1}
H=\hbar \alpha \,n \;\;, n =0,1,2,...
\end{equation}
It is easy to construct an annulator function. For example one may take ($
\hbar^{\prime}=\alpha  \hbar$)
\begin{equation}  \label{harmony3}
F^{\prime}(z)=\frac{e^{-\frac{2Cz}\hbar^{\prime}}}{\Gamma (-\frac{2Cz}%
\hbar^{\prime})}=\prod_{k=1}^\infty \left( 1-\frac{2z}{\hbar^{\prime}k}%
\right) e^{\frac{2z}{\hbar^{\prime}k}},
\end{equation}
where the infinite-product representation for $1/\Gamma(z)$-function is
used:
\begin{equation}
\frac 1{\Gamma (z)}=e^{-Cz}z\prod_{k=1}^\infty \left( 1+\frac zk\right)
e^{-\frac zk}
\end{equation}
for $C$ being the Euler constant.

Therefore, one concludes that the Hamiltonian of harmonic oscillator
corresponding to the background of flat Euclidean metric and certain $x$%
-dependent dilaton field is a solution of full nonlinear equations of motion
determined by the function {(\ref{harmony3})}. Note that this simple example
may be considered as a regularization of flat space Hamiltonian ($\alpha = 0$%
) which is not easy to work out along the lines of formal operator approach
since it has the continuous spectrum.

Now let us take $F(\s)=\theta(\s)$. The equations of motion read
\be
\d(\hat{H})=0.
\ee
$\d(\s)$-function may be considered as a limit of an ordinary function which tends to zero
everywhere except $\s=0$, and to infinity, when $\s=0$. Thus,
the only sensible interpretation of the last equation is that $\hat{H}$ should not have zero eigenvalues in
$L^2$. Note that the main case of this paper, the flat vacuum
\be
H=-p^2
\ee
conforms with this interpretation, as this operator is known to have no normalizable zero modes.
Thus we conjecture that {\it a symbol $H(x,p)$ of an operator $\hat{H}$ is a solution of the equations of motion governed by the
"quantized volume" action, if $\hat{H}$ has no normalizable zero modes in $L^2({\bf R}^d)$}.
We stress however that as far as it concerns the actual dynamics of conformal higher spin fields,
this interpretation, obtained in the formal operator approach, is fragile
and should be used only as a motivation for more explicit study.

\section{Gauge transformations for traceless tensors.}\label{ApC}

\noindent Here we rewrite gauge transformations \re{gtr111}
\begin{equation}\label{lintotalap}
\delta h(x,p)= -2 a(x,p)(p^2+\m^2) -2 p_m \eta^{mn}
\partial_n \epsilon (x,p)
\end{equation}
in terms of $a$-invariant traceless tensors.

We need some simple tools to handle traces of tensor coefficients
of arbitrary functions. Let us note that, given any function
$$F(x,p)=\sum\limits_{k=0}^\infty F^{m(k)} (x)
p_{m_1}...p_{m_k},$$ one can unambiguously represent it in the
form \be \label{decom} F(x,p)= \sum\limits_{l=0}^\infty
\sum\limits_{k=0}^\infty F_{(l)}^{m(k)} (p^2)^l p_{m_1}...p_{m_k},
\ee where $F_{(l)}^{m(k)}$ are traceless,
${{F_{(l)}}{}_n}^{nm(k-2)}=0$. This is easily done by decomposing
each $F^{m(k)}$ to its traceless part and the traces
$F^{m(k)}=F_{(0)}^{m(k)}+\eta^{m(2)}F_{(1)}^{m(k-2)}
+\eta^{m(2)}\, \eta^{m(2)} F_{(2)}^{m(k-4)} +...$, then summing up
the power series by momenta and noting that the trace parts give
the powers of $p^2$. The decomposition \re{decom} is then
rewritten as \be F(x,p)= \sum\limits_{k=0}^\infty F^{m(k)} (p^2)
p_{m_1}...p_{m_k}, \ee where $F^{m(k)} (p^2)=
\sum\limits_{l=0}^\infty F_{(l)}^{m(k)} (p^2)^l. $ Decomposing
the power series $F^{m(k)} (\sigma)$ at the point $\sigma=-\m^2$
one gets \be \label{decom2}\begin{array}{c}
 F(x,p)=\sum\limits_{k=0}^\infty  F_{[-\m^2]}^{m(k)}(p^2+\m^2)
p_{m_1}...p_{m_k}= \\ \\\sum\limits_{l=0}^\infty
\sum\limits_{k=0}^\infty (p^2+\m^2)^l F_{[-\m^2](l)}^{m(k)}
p_{m_1}...p_{m_k}= \sum\limits_{l=0}^\infty(p^2+\m^2)^l
F_{[-\m^2](l)},
\end{array}
\ee where the power series $F_{[-\m^2](l)}$ contain only traceless
coefficients. Given $\m^2$, we will say that the $F_{[-\m^2](0)}$
term is the traceless part of the function $F(x,p)$ and the
first, second and further traces of $F$ are represented by
$(p^2+\m^2)^l F_{[-\m^2](l)} $ with $l=1,2,...$ forming altogether
the traceful part of $F$. The function is traceless if it is
equal to its traceless part. In this sense each coefficient
$F_{[-m^2](l)}$ is a traceless function.

Now represent all the entries of the gauge transformation laws
\re{lintotalap} in the form \re{decom2} to get \be \label{Glaws2}
\de \sum\limits_{l=0}^{\infty}  (p^2+\m^2)^l\,h_{[-\m^2]\,(l)} =
\sum\limits_{l=0}^{\infty} \left\{-2 (p^2+\m^2)^{l+1} a_{[-\m^2]\,(l)}
 -2 (p^2+\m^2)^l p^m \pa_m \e_{[-\m^2]\,(l)} \right\}
\ee wherefrom it is seen that all the traces of $h$ may be gauged
away by $a$-transformations. In fact, the very destination of $a$
is to gauge away the traces of $h$. It is worth noting that the
traceful part of $\e$ is already contained in $a$ as the gauge
transformations \re{Glaws2} do not change if one redefines $\e,a$
according to \be \de \e= (p^2+\m^2)\nu\,,\,\de a= - p^m
\pa_m \nu. \ee Therefore without loosing  generality one may set
$\e$ traceless \be \label{etr1}\e= \e_{[-\m^2]0} \equiv -2\ve
=-2\sum\limits_{k=0}^{\infty} \ve^{m(k)} p_{m_1}...p_{m_k}\,;\,
{\ve_n}^{nm(k-2)}=0. \ee
For any action ${\cal A}[h]$ invariant w.r.t. gauge
transformations \re{lintotalap}, $h$ should enter ${\cal A}[h]$ in
$a$- and $c$-invariant combinations only as far as $a$
transformations are purely algebraic. It is easy to see that the
only $a$-invariant is the traceless function
 \be \label{5} \varphi=h_{[-\m^2](0)}. \ee
It is easy to derive the $\ve$ transformation laws for the
coefficients of $\varphi$ which read \be \label{gtra}  \de
\varphi^{\,m(s)} =( {\mbox{Traceless part of}}\; \pa^{m}
\ve^{m(s-1)}) -\m^2\, \frac{s+1}{2s+d}\, \pa_n {\ve^{nm(s)}}.\ee
For $\m^2=0$, these are the gauge transformations of conformal
higher spin theories, which are seen to decay into independent
subsystems described in terms of rank-$s$ traceless tensor and
rank-$(s-1)$ traceless parameter. For $\m^2 \neq 0$, as it is
proved in the main text, for each spin $s$ there arises
a deformation which appears to be related to its $\m=0$
counterpart by undressing and reconstruction maps studied in Sec. \re{dresundres},
\re{reconstruct}.

\section{Generating functions
for integration over $d$-ball and $d-1$-sphere}\label{ApD}

Given the Euclidean metric $g^{mn}=\d^{mn}$
in a $d$-dimensional space parameterized by $p_m$, one defines the "constraint
surface" as a $d-1$-sphere \be p^2 \equiv g^{mn} p_m p_n =
\m^2,\;\; \m \in {\bf R} \ee The integrals to be calculated are \be
\label{gen11} I^{(k)}_{m_1...m_s}(x)=\int d^dp\;
\theta^{(k)}(\m^2-p^2)p_{m_1}...p_{m_s} , \ee where $\theta^{(k)}$ is the
$k$-th derivative of the "step" $\theta$-function $\theta(\sigma)$ :
\be \ba
\theta(\sigma)=0\;\;,\;\; \sigma < 0 \\ \\
\theta(\sigma)=\fr\;\;,\;\; \sigma = 0\\ \\
\theta(\sigma)=1\;\;,\;\; \sigma > 0.
\ea
\ee
In fact only even-$s$ integrals are to be calculated, because
odd-$s$ integrals equal zero.

The first derivative of $\theta$-function is $\delta$-function, etc.,
\be  \label{gen12}
\pa^k_\s \theta(\s) =\delta^{(k-1)} (\s)
\ee
Therefore, all the integrals \re{gen11} are well-defined: for $k=0$ the actual
integration goes over the $d$-ball inside the constraint surface,
while for $k>0$ the integrals are localized on the constraint surface
being the $d-1$-sphere. The simplest case is
\be \label{gen14}
I^{(0)}=\int d^dp\; \theta (\mu^2 - p^2) =v_d \mu^d,
\ee
where
\be
v_d=\frac{\pi^{\hd}}{\G(\hd+1)}
\ee
is the volume of a unit ball in $d$ dimensions.

Due to \re{gen12}, the integrals for different $k$ are related by
differentiation by $\m^2$:
\be
I^{(k)}_{m_1...m_s}=(\frac{\pa}{\pa \m^2})^k I^{(0)}_{m_1...m_s}.
\ee
The collection of all the integrals \re{gen11} is described by the
generating functions
\be\label{genf1}
G^{(k)}_d (l) =\sum \limits_{s=0}^{\infty} \frac{i^s}{s!}
\;I^{(0)}_{m_1...m_s}(x) l^{m_1} ... l^{m_s}=\int d^dp\; \theta^{(k)} (\m^2-p^2)
e^{ip_m l^m}, \ee which depend on the auxiliary vector variable $l^m$.
Likewise $I^{(k)}_{m_1...m_s}$ the generating functions $G^{(k)}_d (l)$
are related by differentiation by $\m^2$,
\be
G^{(k)}_d (l)=(\frac{\pa}{\pa \m^2})^k G^{(0)}_d (l),
\ee
so all generating functions may be obtained from any one either by
differentiation or by integration. It is convenient to calculate
$G^{(1)}_d (l)$,
\be
G^{(1)}_d (l) =\int d^dp\; \d (\m^2-p^2)
e^{ip_m l^m}, \ee which satisfies differential equation
\be \label{gen13}
g^{mn} \frac{\pa}{\pa l^n}\frac{\pa}{\pa l^m} G^{(1)}_d(l)
=-\m^2 G^{(1)}_d (l) ,
\ee
as $\delta(\m^2-p^2)(\m^2- p^2)=0$.
The integrals exhibit $so(d)$ invariance. Hence the result of
integration should depend only on $l^2=l^nl^n$, the generating function
depends on $l^m$ only via single variable $l=\sqrt{l^2}$.
Then the equation \re{gen13} reads
\be l^{1-d}\; \pa_l \;l^{d-1} \;\pa_l\; G^{(1)}_d (l) = -\m^2 G^{(1)}_d (l). \ee
The substitution
\be
G^{(1)}_d (l)=l^{1-\hd} f(\m l)
\ee
turns this equation into the Bessel's equation,
\be \label{gen15}
(\s^2 \pa_\s^2 +\s \pa_\s +\s^2  -(\hd-1)^2) f(\s)=0.
\ee
The solution should be regular at $\s=0$ as
\be  \label{gen16}
G^{(1)}_d (0)=(\frac{\pa}{\pa \m^2}) \int d^dp\; \theta (\m^2-p^2) =
 v_d \frac{d}{2} \m^{d-2}.
\ee
Among the solutions with integer or half-integer
indices, only the Bessel's function of the first kind are regular at the
origin, therefore the correct solution for \re{gen15} is
\be
f(\s)=\k J_{\hd-1}(\s),
\ee
where $\k$ is a $\m$-dependent constant,
$J_{\hd-1}(\s)$ is the Bessel's function of the first kind, of index
$\hd-1$, see App. \re{ApE}.
Matching the behavior of $J_{\hd-1}(\s)$ at the origin with the
"boundary condition" \re{gen16} one gets the final result
\be
G^{(1)}_d (l)= 2^{\hd-1} \pi^{\hd} \m^{\hd-1} l^{1-\hd}
J_{\hd-1}(\m l)
\ee
This formula contains only even powers of $l$ as it should.
$G^{(k)}_d$'s can be easily obtained by making use of the identities \re{uuret}.
\be \label{genf4}
G^{(k)}_d (l) =2^{\hd-k}\pi^{\hd} \m^{\hd-k} l^{-\hd+k} J_{\hd-k} (\m l)
\ee
Rewriting the generating functions as power series in $l^m$ provides all the
integrals \re{gen11} as tensor powers of the metric $\d^{mn}$ multiplied by
certain $\m$-monomials.

\section{Bessel's functions}\label{ApE}

Here we collect the properties of Bessel's functions of the first kind
\cite{vilenkin} used in the main text.  Bessel's function of the complex variable $z \in
{\bf C}$, of the first kind, of index $p\in {\bf C}$ is \be \label{bes11}
J_\n(z)=(\frac{z}{2})^\n \sum \limits_{k=0}^{\infty}
\frac{(-)^{k}}{k!  \Gamma(\n+1+\k)} (\frac{z}{2})^{2k}.
\ee
The series \re{bes11} converges for any finite $z$ and $\n$.
It satisfies the differential equation
\be (z^2 \pa_z^2 +z \pa_z +z^2  -\n^2 ) J_\n(z)=0. \ee
The useful identities used in the main text are
\be\label{uuret}
zU''_\n(z)+(1+2\n)U'_\n(z)+zU_\n(z)=0,\;U_\n(z)\equiv z^{-\n} J_\n(z).
\ee
\be\label{bess889}
\pa_z(z^\n J_\n(z))=z^\n J_{\n-1}(z).
\ee
\be \label{bess890}
\pa_z(z^{-\n} J_\n(z))=-z^{-\n} J_{\n+1}(z).
\ee
\be\ba\label{bessgood}
\int \limits_{0}^{\frac{\pi}{2}} d\q J_{p-q} (r_1 \cos \q) J_{q-1} (r_2 \sin\q)
(\cos \q)^{p-q+1} (\sin\q)^q = r_1^{p-q} r_2^{q-1} (r_1^2+r_2^2)^{-\frac{p}{2}}
J_p(\sqrt{r_1^2+r_2^2})\\ \\
p,q=0,\frac{1}{2},1,\frac{3}{2},2,...
\ea\ee
\be \ba \label{Bess4}
r^{-\frac{n-2}{2}} J_{\frac{n-2}{2}}(r)=
2^{\frac{n-2}{2}} \G (\frac{n}{2}-1) \sum \limits_{k=0}^{\infty}
(-)^k (k+\frac{n}{2} -1) r_1^{\frac{n-2}{2}} r_2^{\frac{n-2}{2}}J_{\frac{n-2}{2}}(r_1)  J_{\frac{n-2}{2}}(r_2)
\cC_k^{\frac{n-2}{2}} (\cos \vp)\\ \\
r^2 = r_1^2 +r_2^2 +2 r_1 r_2 \cos \vp
\ea
\ee
where $n-2=0,1,2,...$ and $\cC_m^p(t)$ are Gegenbauer polynomials
\be \ba
\cC_m^p(t) =\frac{2^m \G(p+m)}{m! \G(p)} \left[t^m -\frac{m(m-1}{2^2 (p+m-1)} t^{m-2} +
\frac{m(m-1)(m-2)(m-3)}{2^4 \times 1\times 2(p+m-1)(p+m-2)} t^{m-4}+...\right].
\ea\ee
\be
J_{-\fr} (z)=\sqrt{\frac{2}{\pi z}} \cos z
\ee
\be  \label{form25}
\G(\frac{n-2}{2}) \sum \limits_{m=0}^{\infty} i^m (m+\frac{n-2}{2})
\frac{J_{m+\frac{n-2}{2}}(t)}{(\frac{t}{2})^{\frac{n-2}{2}}}
\cC^{\frac{n-2}{2}}_m(x) =e^{itx}.  \ee

\section{Tensor fields of Weyl invariant dilaton gravity}\label{ApF}

The transformation laws \re{qtrans1} mix dilaton with metric and
therefore the dilaton $D(x)$ is not a conventional scalar field $D'(x)$ which
should transform under the Weyl dilations and $x$-diffeomorphisms
as follows
\be
\d D'(x) = (-\xi^m \pa_m + 2\a) D'(x)
\ee
As a consequence, the ``quantized volume'' action \re{actires} being expressed in terms of the original fields
which include dilaton $D$, though being by construction reparametrization and Weyl-invariant,
does not have the conventional form of gravity + matter systems. However, there exists the redefinition
\re{redef75} after which the action takes the conventional form. Here we prove the last statement.

Consider the ``quantized volume'' action \re{acti} in the quadratic ansatz case \re{quadr111}.
It is expanded in semiclassical power series \re{semicl3}, written schematically as
\be\label{act777}
\cA[H]=\cA_0[H]+\hbar^2\cA_2[H]+o(\hbar^4)
\ee
By the very construction, it is invariant w.r.t. similarity
transformations \re{qanstr1} which are expanded in power series in $\hbar$,
\be\label{var52}
\d H_2 (x,p) = \d_0 H_2 +\hbar^2 \d_2 H_2
\ee
Comparing two last formulae one finds
\be\ba\label{act7768}
\d_0 \cA_0 =0\\ \\ \d_0 \cA_2 +\d_2 \cA_0=0,
\ea\ee
where $\d_0,\d_2$ demote the variation of the action due to $\d_0, \d_2$ variations in \re{var52}.
The first row is just a classical invariance of the cosmological term. The second row is more interesting. It tells us
that the classical, $\d_0$, variation of $\cA_2$ is not a boundary term. This is because $\cA_2$ contains the ``bad term''
which does not transform like $``\sqrt{g}\times \mbox{(scalar of correct Weyl weight)} "$ w.r.t. diffeomorphisms $\xi^m(x)$ and Weyl dilations $\a(x)$.

Let us consider for the moment the diffeomorphisms only.
Then the dilaton's kinetic term $\sim\hbar^2 \int d^dx \sg D^{\hd-3} g^{mn} D_{,m} D_{,n}$ is
invariant, up to higher degrees in $\hbar^2$. Therefore, the only contribution to $\d_0 \cA_2$ comes from the $Y$-term which reads
\be  \ba\label{acy}
v_d \int d^dx \sg\left\{ -\frac{\hbar^2 d}{8}  D^{\hd-1} Y, \right\}
\ea \ee
($v_d$ is the volume of a unit ball) where $Y$ is an expression built entirely in terms of the metric and its derivatives
\re{yterm}.
Let us write down the variation of $Y$ w.r.t. diffeomorphisms as
\be
\d_0 Y =-\xi^m\pa_m Y +\tilde{\d} Y,
\ee
such that $\tilde{\d} Y$ measures ``non-scalarity'' of $Y$. Then the only contribution to $\d_0 \cA_2$ comes from
$\tilde{\d} Y$.

Despite of the non-invariance of $\cA_2$ w.r.t. classical gauge transformations $\d_0$ the
full action \re{act777} is invariant, that is reflected by the second row of \re{act7768}.
This means that the non-invariant, due to $\tilde{\d}Y $,
term in the classical variation of $\cA_2$ is compensated by the contribution coming from the
$\d_2$-variation of the classical term $\cA_0$.
In fact, the only $\d_2$ variations are due to $\hbar^2$-terms in the variation of $D(x)$ in \re{qtrans1}.
They cancel the $\d_0$-variation of $\cA_2$ , i.e. they cancel (up to a boundary term) the variation of \re{acy}
due-to non-scalarity of the $\d_0 Y$.
Writing the second row of \re{act7768} explicitly one gets
\be\ba \label{act8512}
0=\int d^dx \sg\left\{\hd D^{\hd-1} \d_2 D -\frac{d}{8}
D^{\hd-1} \d_0 Y \right\}+\mbox{boundary terms}=\\ \\=\int d^dx \sg\hd D^{\hd-1} \left\{
\d_2 D -\frac{1}{4}
\tilde{\d} Y \right\}+\mbox{boundary terms}
\ea\ee
The expression in braces in the second row does not depend on $D(x)$, therefore it is identically zero,
\be \label{act8513}
0=\d_2 D -\frac{1}{4}\tilde{\d}Y.
\ee
This is the result we need. It tells us that the $\d_2$ variation of the dilaton w.r.t. diffeomorphisms
may be cancelled by the $\tilde{\d}$ variation of $Y$. But as $Y$ depends on metric and its derivatives only while
for metric the full variation consists of $\d_0$ term, it appears that one can construct the modified scalar
$\tilde{D}(x)$ which transforms as a conventional scalar field w.r.t. diffeomorphisms,
\be\label{ch446}
\tilde{D}(x)=D(x)-\frac{\hbar^2}{4} Y,
\ee
\be
\d \tilde{D}(x)= -\xi^m \pa_m \tilde{D}(x).
\ee
Indeed, the $\d_2$ variation of $D(x)$ is cancelled by the $\tilde{\d}$ variation of $Y$.

Next, by direct calculations one gets the transformations properties w.r.t. Weyl dilations,
\be\ba
\d g^{mn}=2\a g^{mn},\;\;\d D =2\a D +\frac{\hbar^{2}}{2}g^{mn} \a_{,mn},\\ \\
\d Y=\frac{2}{3}(d-1)g^{mn} \a_{,mn}+ \frac{2}{3}(d-4)\a_{,n}(\frac{1}{\sg}\pa_m \sg g^{mn}),\\ \\
\d R =2(d-1)g^{mn} \a_{,mn}+2(d-1)\a_{,n}(\frac{1}{\sg}\pa_m \sg g^{mn})\\ \\
\d D-\frac{\hbar^2}{4}(Y+\k R)=2\a (D-\frac{\hbar^2}{4}(Y+\k R))+\\ \\
+\frac{\hbar^2}{2}g^{mn} \a_{,mn}(1-\frac{1}{3}(d-1)-\k(d-1)) +(\frac{1}{\sg}\pa_m \sg g^{mn})
(-\frac{1}{4})(\frac{2}{3}(d-4)+2\k(d-1)),
\ea\ee
for $R$ being the scalar curvature and $\k$ is an arbitrary, wherefrom it follows that if one takes
\be
\k=\frac{4-d}{3(d-1)}
\ee
then the modified scalar
\be
D'=D-\frac{\hbar^2}{4}(Y-\frac{d-4}{3(d-1)} R)
\ee
transforms as it is required in \re{qtrans2}.

\section{Free massless scalar and conformal group.}\label{ApG}

It is useful to describe free massless scalar, satisfying the Klein-Gordon
equation
\be\label{kg1}
\Box \psi(x) (x^m)=0,\; m=0,1,...d-1, x^m \in {\bf R}^{d-q,q}
\ee
by lifting this system to $d+2$ dimensional pseudo-Euclidean space with
signature $d-q+1,\;q+1$ parameterized by coordinates
$Y^A \in {\bf R}^{d-q+1,q+1},\; A=0,...d+1$.
The original $d$-dimensional space is realized as a factor space of the light
cone
\be Y^2\equiv Y^A Y_A =0 \ee by the equivalence relation \be \label{conedilat} Y^A \sim \l
Y^A, \l \in {\bf R}_+, \ee
The $so(d-q+1,q+1)$ generators are \be
M_{AB}=-i\hbar(Y_A \pa_B -Y_B \pa_A).
\ee
Let $U$ be an open domain in ${\bf R}^{d-q+1,q+1}$ which includes the light cone.
Consider the space of equivalence classes $Q_k$ of scalar fields on $U$ of the
form
\be\ba\label{eq753}
\Psi(Y)\sim \Psi(Y) +Y^2 c(Y),\\ \\ Y\pa_Y\; \Psi(Y)=k_d \Psi(Y),\;\;Y\pa_Y c(Y)=(k_d
-2)c(Y) \ea\ee
where $\f$ and $c$ are regular in $U$.
Clearly, the space of equivalence classes \re{eq753} is isomorphic to the space
of scalar densities on $\cM_{d-q,q}$ of a definite conformal weight.
After representing $Y^A, Y^2=0$ as $Y^A=v Y^A (x)$, where
$x$ are invariants of dilations \re{conedilat}, the canonical projection to the space of scalar densities is given by the
formula
\be
\p(x)=\Psi(Y)|_{v=1}.
\ee
Consider the equation
\be\label{dal45}
\pa_Y^2 \Psi(Y) =0.
\ee
Due to the identity
\be
\pa_Y^2 Y^2 c(Y)=(4Y\pa_Y+2(d+2)+Y^2\pa^2_Y )c(Y),
\ee
the equation \re{dal45} is well-defined on the in the space of
equivalence classes $Q_k$ if
\be
k_d=\frac{2-d}{2},
\ee
what we suppose henceforth. Therefore, the equation \re{dal45} may be expressed
in terms of the scalar density $\psi(x)$. It easy to get that
it is exactly the massless Klein-Gordon equation \re{kg1}. Thus, one has
the conformally covariant description of the free massless field in $d$
dimensions in terms of $d+2$ dimensional space of equivalence classes
$Q_{\frac{2-d}{2}}$. In this realization it is easy to derive the
identity
the generators of the conformal group satisfy,
\be\ba
M_{AB} {M^B}_C+M_{CB} {M^B}_A=\hbar^2\left(\eta_{AC}(d-2) +Y_A Y_C \pa_Y^2
+Y^2 \pa_A \pa_C \right) \Rightarrow \\ \\
(M_{AB} {M^B}_C+M_{CB} {M^B}_A )\Psi(Y) \sim \hbar^2 \eta_{AC}(d-2) \Psi(Y) \Rightarrow \\ \\
(M_{AB} {M^B}_C+M_{CB} {M^B}_A) \psi(x) = \hbar^2 \eta_{AC}(d-2) \psi(x)\Rightarrow
N_{AB} \psi(x)=0
\ea\ee

\section{Coordinate representation of an operator with a given Weyl symbol}\label{ApH}
\label{operweyl}
According to the textbook formulae \cite{berezin}, given the operator $\hat{f}$ with the Weyl symbol
$f(x,p)$, its action on the wave function $u(x)$ is represented by the integral formula
\be\label{weyl98}
(\hat{f}u)(x)=\iww \int d^d y d^dp\; e^{\frac{i}{\hbar}(x-y)p} f(\frac{x+y}{2}) u(y).
\ee
Let $f(x,p)$ be a power series in momenta,
\be
f(x,p)=\sum \limits_{s=0}^{\infty} f^{m(s)}(x) p_{m_1}...p_{m_s}.
\ee
Let us find its action on $u(x)$.
Introducing the variable $q=x-y$ and using integral representation for the $\d$-function and its derivatives, one represents the expression
\re{weyl98} in the form
\be\ba\label{weyl988}
(\hat{f} u)(x)= \sum \limits_{s=0}^{\infty}(i\hbar)^s \frac{\pa}{\pa q^{m_1}}...\frac{\pa}{\pa q^{m_s}}
[f^{m(s)}(x-\frac{q}{2})u(x-q)]|_{q=0}=\\ \\ \sum \limits_{s=0}^{\infty} \sum\limits_{k=0}^{s}
(-i\hbar)^s C^k_s (\fr)^k f^{m(k)m(s-k)}_{,m(k)}(x)u_{,m(s-k)}(x),
\ea\ee
where $C^k_s=\frac{s!}{k!(s-k)!}$ are binomial coefficients.

\section{Weyl spinor formalism in six dimensions}\label{ApI}

This appendix is taken from the paper \cite{Lyakhovich:1996ne}.
Notation is as follows: capital Latin letters are used for Minkowski
space indices and small Latin letters for spinor ones. The metric is chosen
in the form: $\eta _{AB}=diag(-,+,...,+)$. The Clifford algebra of $8\times
8 $ Dirac matrices $\Gamma _A$ reads: $\left\{ \Gamma _A,\Gamma _B\right\}
=-2\eta _{AB}$. The suitable representation for $\Gamma _A$ is
\begin{equation}
\label{ap1}\Gamma _A=\left(
\begin{array}{cc}
0 & (\sigma _A)_{a
\stackrel{.}{a}} \\ (\widetilde{\sigma }_A)^{\stackrel{.}{a}a} & 0
\end{array}
\right) \ ,\quad
\begin{array}{c}
\sigma _A=\left\{ 1,\gamma _0,i\gamma _1,i\gamma _2,i\gamma _3,\gamma
_5\right\} \\
\widetilde{\sigma }_A=\left\{ 1,-\gamma _0,-i\gamma _1,-i\gamma _2,-i\gamma
_3,-\gamma _5\right\}
\end{array}
\end{equation}
where $\gamma _i,i=0,1,2,3,5$ are the ordinary Dirac matrices in four
dimensions. The charge conjugation matrix is defined as
\begin{equation}
\label{ap2}C=\Gamma _2\Gamma _4=\left(
\begin{array}{cc}
I & 0 \\
0 & \widetilde{I}
\end{array}
\right) \ ,\quad I=\widetilde{I}=\left(
\begin{array}{ccc}
\begin{array}{cc}
0 & 1 \\
-1 & 0
\end{array}
& | & 0 \\
--- & | & --- \\
0 & | &
\begin{array}{cc}
0 & 1 \\
-1 & 0
\end{array}
\end{array}
\right)
\end{equation}

The spinor representation of $SO(5,1)$ on Dirac spinors $\Psi =\left(
\begin{array}{c}
\lambda _a \\
\overline{\pi }^{\stackrel{.}{b}}
\end{array}
\right) $ is generated by
\begin{equation}
\label{ap3}
\begin{array}{c}
\Sigma _{AB}=-\frac 14\left[ \Gamma _A,\Gamma _B\right] =\left(
\begin{array}{cc}
(\sigma _{AB})_a{}^b & 0 \\
0 & (\widetilde{\sigma }_{AB})^{\stackrel{.}{a}}{}_{\stackrel{.}{b}}
\end{array}
\right) = \\
\\
=\left(
\begin{array}{cc}
-\frac 14\left( \sigma _A{}_{a\stackrel{.}{a}}\widetilde{\sigma }_B{}^{%
\stackrel{.}{a}b}-\sigma _B{}_{a\stackrel{.}{a}}\widetilde{\sigma }_A{}^{%
\stackrel{.}{a}b}\right) & 0 \\
0 & -\frac 14\left( \widetilde{\sigma }_A{}^{\stackrel{.}{a}b}\sigma _B{}_{b%
\stackrel{.}{b}}-\widetilde{\sigma }_B{}^{\stackrel{.}{a}b}\sigma _A{}_{b%
\stackrel{.}{b}}\right)
\end{array}
\right)
\end{array}
\end{equation}
The representation is decomposed into two irreducible ones corresponding to
the left- and right-handed Weyl spinors. It turns out that the
representation (\ref{ap3}) and its complex conjugated are equivalent: ($%
\sigma _{AB}^{*})_{\stackrel{.}{a}}{}^{\stackrel{.}{b}}=I_{\stackrel{.}{a}%
}{}^a(\sigma _{AB})_a{}^bI_b{}^{\stackrel{.}{b}},\ (\widetilde{\sigma }%
_{AB}^{*})^a{}_b=\widetilde{I}{}^a{}_{\stackrel{.}{a}}(\widetilde{\sigma }%
_{AB})^{\stackrel{,}{a}}{}_{\stackrel{.}{b}}\widetilde{I}{}^{\stackrel{.}{b}%
}{}_b$. So, one can convert the dotted spinor indices into undotted ones%
$$
\overline{\lambda }_a=I_a{}^{\stackrel{.}{a}}\stackrel{*}{\lambda }_{%
\stackrel{.}{a}}\quad ,\qquad \overline{\pi }^a=\widetilde{I}{}^a{}_{%
\stackrel{.}{a}}\stackrel{*}{\pi }{}^{\stackrel{.}{a}}
$$
While the gradient and contragredient representations are inequivalent
because of absence of an object raising and/or lowering spinor indices as
distinguished from the four-dimensional case. It is convenient to turn from
the matrices ($\sigma _A)_{a\stackrel{.}{a}},(\widetilde{\sigma }_A)^{%
\stackrel{.}{a}a}$ to ($\sigma _A)_{ab}=(\sigma _A)_{a\stackrel{.}{a}}%
\widetilde{I}{}^{\stackrel{.}{a}}{}_b,(\widetilde{\sigma }_A)^{ab}=%
\widetilde{I}{}^a{}_{\stackrel{.}{a}}(\widetilde{\sigma }_A)^{\stackrel{.}{a}%
a}$. They possess a number of relations
\begin{equation}
\label{ap4}
\begin{array}{c}
\begin{array}{cc}
(\sigma _A)_{ab}=-(\sigma _A)_{ba}{} & {}(\widetilde{\sigma }_A)^{ab}=-(%
\widetilde{\sigma }_A)^{ba}
\end{array}
\\
\\
\begin{array}{cc}
(\sigma _A)_{ab}{}(\sigma ^A)_{cd}=-2\epsilon _{abcd}{} & {}(\widetilde{%
\sigma }_A)^{ab}(\widetilde{\sigma }^A)^{cd}=-2\epsilon ^{abcd}
\end{array}
\\
\\
\begin{array}{cc}
(\sigma _A)_{ab}=-\frac 12\epsilon _{abcd}(\widetilde{\sigma }_A)^{cd}{} &
{}(\widetilde{\sigma }_A)^{ab}=-\frac 12\epsilon ^{abcd}(\sigma ^A)_{cd}
\end{array}
\\
\\
(\sigma _A)_{ab}(
\widetilde{\sigma }^A)^{cd}=2\left( \delta _a{}^c\delta _b{}^d-\delta
_a{}^d\delta _b{}^c\right) \ ,\quad (\sigma _A)_{ab}(\widetilde{\sigma }%
_B)^{ba}=-4\eta _{AB} \\  \\
(\sigma _A)_{ab}(
\widetilde{\sigma }_B)^{bc}+(\sigma _B)_{ab}(\widetilde{\sigma }%
_A)^{bc}=-2\eta _{AB}\delta _a{}^c \\  \\
(\widetilde{\sigma }_A)^{ab}(\sigma _B)_{bc}+(\widetilde{\sigma }%
_B)^{ab}(\sigma _A)_{bc}=-2\eta _{AB}\delta ^a{}_c
\end{array}
\end{equation}
Here we introduced two invariant tensors $\epsilon _{abcd}$ and $\epsilon
^{abcd}$, totally antisymmetric in indices and $\epsilon _{1234}=\epsilon
^{1234}=1$. With the aid of introduced objects one may convert the vector
indices into antisymmetric pairs of spinor ones. E.g. for a given vector $%
p_A$%
\begin{equation}
\label{ap5}p_A\rightarrow p_{ab}=p_A(\sigma ^A)_{ab}\ ,\quad p^{ab}=p_A(%
\widetilde{\sigma }^A)^{ab}\ ,\quad p_A=-\frac 14p_{ab}(\widetilde{\sigma }%
_A)^{ba}=-\frac 14p^{ab}(\sigma _A)_{ba}
\end{equation}

\end{document}